\documentclass[aps,prd,reprint,superscriptaddress,nofootinbib]{revtex4-1}
\usepackage[all]{xy}
\usepackage{amsmath,amsthm,amssymb}
\usepackage[dvips]{graphicx}
\usepackage{comment}
\usepackage{array}
\usepackage{bm}
\allowdisplaybreaks[1]

\newtheorem*{lemma}{Lemma}

\usepackage{color}
\usepackage{hyperref}

\theoremstyle{plain}

\renewcommand{\P}{\mathsf{P}}
\newcommand{\Q}{\mathsf{Q}}

\newcommand{\beq}{\begin{equation}}
\newcommand{\eeq}{\end{equation}}

\newcommand{\rmin}{r_{\rm min}}
\newcommand{\rmax}{r_{\rm max}}

\renewcommand{\BibitemShut}[1]{}

\newcommand{\SP}[1]{\left(#1\right)}
\newcommand{\bP}[1]{\big(#1\big)}
\newcommand{\BP}[1]{\Big(#1\Big)}

\newcommand{\BB}[1]{\Big[#1\Big]}

\newcommand{\tet}[2]{e_{#1}^{#2}}				

\begin{document}
\title{Completion of metric reconstruction for a particle orbiting a Kerr black hole}
\author{Cesar Merlin}
\affiliation{Mathematical Sciences, University of Southampton, Southampton, SO17 1BJ, United Kingdom}
\author{Amos Ori}
\affiliation{Department of Physics, Technion-Israel Institute of Technology, Haifa 32000, Israel}
\author{Leor Barack}
\affiliation{Mathematical Sciences, University of Southampton, Southampton, SO17 1BJ, United Kingdom}
\author{Adam Pound}
\affiliation{Mathematical Sciences, University of Southampton, Southampton, SO17 1BJ, United Kingdom}
\author{Maarten van de Meent}
\affiliation{Mathematical Sciences, University of Southampton, Southampton, SO17 1BJ, United Kingdom}

\begin{abstract}
Vacuum perturbations of the Kerr metric can be reconstructed from the corresponding perturbation in either of the two Weyl scalars $\psi_0$ or $\psi_4$, using a procedure described by Chrzanowski and others in the 1970s. More recent work, motivated within the context of self-force physics, extends the procedure to metric perturbations sourced by a particle in a bound geodesic orbit. However, the existing procedure leaves undetermined a certain stationary, axially-symmetric piece of the metric perturbation. In the vacuum region away from the particle, this ``completion'' piece corresponds simply to mass and angular-momentum perturbations of the Kerr background, with amplitudes that are, however, a priori unknown. Here we present and implement a rigorous method for finding the completion piece. The key idea is to impose continuity, off the particle, of certain gauge-invariant fields constructed from the full (completed) perturbation, in order to determine the unknown amplitude parameters of the completion piece. We implement this method in full for bound (eccentric) geodesic orbits in the equatorial plane of the Kerr black hole. Our results provide a rigorous underpinning of recent results by Friedman {\it et al.}\ for circular orbits, and extend them to non-circular orbits.
\end{abstract}

\date{\today}
\maketitle
\section{Introduction}

Gravitational perturbations of the Kerr geometry are often studied within the null-tetrad framework of Newman and Penrose, using Teukolsky's formalism \cite{Teuk}. In this approach one does not work with the metric perturbation directly, but instead one considers the perturbations in the Weyl curvature scalars $\psi_0$ or $\psi_4$ as proxies. The perturbation equations governing these scalars are fully separable by means of a (spin-weighted) spheroidal-harmonic and Fourier decomposition, and thus conveniently reduce to a set of decoupled ordinary differential equations. In some problems, however, one is interested in the metric perturbation itself. One such problem of contemporary interest is that of calculating the gravitational self-force acting on an orbiting particle \citep{poisson,Barack:2009ux}, in which knowledge of the full local metric perturbation near the particle is required. In such problems one faces the challenge of {\em metric reconstruction}: Given the (harmonic modes of the) perturbation in $\psi_0$ or $\psi_4$, how does one recover the corresponding metric perturbation?

A reconstruction procedure for {\em vacuum} perturbations was developed long ago in papers by Chrzanowski \citep{chrza}  and Cohen and Kegeles \citep{cohen79}, with further contributions from Wald \citep{Waldrec}, Stewart \citep{Stewart:1978tm}, and (more recently) Lousto and Whiting \citep{Lousto:2002em}; in keeping with common nomenclature we shall refer to it here as the CCK procedure. The procedure yields a vacuum metric perturbation in (one of two) particular, traceless ``radiation'' gauges [cf.\ Eq.\ (\ref{gaugecondition})]. The reconstructed perturbation is determined only up to a 4-parameter family of Petrov type D vacuum perturbations \citep{waldtheo}, representing (i) perturbations into Kerr geometries of a different mass or (ii) a different angular-momentum, and perturbations away from Kerr into (iii) Kerr-Newman-Tamburino-Unti (Kerr-NUT) or (iv) C-metric geometries. These perturbations are all stationary and axisymmetric. In the vacuum case, Kerr-NUT and C-metric perturbations are ruled out based on regularity \citep{waldtheo}, but the mass and angular-momentum perturbations remain arbitrary within the CCK procedure. These two ``missing'' pieces of the metric perturbation must be determined separately [e.g., in the vacuum problem, through conditions imposed on the total Arnowitt--Deser--Misner (ADM) mass and angular momentum of the spacetime].  We shall refer to the task of fixing the missing pieces as the {\em completion} of the reconstruction procedure, and to the missing pieces themselves as the ``completion'' part of the perturbation.

The CCK procedure is no longer directly applicable in the non-vacuum case, with the root cause of complication being the inconsistency of the (traceless) radiation gauge condition with the linearized Einstein's equations when matter sources are present \citep{PriceThesis,Price:2006ke}. Notably, in the presence of sources, the (mode-sum based) CCK procedure fails to return a valid solution not only within the matter region but also at vacuum points {\em away} from any sources \cite{barack1,Ori:2002uv,BMP1}. With the self-force problem as a prime motivation, Ori \citep{Ori:2002uv} devised a reconstruction procedure for perturbations sourced by a point particle in a bound orbit around a Kerr black hole. Specifically, he prescribed the reconstruction of a (radiation-gauge) metric perturbation in the vacuum regions $r>r_{\rm p}(t)$ and $r_{+}<r<r_{\rm p}(t)$, where $r=r_{\rm p}(t)$ is the radial location of the particle and $r=r_{+}$ the horizon's radius; we hereafter adopt standard Boyer-Linquist coordinates $\{t,r,\theta,\varphi\}$. Ori showed that the analytical extension of the solution from either vacuum region across $r=r_{\rm p}(t)$ produces a string-like gauge singularity that extends radially from the particle into the opposite vacuum domain. 

Later, Friedman, Keidl, Shah (FKS) and collaborators \citep{Keidlhom,friedman1,friedman2,friedman3} prescribed an alternative reconstruction, specialized to circular equatorial orbits of radius $r=r_0$, in which the singularities were replaced with a gauge discontinuity (and a delta function) on the sphere $r=r_0$.\footnote{The irregularity of the FKS reconstructed metric on the sphere $r=r_0$ was highlighted in Ref.\ \citep{BMP1}, referring to the FKS gauge as the ``no-string'' gauge.} The procedure was recently generalized by Van de Meent and Shah to any bound equatorial orbits \cite{vandeMeent:2015lxa}, using the method of extended homogeneous solutions \cite{Barack:2008ms}.  Motivated by these developments, Pound {\it et al.}~\citep{BMP1} obtained a rigorous formulation of the self-force, complete with a practical mode-sum calculation formula, starting from a reconstructed metric perturbation in either Ori's or FKS's approach.

The self-force formulation of Ref.\ \citep{BMP1} assumes that one knows how to complete the metric reconstruction; in general, the completion piece has an important contribution to the local self-force experienced by the particle. However, how to obtain the completion piece remains an open problem, in general.\footnote{The two recent numerical implementations of the Pound {\it et al.} formulation---by Merlin and Shah \cite{MerSha} in Schwarzschild and by Van de Meent \cite{vandeMeent:2016pee} in Kerr---apply the completion determined in the current paper.} Keidl {\it et al.} show, in \citep{Keidlhom,friedman1}, that Kerr-NUT and C-metric perturbations must be excluded for regularity reasons even in the particle case; and they derive the remaining, physical completion piece in the case of circular equatorial geodesic orbits. However, their calculation is restricted to that class of orbits, and their method relies on certain assumptions that are yet to be confirmed (see below). Our goal here is to describe a general, rigorous method for deriving the completion piece for bound orbits in Kerr geometry, and we will go on to implement it for generic (bound) orbits in the equatorial plane. We will thereby confirm and extend the results of Keidl {\it et al.}, and supply a necessary ingredient to enable self-force calculations from a reconstructed metric. 

For a particle in a bound orbit, the task of completion takes the following simple form. Let ${\cal S}^+$ and ${\cal S}^-$ denote, respectively, the two vacuum regions $r>r_{\rm p}(t)$ and $r_+\leq r<r_{\rm p}(t)$, and let $h^{\rm rec\pm}_{\alpha\beta}$ represent the piece of the metric perturbation obtained by applying the reconstruction procedure in the respective domains ${\cal S}^{\pm}$ (with the usual, retarded boundary conditions).  We refer here specifically to an FKS-like ``no-string'' reconstruction (as implemented most recently in \cite{MerSha,vandeMeent:2015lxa,vandeMeent:2016pee}), in which $h^{\rm rec\pm}_{\alpha\beta}$ are each regular in their respective vacuum domains. The full, completed metric perturbation in each of ${\cal S}^{\pm}$ is given by
\begin{equation}
h^{\pm}_{\alpha\beta}=h^{\rm rec\pm}_{\alpha\beta}+h^{\rm comp\pm}_{\alpha\beta},
\end{equation}
where $h^{\rm comp\pm}_{\alpha\beta}$ are the completion pieces in the respective domains. The latter have the form 
\begin{equation}\label{h_comp}
h^{\rm comp\pm}_{\alpha\beta}= {\cal E}^{\pm}h^{(\delta M)}_{\alpha\beta}+{\cal J}^{\pm}h^{(\delta J)}_{\alpha\beta},
\end{equation}
where ${\cal E}^{\pm}$ and ${\cal J}^{\pm}$ are constant coefficients (depending only on the details of the orbit), and
$h^{(\delta M)}_{\alpha\beta}$ and $h^{(\delta J)}_{\alpha\beta}$ are certain homogeneous, stationary and axisymmetric perturbations representing, respectively, mass and angular-momentum perturbations of the Kerr geometry. These two perturbations can be readily written down in analytic form (fixing the gauge and the overall normalization), as we do in Eqs.\ (\ref{eq:dMexplicit}) and (\ref{eq:dJexplicit}) below. The problem of completion thus reduces to that of determining the values of the four coefficients ${\cal E}^{\pm},{\cal J}^{\pm}$. In fact, ${\cal E}^+$ and ${\cal J}^+$ may be readily deduced from global conditions on the total mass and angular-momentum contents of the system (this will be described in Sec.\ \ref{s:asymp}), so the problem further reduces to that of determining ${\cal E}^{-}$ and ${\cal J}^{-}$ alone, or, equivalently, the two differences 
\begin{equation}\label{diffEJ}
[{\cal E}]:={\cal E}^{+}-{\cal E}^{-},
\quad\quad
[{\cal J}]:={\cal J}^{+}-{\cal J}^{-}.
\end{equation}

In this work we propose and implement a new strategy for determining $[{\cal E}]$ and $[{\cal J}]$. The basic idea is as follows. Let $\cal S$ represent the (2+1-dimensional) surface $r=r_p(t)$ that is the interface between ${\cal S}^+$ and $\cal S^{-}$. The particle's orbit traces a timelike curve $\gamma$ in $\cal S$, and we let $\check{\cal S}:={\cal S}-\gamma$, i.e.\ $\check{\cal S}$ is the part of $\cal S$ excluding the particle's orbit. 
{\em Our strategy is based on the expectation that gauge-invariant fields constructed from the full, physical perturbation must be smooth everywhere but on the particle, and, in particular, they must  be smooth on $\check{\cal S}$}.  Thus, we construct a suitable set of (real) invariant fields ${\cal I}_n^{\pm}$ ($n=1,2,\ldots$) corresponding to the full perturbation $h^{\pm}_{\alpha\beta}$, and require that ${\cal I}_n^{+}={\cal I}_n^{-}$ on $\check{\cal S}$, for each $n$.
This continuity requirement translates to a set of simple algebraic equations for $[{\cal E}]$ and $[{\cal J}]$, which are then solved. Since there are two unknowns, we require two independent matching conditions. This can be achieved by imposing ${\cal I}_n^{+}={\cal I}_n^{-}$ for a pair of independent invariants (say ${\cal I}_1$ and ${\cal I}_2$) at an arbitrarily chosen point of $\check{\cal S}$; or, possibly, by imposing continuity of a single invariant (say ${\cal I}_1$) at two different longitudinal points of $\check{\cal S}$. We shall confirm that the two procedures give identical results, and, indeed, that they each automatically guarantee the continuity of all invariants ${\cal I}_n$ on the entire surface $\check{\cal S}$.

Since the completion piece $h^{\rm comp\pm}_{\alpha\beta}$ is stationary and axisymmetric, in the above calculation we need only concern ourselves with the stationary and axisymmetric piece of $h^{\pm}_{\alpha\beta}$. Since $h^{\rm comp\pm}_{\alpha\beta}$ is given in a simple analytic form, the main calculation task, therefore, is to derive the stationary and axisymmetric piece of the reconstructed metric $h^{\rm rec\pm}_{\alpha\beta}$. The reconstruction procedure yields individual multipole ($\ell$-)modes of $h^{\rm rec\pm}_{\alpha\beta}$, and the main challenge is in the evaluation of the sum of multipole contributions. We show how this can be done analytically. In fact, the stationarity and axial symmetry of the relevant perturbation enable us to perform the entire calculation analytically, even for non-circular orbits.   

We note the distinction between the task of completion and the (more ambitious) task of constructing a metric perturbation $h_{\alpha\beta}$ in a gauge in which it is globally smooth (except on the particle). Even after completion, our perturbation will in general fail to be continuous on $\check{\cal S}$. This discontinuity can, in principle, be removed with a suitable gauge transformation, but here we do not pursue this additional task of ``gauge regularization''. Whether a gauge regularization is required in practice depends on the particular application, and sometimes it suffices to gauge-regularize only some relevant piece of the perturbation; we shall discuss a few examples in the concluding section of this paper. 
We intend to present a systematic treatment of gauge regularization in a future work. 

Finally, we note that our calculation, and the completion perturbation that comes out of it, apply specifically for a reconstruction done in the so-called ``ingoing'' radiation gauge [see Eq.\ (\ref{gaugecondition})]. To determine the completion for a reconstruction in the companion ``outgoing'' gauge would require a separate calculation, which we have not carried out (though we expect it to be entirely analogous to the calculation presented here). 

The structure of this paper is as follows. In Sec.\ \ref{s:transpsi2} we present our set of auxiliary gauge-invariant quantities ${\cal I}_n$. In Sec.\ \ref{s:Schcir}, as a warm-up exercise, we perform our completion calculation and determine $[{\cal E}]$ and $[{\cal J}]$ for circular geodesic orbits in Schwarzschild spacetime. Section \ref{s:Kerrcirc} extends the calculation to circular equatorial geodesic orbits in Kerr spacetime, and Sec.\ \ref{s:ecceqorbits} extends it further to all bound (eccentric) geodesic orbits in the equatorial plane in Kerr. In Section \ref{s:asymp} we use asymptotic analysis at spatial infinity in order to determine the completion amplitudes ${\cal E}^+$ and ${\cal J}^+$, and consequently, using our now-known values of $[{\cal E}]$ and $[{\cal J}]$, also the amplitudes ${\cal E}^-$ and ${\cal J}^-$.  Section \ref{s:summary} contains a summary and a discussion of remaining issues and generalizations. Some of the technical details of our calculation are relegated to appendices. 

Our conventions for the Newman-Penrose formalism and for the reconstruction procedure follow those of Ref.\ \cite{vandeMeent:2015lxa}. In particular, we adopt the metric signature ${-}{+}{+}{+}$ (unlike, e.g., FKS and much of the early Newman-Penrose literature). 
For convenience, we give in Appendix \ref{a:KerrBack} a full review of vacuum reconstruction using our conventions. We use geometrized units with $G=c=1$ throughout. 

In the rest of this introduction we review previous attempts at the completion problem, and describe some other relevant work. We highlight the way in which our method differs from that of earlier work.

\subsection{Survey of previous, related work}

An initial investigation of the completion problem for particle sources was carried out by L.\ Price (unpublished thesis, \citep{PriceThesis}). Specializing to a Schwarzschild background, Price attempted to determine the completion piece through the requirement that $h^{\rm comp+}_{\alpha\beta}$ matched smoothly with $h^{\rm comp-}_{\alpha\beta}$ on $\check{\cal S}$ (allowing for arbitrary gauge transformations on either sides of the surface). In Kerr, this procedure only makes sense under the unproven assumption that the reconstructed part $h^{\rm rec}_{\alpha\beta}$ is itself smooth on $\check{\cal S}$ (up to a gauge transformation). In our method we instead impose continuity (up to gauge) of the {\it full} (completed) perturbation, so need not resort to making such an assumption. Also, as described above, we impose continuity of certain invariant fields and not of the (gauge dependent) metric perturbation. This way we evade the arduous task of gauge regularization, which is unnecessary for the sole purpose of determining $h^{\rm comp}_{\alpha\beta}$. 

In their series of papers pioneering the radiation-gauge approach to the self-force, FKS have tackled the problem of determining the completion piece for circular geodesic orbits in the equatorial plane (first in Schwarzschild \citep{Keidlhom,friedman1,friedman2} and later in Kerr \citep{friedman3}). Their treatment invokes the Komar definitions of energy and angular momentum as applied to the stationary and axisymmetric piece of the perturbed spacetime: The amplitudes ${\cal E}^{\pm}$ and ${\cal J}^{\pm}$ are determined (essentially) by fixing the Komar mass and angular momentum of the perturbed spacetime at $r\to\infty$ and on the black hole's horizon. It is implicitly assumed, however, that the reconstructed piece $h^{\rm rec}_{\alpha\beta}$ has no contribution to the Komar quantities. This is readily justified in the Schwarzschild case, where the mass and angular momentum content of the perturbation is contained entirely in its monopole and dipole modes (which have no contribution from $h^{\rm rec}_{\alpha\beta}$). But, to the best of our knowledge, the assumption remains unproven in the Kerr case. The calculation to be presented in the current paper will indirectly establish the validity of FKS's assumption. 

In a slightly different context, Dolan and Barack \citep{dolan3}\footnote{Ref.\ \citep{dolan3} discusses a direct calculation of the metric perturbation (in the Lorenz gauge) via numerical time evolution of the linearized Einstein's equations. The problem of completion takes a different form within this treatment, the main issue being the mitigation of gauge instabilities that affect the stationary and axisymmetric part of the perturbation.} recently discussed an alternative method for determining the mass and angular-momentum content of an arbitrary region of perturbed space, building on work by Abbott and Deser \citep{AbbDes}. The Abbott--Deser formulation relies only on the existence of time-translation and rotational Killing symmetries in the {\it background} spacetime, and is thus applicable to a general perturbed Kerr geometry. The method prescribes certain conserved quantities (one for each background Killing field), which are constructed from the metric perturbation and its first derivatives, integrated over a closed 2-surface on a spacelike hypersurface. This provides a quasi-local definition of the energy and angular-momentum content of the volume enclosed within the surface, which can be shown to coincide with standard definitions (e.g., ADM's) in the appropriate limits.  One can imagine using this method to determine the completion amplitudes ${\cal E}^{\pm}$ and ${\cal J}^{\pm}$ by fixing the Abbott--Deser mass and angular momentum of the completed perturbation at infinity and on the horizon.
We have attempted this approach, but found the necessary surface integrals, and summation over modes, very hard to evaluate in practice (except at infinity). Thus, we have not been able to use this method for determining ${\cal E}^{\pm}$ and ${\cal J}^{\pm}$. Nonetheless, we think that, with some further development, the approach may provide a viable alternative to (and a check on) our method. 

An essentially equivalent completion problem was recently studied by Sano and Tagoshi, who considered the stationary and axisymmetric configuration of a rotating circular mass ring around a Schwarzschild \citep{SaTa} or a Kerr \citep{SaTa2} black hole.
Their analysis, like ours, seeks to obtain $[{\cal E}]$ and $[{\cal J}]$ from continuity conditions imposed outside the matter source. However, Sano and Tagoshi do not employ gauge-invariant quantities as in our method, and instead require continuity of the metric perturbation and of the (gauge dependent) Weyl scalars $\psi_1$, $\psi_2$ and $\psi_3$. In their construction, the completed metric perturbation and Weyl scalars are smooth on the sphere $r=r_0$ (where $r_0$ is the ring's radius), off the ring itself, but are singular on the equatorial plane outside the ring. Due to the remaining singularity, it remains unclear whether the prescribed completion is unique. As we will demonstrate in the current paper (for a point particle source), the completion is determined uniquely by looking at invariant quantities that must be smooth everywhere in the vacuum region.

\section{Auxiliary gauge invariants}\label{s:transpsi2}

In this section we prescribe several useful gauge-invariant quantities ${\cal I}_n(h_{\alpha\beta})$ ($n=1,2,\ldots$) constructed from a generic metric perturbation $h_{\alpha\beta}$ given in an arbitrary gauge.  Each of the fields ${\cal I}_n$ is a (real-valued) differential functional of the metric perturbation, involving at most third derivatives of $h_{\alpha\beta}$. Our invariants (unlike the ``radiative'' Weyl scalars $\psi_0$ and $\psi_4$) encode information about the mass and angular-momentum content of the perturbation, in a way that makes them useful for our purpose of determining the completion piece---as will be described in subsequent sections.  Our construction assumes a Kerr background with mass parameter $M$ and spin parameter $a\ne 0$. The Schwarzschild case, $a=0$, requires a separate treatment and will be considered in subsection \ref{s:InvSch}.

Of the five (complex) Weyl curvature scalars [see Eq.\ \eqref{eq:psi} for definitions, and Appendix \ref{a:KerrBack} for a review], only $\psi_2$ is nonzero in the background Kerr geometry:
\begin{equation}\label{psi20}
\psi^{(0)}_2=\varrho^3 M,
\end{equation}
where 
\begin{equation}\label{varrho}
\varrho=-(r-ia\cos\theta)^{-1},
\end{equation}
and hereafter a superscript `(0)' denotes the background value of a field defined in the perturbed spacetime. The linear perturbation of $\psi_2$, which we denote by $\psi_2^{(1)}$, is gauge-dependent. Under a first-order gauge transformation $ x^\alpha\rightarrow x^\alpha+\xi^\alpha$ it transforms according to 
\begin{equation}\label{psi2trans}
\psi_2^{(1)}\to \psi_2^{(1)}-\xi^{\alpha}\psi^{(0)}_{2,\alpha},
\end{equation}
where a comma denotes partial differentiation.

Our construction is based on identifying a {\it reference gauge} in which the linear perturbation of $\psi_2$ vanishes: $\tilde\psi_2^{(1)}=0$; we hereafter use an overtilde to indicates values in the reference gauge.\footnote{Our reference gauge has been employed at least once earlier in the literature (for a different purpose)---see Sec.\ 82 of Chandrasekhar's monograph \cite{chand}.} For a perturbation $h_{\alpha\beta}$ in some given (but arbitrary) gauge, let $\tilde \xi$ be the generator of a transformation to the reference gauge. By our definition of the reference gauge, $\tilde\xi$ satisfies 
\begin{equation}\label{xieq}
\tilde\xi^{\alpha}\psi^{(0)}_{2,\alpha}=\psi_2^{(1)},
\end{equation}
where, on the right-hand side, $\psi_2^{(1)}$ is the perturbation associated with $h_{\alpha\beta}$ in the original gauge. Recalling Eq.\ (\ref{psi20}), and that $\psi^{(0)}_{2}$ and $\psi_2^{(1)}$ are complex, we observe that Eq.\ (\ref{xieq}) constitutes a complex algebraic equation for the two real components $\tilde\xi^r$ and $\tilde\xi^\theta$. The solutions read
\begin{equation}
\tilde{\xi}^{r}=\mathrm{Re}(\Phi),\qquad\quad\tilde{\xi}^{\theta}=\frac{\mathrm{Im}(\Phi)}{a\sin\theta}, \label{eq:gauge9}
\end{equation}
where $\Phi:=\psi_{2}^{(1)}/(3M\varrho^{4})$.
This prescribes the gauge transformation from an arbitrary original gauge to our reference gauge; the components $\tilde\xi^t$ and $\tilde \xi^\varphi$ remain arbitrary.  An important consequence is that the condition $\tilde\psi_2^{(1)}=0$ can be said to {\em fix} the reference gauge, up to gauge transformations in the $t\varphi$ plane. 

Now consider the components $\tilde h_{\alpha\beta}$ of the metric perturbation in the reference gauge. Four of the components, namely $\tilde h_{rr}$, $\tilde h_{r\theta}=\tilde h_{\theta r}$ and $\tilde h_{\theta\theta}$, are completely determined by $\tilde\xi^r$ and $\tilde \xi^\theta$ (independently of $\tilde\xi^t$ and $\tilde \xi^\varphi$): We have 
\begin{equation}
\tilde{h}_{ab}=h_{ab}-2\tilde{\xi}_{(a,b)}+2\Gamma_{ab}^{(0)c}\tilde{\xi}_{c}, 
\label{eq:hab00}
\end{equation}
where henceforth the indices $a,b,c$ run over $\{r,\theta\}$ only, and parenthetical indices are symmetrized 
[in this example, $\tilde\xi_{(a,b)}=(\tilde\xi_{a,b}+\tilde\xi_{b,a})/2$].  In Eq.\ (\ref{eq:hab00}), $\Gamma_{ab}^{(0)c}$ are Christoffel symbols associated with the background (Kerr) metric $g_{\alpha\beta}^{(0)}$, and we have used $\Gamma_{ab}^{(0)t}= 0=\Gamma_{ab}^{(0)\varphi}$. The covariant components $\tilde\xi_a$ are given by
$\tilde\xi_r=g^{(0)}_{rr}\tilde\xi^r=(\Sigma/\Delta)\tilde\xi^r$ and
$\tilde\xi_\theta=g^{(0)}_{\theta\theta}\tilde\xi^\theta=\Sigma\tilde\xi^\theta$,
where $\Sigma:=r^2+a^2\cos^2\theta$ and $\Delta:=r^2-2Mr+a^2$. Note that the right-hand side of Eq.\ (\ref{eq:hab00}) is, implicitly, a linear combination of the metric perturbation $h_{\alpha\beta}$ and its first, second and third derivatives. This can be made explicit using Eq.\ (\ref{eq:gauge9}) and the second-order differential operation that produces $\psi_2^{(1)}$ out of ${h}_{\alpha\beta}$ (and $g_{\alpha\beta}^{(0)}$).

Note further that the values of the components $\tilde{h}_{ab}$ are completely fixed (because the reference gauge is fixed up to transformations in the $t\varphi$ plane, which, however, do not affect $\tilde{h}_{ab}$). In other words, the right-hand side of Eq.\ (\ref{eq:hab00}) describes {\em gauge-invariant} combinations of the perturbation $h_{\alpha\beta}$ and its derivatives. There is one such invariant combination for each component $\tilde{h}_{ab}$, i.e., three independent invariants in total: $\tilde h_{rr}$, $\tilde h_{\theta\theta}$ and (say) $\tilde h_{r\theta}$.
It may sound confusing that components of the metric perturbation in a particular gauge are said to be gauge-invariant. To avoid such confusion, it is useful at this point to dispose with the notion of a reference gauge and simply think of $\tilde h_{ab}$ as gauge-invariant functionals of $h_{\alpha\beta}$, i.e., the metric perturbation in an arbitrary gauge.
To reinforce this perspective, we introduce the renaming 
\begin{equation}\label{eq:I_Kerr}
\{{\cal I}_1,{\cal I}_2, {\cal I}_3\}:=\{\tilde h_{rr},\tilde h_{\theta\theta},\tilde h_{r\theta} \},
\end{equation}
and recall that the fields ${\cal I}_n$ are constructed from $h_{\alpha\beta}$ using Eqs.\ (\ref{eq:gauge9}) and (\ref{eq:hab00}).

It is straightforward to confirm the gauge invariance of ${\cal I}_n(h_{\alpha\beta})$ with a direct calculation, as follows. Under an arbitrary gauge transformation $x^\alpha\rightarrow x^\alpha+{\xi}^{\alpha}$, the relevant components of the metric perturbation in the original gauge transform according to $h_{ab}\rightarrow h_{ab}+\delta_\xi h_{ab}$, with  
\begin{equation}
\delta_\xi h_{ab}=-2{\xi}_{(a,b)}+2\Gamma_{ab}^{(0)c}{\xi}_{c},
\label{eq:hab11}
\end{equation}
where we have again used $\Gamma_{ab}^{(0)t}= 0=\Gamma_{ab}^{(0)\varphi}$.
The perturbation in $\psi_2$ transforms as in Eq.\ (\ref{psi2trans}), namely $\psi_2^{(1)}\rightarrow\psi_2^{(1)}+\delta_\xi \psi_2^{(1)}$, with 
\begin{equation}\label{eq:psi2trans2}
\delta_\xi \psi_2^{(1)}=-{\xi}^{\alpha}\psi^{(0)}_{2,\alpha}.
\end{equation}
The quantities $\tilde h_{ab}$ in Eq.\ (\ref{eq:hab00}) transform, in turn, as
$\tilde h_{ab}\rightarrow \tilde h_{ab}+\delta_\xi\tilde h_{ab}$, with
 \begin{equation}
\delta_\xi\tilde{h}_{ab}=\delta_\xi h_{ab}-2\delta_\xi\tilde{\xi}_{(a,b)}+2\Gamma_{ab}^{(0)c}\delta_\xi\tilde{\xi}_{c}, 
\label{eq:hab22}
\end{equation}
where 
$\delta_\xi\tilde{\xi}_{c}$ is the gauge change in $\tilde{\xi}_{c}$, and we have once more used $\Gamma_{ab}^{(0)t}= 0=\Gamma_{ab}^{(0)\varphi}$. 
To calculate $\delta_\xi\tilde{\xi}_{c}$, use Eq.\ (\ref{xieq}) to obtain $\delta_\xi\tilde\xi^{\alpha}\psi^{(0)}_{2,\alpha}=\delta_\xi\psi_2^{(1)}$, which, combined with Eq.\ (\ref{eq:psi2trans2}), then gives
\begin{equation}
\delta_\xi\tilde\xi^{\alpha}\psi^{(0)}_{2,\alpha}=-{\xi}^{\alpha}\psi^{(0)}_{2,\alpha}.
\end{equation}
This equation admits a unique solution for the two components $\delta_\xi\tilde\xi^{a}$, given by $\delta_\xi\tilde\xi^{a}=-\xi^a$. Hence also 
\begin{equation}\label{deltaxi}
\delta_\xi\tilde\xi_{a}=-\xi_a.
\end{equation}
Substituting from Eqs.\ (\ref{deltaxi}) and (\ref{eq:hab11}) into (\ref{eq:hab22}) gives
\begin{equation}
\delta_\xi\tilde{h}_{ab}=0,
\end{equation}
which establishes the invariance of ${\cal I}_n$ under arbitrary gauge transformations.

\subsection{Schwarzschild case}\label{s:InvSch}

In the case of a Schwarzschild background, $a=0$, Eq.\ (\ref{psi20}) gives $\psi_2^{(0)}=-M/r^3$, and $\psi_2^{(0)}$ is a real field. It follows immediately (recalling also the general invariance of $\psi_2^{(1)}$ under infinitesimal tetrad rotations---see App.\ \ref{a:KerrBack}) that $\mathrm{Im}(\psi_{2}^{(1)})$ is a {\em gauge-invariant} field. This means that our reference gauge, as defined in the Kerr case, generally does not exist: no gauge transformation can nullify $\psi_2^{(1)}$, because its imaginary piece is invariant (and generally nonzero). Instead, we shall choose our reference gauge to be one in which $\mathrm{Re}(\tilde\psi_{2}^{(1)})=0$. The generator $\tilde\xi^\alpha$ of a gauge transformation to the reference gauge then satisfies [in analogy with Eq.\ (\ref{xieq})] $\tilde\xi^{\alpha}\psi^{(0)}_{2,\alpha}=\mathrm{Re}(\psi_2^{(1)})$. Since $\psi_2^{(0)}$ depends only on $r$, the components $\tilde\xi^t$, $\tilde\xi^\theta$ and $\tilde\xi^\varphi$ remain undetermined. However, $\tilde\xi^r$ is still uniquely determined. It is given by 
\begin{equation}\label{xir}
\tilde{\xi}^{r}=\frac{r^{4}}{3M}\mathrm{Re}(\psi_{2}^{(1)}),
\end{equation}
which coincides with the $a=0$ reduction of the general Kerr value given in Eq.\ (\ref{eq:gauge9}).

We see that, in the Schwarzschild case, the reference gauge is fixed only up to arbitrary transformations in the space spanned by $\tilde\xi^t$, $\tilde\xi^\theta$ and $\tilde\xi^\varphi$. However, there is still a certain component of the metric perturbation in the reference gauge that is completely determined by $\tilde\xi^r$ alone, namely 
\begin{equation}
\tilde{h}_{rr}=h_{rr}-2\tilde{\xi}_{r,r}+2\Gamma_{rr}^{(0)r}\tilde{\xi}_{r},\label{eq:hrr}
\end{equation}
where $\tilde\xi_r=(1-2M/r)^{-1}\tilde\xi^r$ and we have used $\Gamma_{rr}^{\gamma}=0$ for $\gamma=t,\theta,\varphi$ in the Schwarzschild case. The gauge invariance of $\tilde h_{rr}$ follows in exactly the same way as in the Kerr case.

For our completion calculation we shall require {\it two} auxiliary invariants.
Since $\tilde{h}_{rr}$ is the only invariant component of $\tilde h_{\alpha\beta}$, we must look elsewhere. Fortunately, a second useful invariant immediately suggests itself in the Schwarzschild case, and has already been mentioned: the field $\mathrm{Im}(\delta\psi_{2}^{(1)})$ itself. Thus, for our completion analysis in Schwarzschild, we shall utilize the two invariants
\begin{equation}\label{eq:I_Schw}
\{{\cal I}_1,{\cal I}_2\}_{\rm Schw}:=\{\tilde h_{rr},\mathrm{Im}(\psi_{2}^{(1)})\}.
\end{equation}
Note that, in the Schwarzschild case, our ${\cal I}_2$ involves only up to {\em second} derivatives of the original metric perturbation $h_{\alpha\beta}$. The invariant ${\cal I}_1$, and all three of our invariants in the Kerr case, involve up to third derivatives.

\section{Circular orbits in Schwarzschild spacetime}\label{s:Schcir}

We start, in this section, by calculating the completion piece of the metric perturbation for a configuration consisting of a circular geodesic orbit around a Schwarzschild black hole. This will serve to illustrate (and test) our method in a relatively simple setting. 

Thus, we consider a particle of mass $\mu$ moving in a circular geodesic orbit of radius $r=r_0$ around a Schwarzschild black-hole of mass $M\gg\mu$. The gravitational self-force acting on the particle is ignored. Without loss of generality, we let the orbit lie in the equatorial plane, $\theta=\pi/2$. The particle's energy-momentum is given by
\begin{equation}\label{Tmunu_general}
T^{\alpha\beta}=\mu \int_{-\infty}^{\infty}u^{\alpha}u^\beta\delta^4(x-x^\mu_{\rm p}(\tau)) (-g^{(0)})^{-1/2} d\tau,
\end{equation}
where $g^{(0)}:=\det(g_{\alpha\beta}^{(0)})=-r^4\sin^2\theta$, $x^\mu_{\rm p}(\tau)$ denotes the particle's worldline (parametrized by proper time $\tau$), and $u^{\alpha}:=dx^\alpha_{\rm p}/d\tau$ is the particle's 4-velocity. For our circular equatorial orbits, this reduces to
\begin{equation}\label{Tmunu}
T^{\alpha\beta}=\frac{\mu u^{\alpha}u^{\beta}}{r_0^2 u^t}\delta(r-r_0)\delta(\cos\theta)\delta(\varphi-\Omega t),
\end{equation}
where $\Omega=u^\varphi/u^t$ is the particle's angular velocity. The conserved energy and angular momentum along the geodesic are, respectively, 
\begin{eqnarray}
E&:=&-\mu u_t=\mu (1-2M/r_0)\left(1-3M/r_0\right)^{-1/2},
\nonumber\\
L&:=&\mu u_{\varphi}=\mu (Mr_0)^{1/2}\left(1-3M/r_0\right)^{-1/2},
\end{eqnarray}
where $u_{\alpha}=g^{(0)}_{\alpha\beta}u^{\beta}$.
The surface $\cal S$ defined in the introduction is now the (2+1-dimensional) sphere $r=r_0$, and (in what is a slight redefinition) we use $\check{\cal S}$ to denote $\cal S$ minus the (1+1-dimensional) equatorial ring $(r,\theta)=(r_0,\pi/2)$. We use superscripts $`+'$ or $`-'$ to denote fields defined on $r>r_0$ or $r<r_0$, respectively, or otherwise quantities defined through the respective limits $r\to r_0^+$ or $r\to r_0^-$.

Our workplan is as follows. In Sec.\ \ref{ss:Sch_psi4} we (analytically) solve the relevant Teukolsky equation to obtain the stationary and axisymmetric piece of the Weyl curvature scalar $\psi_4$. This is the starting point for a CCK procedure, which we apply in Sec.\ \ref{ss:Sch_cck}. The end product is $h_{\alpha\beta}^{{\rm rec}\pm}$---the ``reconstructed'' piece of the metric perturbation on either side of $\cal S$, and we also obtain the piece $\psi_2^{{\rm rec}\pm}$, associated with $h_{\alpha\beta}^{{\rm rec}\pm}$, of the Weyl curvature scalar $\psi_2$. In Sec.\ \ref{ss:Sch_rec}, given $h_{\alpha\beta}^{{\rm rec}\pm}$ and $\psi_2^{{\rm rec}\pm}$, we then construct ${\cal I}_1^{{\rm rec}\pm}$ and ${\cal I}_2^{{\rm rec}\pm}$---the corresponding ``reconstructed'' pieces of the two invariants ${\cal I}_1$ and ${\cal I}_2$. In Sec.\ \ref{ss:Sch_comp} we similarly calculate the contributions ${\cal I}_1^{{\rm comp}\pm}$ and ${\cal I}_2^{{\rm comp}\pm}$ due to the completion piece $h_{\alpha\beta}^{{\rm comp}\pm}$ of the metric perturbation, writing the latter as in Eq.\ (\ref{h_comp}), with $h_{\alpha\beta}^{(\delta M)}$ and $h_{\alpha\beta}^{(\delta J)}$ specified in analytic form and the coefficients ${\cal E}^\pm,{\cal J}^{\pm}$ left unknown. Finally, in Sec.\ \ref{ss:Sch_jumps}, we determine the jumps $[\cal E]$ and $[\cal J]$ from the condition that the complete invariants, ${\cal I}_n^{\pm}:={\cal I}_n^{{\rm rec}\pm}+{\cal I}_n^{{\rm comp}\pm}$, satisfy ${\cal I}_n^{+}={\cal I}_n^{-}$ on $\check{\cal S}$.

\subsection{Stationary and axisymmetric piece of $\psi_4$}\label{ss:Sch_psi4}

The stationary and axisymmetric (SAS) piece of $\psi_4$ can be expressed as a sum over multipole-mode contributions, in the form 
\begin{equation}\label{Teq}
\psi_4^{\rm SAS}=r^{-4}\sum_{\ell=2}^\infty R_{\ell}(r)\, {}_{-2}\!Y_{\ell 0}(\theta),
\end{equation}
where ${}_sY_{\ell m}$ are {\it spin-weighted spherical harmonics}---see Appendix \ref{a:KerrBack} for a definition and how to express them in terms of ordinary ($s=0$) spherical harmonics. The factor $r^{-4}$ is conventional. Mode by mode, the functions $R_{\ell}(r)$ satisfy the radial Teukolsky equation 
\begin{equation}\label{eq:TeukolskyCirc}
\Delta^2 \frac{d}{dr}\left(\Delta^{-1}\frac{R_{\ell}}{dr}\right)-\lambda\,R_{\ell}= T_{\ell}(r;r_0),
\end{equation}
which is the $a=0$, $\omega=0=m$ reduction of Eq.\ \eqref{eq:kerrR}. Here 
\begin{equation}
\Delta:=r(r-2M)
\end{equation}
and
\begin{equation}
\lambda:=\lambda_2/\lambda_1=(\ell+2)(\ell-1),
\end{equation}
where we have introduced
\begin{equation}
\lambda_s:=(\ell+s)!/(\ell-s)!.
\end{equation}
For our circular-orbit configuration, the source $T_{\ell}$ is the distribution
\begin{multline}\label{eq:sourceCirc}
T_{\ell}(r;r_0)=\Delta^2(r)
\left[s^{\ell}_0(r_0)\delta(r-r_0)+s^{\ell}_1(r_0)\delta'(r-r_0)\right.
\\
\left. +s^{\ell}_2(r_0)\delta''(r-r_0)\right],
\end{multline}
obtained from the general expression (\ref{eq:kerrsource-2}) with the energy-momentum (\ref{Tmunu}) as input. Here
a prime denotes a derivative with respect to the argument, and the factor $\Delta^2$ has been pulled out for later convenience [specifically, to simplify the appearance of Eq.\ (\ref{Cpm}) below]. The coefficients $s^{\ell}_n(r_0)$ work out as
\begin{eqnarray}\label{sn}
s_0^{\ell}&=& \frac{\pi E}{\Delta_0}\left[{}_{-2}\!Y''_{\ell0}(\theta_0)-2{}_{-2}\!Y_{\ell0}(\theta_0)\right]
-\frac{4\pi i L}{r_0^3}\, {}_{-2}\!Y'_{\ell0}(\theta_0) , \nonumber \\
s_1^{\ell}&=& -\frac{2\pi iL}{r_0^2}\, {}_{-2}\!Y'_{\ell0}(\theta_0)
		-\frac{2\pi M  E}{\Delta_0}\, {}_{-2}\!Y_{\ell0}(\theta_0),\nonumber\\
s_2^{\ell}&=& -\frac{\pi M r_0 E}{\Delta_0}\, {}_{-2}\!Y_{\ell0}(\theta_0),
\end{eqnarray}
where $\Delta_0:=\Delta(r_0)=r_0(r_0-2M)$, primes denote $d/d\theta$, and all angular functions are evaluated at $\theta=\theta_0=\pi/2$. 



Two linearly independent homogeneous solutions to Eq.\ \eqref{eq:TeukolskyCirc} are
\begin{eqnarray}\label{eq:PlQl4}
R_{\ell}^{-}&=& (M\lambda_2)^{-1/2} \Delta(r) {\sf P}_{\ell}^{m=2}(r/M-1),
\nonumber\\
R_{\ell}^{+}&=& (M\lambda_2)^{-1/2}\Delta(r) {\sf Q}_{\ell}^{m=2}(r/M-1),
\end{eqnarray}
where ${\sf P}_\ell^{m}$ and ${\sf Q}_\ell^{m}$ are associated Legendre functions of the first and second kinds, respectively,
and the normalization factors $(M\lambda_2)^{-1/2}$ were inserted so as to render the Wronskian,
\begin{equation}\label{W}
W(r):=\frac{dR_{\ell}^{+}}{dr}\,R_{\ell}^{-}-R_{\ell}^{+}\,\frac{dR_{\ell}^{-}}{dr}=-\Delta(r),
\end{equation}
$\ell$-independent.
To construct the physical inhomogeneous solution to Eq.\ \eqref{eq:TeukolskyCirc}, we need to consider the asymptotic behavior of $R_{\ell}^{\pm}$ at infinity, $r\to\infty$, and at the event horizon, $r=2M$. For stationary physical perturbations, $r^4\psi_4$ should fall off at infinity at least as $1/r$, and $\Delta^{-2}\psi_4$ should be regular (smooth) across the horizon.\footnote{The form of the regularity condition for $\psi_4$ at the horizon comes from assuming regularity of the Weyl curvature tensor (in regular coordinates) and taking into account the singular behavior of the Boyer-Lindquist tetrad; see, for example, Section V.\ of \cite{Barack:1999ya}.} An inspection reveals that, for any $\ell\geq 2$, the solution $R_{\ell}^{-}(r)$ blows up (as $\sim r^{\ell+2}$) at infinity, while $\Delta^{-2} R_{\ell}^{-}$ is smooth on the horizon. On the other hand, the solution $R_{\ell}^{+}(r)$ falls off as $\sim r^{1-\ell}$ at infinity, while $\Delta^{-2} R_{\ell}^{+}$ blows up (like $\Delta^{-2}$) on the horizon. Thus, up to constant multiplicative factors, $R_{\ell}^{-}(r)$ is a unique solution regular at the horizon, and $R_{\ell}^{+}(r)$ is a unique solution regular at infinity.  

It follows that Eq.\ \eqref{eq:TeukolskyCirc} admits a {\em unique} inhomogeneous solution that is regular both at infinity and on the horizon (and anywhere else, except at $r=r_0$). It is given by
\begin{multline}\label{Rlvariation}
R_{\ell}(r;r_0)=R_{\ell}^{+}(r)\int_{2M}^{r}\frac{R_{\ell}^{-}(r')T_\ell(r';r_0)}{\Delta(r')W(r')} dr'
\\
+R_{\ell}^{-}(r)\int_{r}^{\infty}\frac{R_{\ell}^{+}(r')T_\ell(r';r_0)}{\Delta(r')W(r')} dr' .
\end{multline}
Substituting for $T_\ell$ from Eq.\ \eqref{eq:sourceCirc} and evaluating the integrals, we obtain the distributional form
\begin{multline}\label{Rl}
R_{\ell}(r;r_0)=C_\ell^{+}(r_0)R_{\ell}^{+}(r)\Theta(r-r_0)\\
+C_\ell^{-}(r_0)R_{\ell}^{-}(r)\Theta(r_0-r) 
+ C_\ell^{\delta}(r_0)\delta(r-r_0),
\end{multline}
where $\Theta(\cdot)$ is the Heaviside step function, and the coefficients are 
\begin{equation}\label{Cpm}
C_\ell^{\pm}(r_0)=\left.\left(-s_0^\ell R^{\mp}_{\ell}+s_1^\ell \frac{dR^{\mp}_{\ell}}{dr}-s_2^\ell \frac{d^2R^{\mp}_{\ell}}{dr^2}\right)\right|_{r=r_0};
\end{equation}
the explicit form of $C_\ell^{\delta}(r_0)$ will not be needed in our analysis.


The metric reconstruction procedure to be applied below will not require the full distributional solution \eqref{Rl}, but, following FKS's method, only the ``one-sided'' functions 
\begin{equation}\label{psi4ell}
\psi_{4\ell}^{\pm}:=C_\ell^\pm(r_0)R_\ell^\pm(r) ,
\end{equation}
which coincide with $R_{\ell}$ in the respective vacuum domains $\cal S^\pm$ (recall $\cal S^+$ and $\cal S^-$ represent the regions $r>r_0$ and $2M\leq r<r_0$, respectively).
The corresponding one-sided solutions for $\psi_4^{\rm SAS}$ are
\begin{equation}\label{psi4SAS}
\psi_4^{\rm SAS\pm}:=\psi_4^{\rm SAS}({\cal S}^\pm)=r^{-4}\sum_{\ell=2}^{\infty} \psi_{4\ell}^{\pm}(r;r_0) {}_{-2}\!Y_{\ell 0}(\theta).
\end{equation}

\subsection{Metric reconstruction and perturbation in $\psi_2$}\label{ss:Sch_cck}

Given the fields $\psi^{\pm}_{4\ell}(r,\theta;r_0)$, we proceed following FKS's procedure to reconstruct the metric perturbations $h^{{\rm rec}\pm}_{\alpha\beta}$ in the corresponding domains ${\cal S}^{\pm}$. (We hereafter omit the label `SAS' for brevity, but it should be clear that throughout the analysis we restrict attention to the SAS sector of the perturbation.) For no particular reason, we choose to reconstruct the metric in the so-called ``ingoing'' radiation gauge [see Eq.\ (\ref{gaugecondition}) for a definition].
As usual, the reconstruction is done mode-by-mode, and follows three steps. In the first step, given $\psi^{\pm}_{4\ell}$, we algebraically construct a certain Hertz potential $\Psi^{\pm}_{\ell}$, itself a solution to the (spin $-2$) Teukolsky equation. In the second step we obtain the $\ell$-mode contribution to $h^{{\rm rec}\pm}_{\alpha\beta}$ by applying a certain second-order differential operator to $\Psi^{\pm}_{\ell}$. Finally, in the third step, we add up all $\ell$-mode contributions to obtain $h^{{\rm rec}\pm}_{\alpha\beta}$.  For our particular application we would need only the component $h^{{\rm rec}\pm}_{rr}$, as well as the value of the perturbation in $\psi_2$ associated with the reconstructed perturbation. The latter, to be denoted by $\psi_2^{(1){\rm rec\pm}}$, will be obtained directly from $\Psi^{\pm}_{\ell}$, with no need to resort to a knowledge of the full perturbation $h^{{\rm rec}\pm}_{\alpha\beta}$.

We begin by constructing the Hertz potential $\Psi^\pm$ corresponding to $\psi_4^{\rm SAS\pm}$. It admits the multipole expansion 
\begin{equation}\label{Psiexpansion}
\Psi^{\pm}=\sum_{\ell=2}^{\infty} \Psi^{\pm}_{\ell}(r) {}_{-2}\!Y_{\ell 0}(\theta),
\end{equation}
and satisfies the homogeneous Teukolsky equation with $s=-2$, as well as the
differential equation
\begin{equation}\label{inversion}
\bar\eth^4 \bar\Psi^{\pm} = 8r^4\psi_4^{\rm SAS\pm},
\end{equation}
which is the relevant reduction of Eq.\ (\ref{eq:Heqkerra}). Here an overbar denotes complex conjugation,\footnote{Note that $\psi_4^{\rm SAS\pm}$ and $\Psi^\pm$ are complex quantities even for our stationary perturbation, owing to the source coefficients $s_0^\ell$ and $s_1^\ell$ of Eq.\ (\ref{sn}) being complex-valued.} and $\bar\eth$ is the ``spin-lowering'' angular differential operator, given explicitely in Eq.\ (\ref{edth}). The action of $\bar\eth$ on a spin-$s$ spherical harmonic is described in Eq.\ (\ref{lowering}).
We note $\bar\Psi^{\pm}$ admits the multipole expansion
\begin{equation}\label{Psibarexpansion}
\bar\Psi^{\pm}=\sum_{\ell=2}^{\infty} \bar\Psi^{\pm}_{\ell}(r) {}_{+2}\!Y_{\ell 0}(\theta),
\end{equation}
which is obtained by taking the complex conjugate of Eq.\ (\ref{Psiexpansion}), noting the symmetry ${}_{+2}\!Y_{\ell 0}\equiv {}_{-2}\!\bar{Y}_{\ell 0}(\equiv {}_{-2}\!{Y}_{\ell 0})$.

Substituting from Eqs.\ (\ref{psi4SAS}) and (\ref{Psibarexpansion}) into Eq.\ (\ref{inversion}), and using (\ref{lowering}) and the orthogonality property of ${}_{-2}\!Y_{\ell 0}$, one arrives at the simple algebraic relation
\begin{equation}\label{Psiell}
\Psi_\ell^{\pm}=8\lambda_2^{-1}\bar\psi_{4\ell}^{\pm}.
\end{equation}
An explicit expression for $\Psi^{\pm}$ is then obtained by combining Eqs.\ (\ref{eq:PlQl4}), (\ref{Cpm}), (\ref{psi4ell}), (\ref{Psiell}) and  (\ref{Psiexpansion}). Note that the above procedure picks out a particular solution of the differential equation (\ref{inversion}); other solutions of that equation are effectively ruled out by the condition that $\Psi^{\pm}$ is of a pure spin $-2$ (i.e., that its angular part satisfies the angular part of the $s=-2$ Teukolsky equation). It can be checked with an explicit calculation (see, e.g., \citep{SaTa}) that no other solution of (\ref{inversion}) satisfies the additional requirement of being a solution to the relevant Teukolsky equation. 

Next, we turn to the metric perturbation $h^{{\rm rec}\pm}_{\alpha\beta}$. Its reconstruction from $\Psi^{\pm}$ is prescribed in Eq.\ (\ref{eq:kerrh}) of Appendix  \ref{a:KerrBack}, which, in our problem, and for the $rr$ component relevant to us, reduces to
\begin{equation}\label{hrr}
h_{rr}^{{\rm rec}\pm}=-(r/\Delta)^2 {\rm Re}(\bar\eth^2\bar\Psi^{\pm}).
\end{equation}
Recalling that $\bar\Psi^{\pm}$ is of spin $+2$, we see that the reconstructed component $h_{rr}^{\pm}$ is of a pure spin zero, as expected (of this particular component, in the Schwarzschild case). Substituting from Eq.\ (\ref{Psiexpansion}) and using (\ref{lowering}) and (\ref{Psiell}) we get, more explicitly, 
\begin{equation}\label{hrrSum}
h_{rr}^{{\rm rec}\pm}(r,\theta;r_0)=-{\rm Re}\frac{8r^2}{\Delta^2}\sum_{\ell=2}^{\infty}\lambda_2^{-1/2}
\psi_{4\ell}^{\pm}(r;r_0)Y_{\ell0}(\theta).
\end{equation}

We further need the perturbation $\psi_2^{(1){\rm rec}\pm}$ corresponding to $h_{\alpha\beta}^{{\rm rec}\pm}$. This can be readily calculated from the full reconstructed perturbation $h_{\alpha\beta}^{{\rm rec}\pm}$, but, to save us the need to obtain other components of the perturbation (in addition to $rr$), we can take advantage of the relation (\ref{eq:deltapsi2}), which conveniently gives $\psi_2^{(1)}$ directly in terms of the Hertz potential $\Psi$. Specialized to stationary perturbations in Schwarzschild, the relation reduces to
\begin{equation}\label{deltapsi2}
\psi_2^{(1){\rm rec}\pm}=\frac{1}{4}\partial_r^2\left(r^{-2}\bar\eth^2\bar\Psi^{\pm}\right).
\end{equation}
We find that the action of $\bar\eth^2$ on the right-hand side once more produces a spin-0 quantity, as expected. 
Substituting from Eqs.\ (\ref{Psiexpansion}),  (\ref{lowering}) and (\ref{Psiell}), we obtain  
\begin{equation}\label{deltapsi2Sum}
\psi_2^{(1){\rm rec}\pm}(r,\theta;r_0)=2\sum_{\ell=2}^{\infty}\lambda_2^{-1/2}
\frac{d^2}{dr^2}\left(\frac{\psi_{4\ell}^{\pm}}{r^2}\right)Y_{\ell0}(\theta).
\end{equation}



\subsection{Auxiliary invariants}\label{ss:Sch_rec}

Equipped with $h_{rr}^{{\rm rec}\pm}$ and $\psi_2^{(1){\rm rec}\pm}$, we now proceed to deriving the  ``reconstructed'' pieces of each of the two invariant fields  $\{{\cal I}_1,{\cal I}_2\}_{\rm Schw}$ on each of the two domains ${\cal S}^{\pm}$---call these $\{{\cal I}_1^{{\rm rec}\pm},{\cal I}_2^{{\rm rec}\pm}\}$, respectively. The field ${\cal I}_1^{{\rm rec}\pm}$ is obtained using Eqs.\ (\ref{xir}) and (\ref{eq:hrr}), with $\psi_2^{(1)}$ and $h_{rr}$ replaced with $\psi_2^{(1){\rm rec}\pm}$ and $h_{rr}^{{\rm rec}\pm}$, respectively. The field ${\cal I}_2^{{\rm rec}\pm}$ is simply the imaginary part of $\psi_2^{(1){\rm rec}\pm}$.  We find
\begin{multline}
{\cal I}_1^{{\rm rec}\pm}(r,\theta;r_0)=
-\frac{4r^4}{3M\Delta^2}\sum_{\ell=2}^{\infty}\lambda_2^{-1/2}Y_{\ell 0}(\theta){\rm Re}[C_\ell^{\pm}(r_0)]
\\
\times\left[\Delta {R_{\ell}^{\pm}}'''(r)-(2r-3M){R_{\ell}^\pm}''(r)+2{R_\ell^\pm}'(r)\right],
\end{multline}
\begin{multline}
{\cal I}_2^{{\rm rec}\pm}(r,\theta;r_0)=
\frac{2}{r^4}\sum_{\ell=2}^{\infty}\lambda_2^{-1/2}Y_{\ell 0}(\theta){\rm Im}[C_\ell^{\pm}(r_0)]
\\
\times\left[r^2 {R_{\ell}^{\pm}}''(r)-4r{R_{\ell}^\pm}'(r)+6R_\ell^\pm(r)\right],
\end{multline}
where we have substituted for $\psi_{4\ell}^{\pm}$ from Eq.\ (\ref{psi4ell}),
and a prime denotes $d/dr$. Recall the coefficients $C_\ell^{\pm}$, defined in Eq.\ (\ref{Cpm}), are certain linear combinations of $R_{\ell}^{\mp}$ and its first and second derivatives, evaluated at $r_0$.  

To proceed, we recall that it is not the invariants ${\cal I}_{n}^{{\rm rec}\pm}$ themselves we are interested in here, but rather their {\em difference} across $\cal S$,
\begin{equation}\label{difference}
 [{\cal I}_n^{\rm rec}](\theta;r_0):=\left.\left({\cal I}_n^{{\rm rec}+}-{\cal I}_n^{{\rm rec}-}\right)\right|_{r=r_0}.
\end{equation}
We have found that a great deal of simplification occurs if one evaluates the difference {\em prior} to the summation over $\ell$ (and in Appendix \ref{a:ProofInterchange} we establish that such an interchange of summation and limit is mathematically legitimate in our case).  The simplification owes itself to the following set of identities, which are satisfied mode by mode for each $\ell\geq 2$:
\begin{eqnarray}\label{Widentities}
R_{+}'R_{-}-R_{+}R_{-}'&=&W=-\Delta, \nonumber\\
R_{+}''R_{-}-R_{+}R_{-}''&=&W'=-2(r-M), \nonumber\\
R_{+}''R_{-}'-R_{+}'R_{-}''&=&\lambda, \nonumber\\
R_{+}'''R_{-}-R_{+}R_{-}'''&=&W''-\lambda=-\lambda_1, \nonumber\\
R_{+}'''R'_{-}-R'_{+}R_{-}'''&=&0 ,\nonumber\\
R_{+}'''R_{-}''-R_{+}''R_{-}'''&=&-\lambda_2/\Delta
\end{eqnarray}
(omitting subscripts $\ell$ and relocating the $\pm$ for improved readability). Here, the first identity is the Wronskian relation of Eq.\ (\ref{W}), and the third identity is obtained by replacing $R_\pm''$ in favor of $R_{\pm}'$ and $R_{\pm}$ using Teukolsky's equation (\ref{eq:TeukolskyCirc}). Other relations are readily obtained by differentiating lower-order identities and again using Teukolsky's equation. Thanks to these relations, the jumps $[{\cal I}_n^{\rm rec}]$ turn out to involve no reference to the (transcendental) functions $R_\ell^{\pm}$ themselves. These functions enter $[{\cal I}_n^{\rm rec}]$ only through their Wronskian, which is elementary and simple.

With the aid of (\ref{Widentities}), and substituting the explicit values of the source coefficients $s_n^\ell$ from Eq.\ (\ref{sn}), we now obtain
\begin{multline}
\label{eq:sum[I1]}
\left[{\cal I}_{1}^{{\rm rec}}\right](\theta;r_0)=\frac{8 \pi  E r_0^4}{3M\Delta_0^2}\sum_{\ell=2}^{\infty} Y_{\ell0}(\theta) Y_{\ell 0}(\theta_0)
\\ 
+\frac{4 \pi E r_0^5(r_0-M)}{3M\Delta_0^3} \sum_{\ell=2}^{\infty}Y_{\ell0}(\theta) Y''_{\ell0}(\theta_0),
\end{multline}
\begin{equation}
\label{eq:sum[I2]}
\left[{\cal I}^{{\rm rec}}_2\right](\theta;r_0)=\frac{4\pi L}{r_0^4 } \sum_{\ell=2}^{\infty}Y_{\ell0}(\theta) Y'_{\ell0}(\theta_0),
\end{equation}
where use has also been made of the relations 
\begin{eqnarray}
{}_{-2}Y_{\ell0}(\theta_0)&=&\lambda_2^{-1/2} Y''_{\ell0}(\theta_0)=-(\lambda_1/\lambda)^{1/2}Y_{\ell0}(\theta_0), \nonumber\\
{}_{-2}Y'_{\ell0}(\theta_0)&=&-(\lambda/\lambda_1)^{1/2} Y'_{\ell0}(\theta_0), \nonumber\\
{}_{-2}Y''_{\ell0}(\theta_0)&=&(\lambda_1/\lambda)^{1/2}(\lambda_1-4) Y_{\ell0}(\theta_0) 
\end{eqnarray}
[derived using (\ref{lowering})] in order to express $s=-2$ harmonics and their derivatives at $\theta_0=\pi/2$ in terms of standard ($s=0$) spherical harmonics and their derivatives there. 
The mode sums in Eqs.\ (\ref{eq:sum[I1]}) and (\ref{eq:sum[I2]}) are readily evaluated in distributional form using the completeness relation
\begin{multline}\label{sumgeneral}
\sum_{\ell=2}^{\infty} Y_{\ell0}(\theta) Y_{\ell0}(\theta_0)=
\frac{\delta(\cos\theta-\cos\theta_0)}{2\pi}-\sum_{\ell=0}^{1} Y_{\ell0}(\theta) Y_{\ell0}(\theta_0)
\\
=\frac{\delta(\cos\theta-\cos\theta_0)}{2\pi}-\frac{1+3\cos\theta\cos\theta_0}{4\pi}, 
\end{multline}
and term-by-term derivatives thereof with respect to $\theta_0$. 
With the sums thus evaluated (and setting $\theta_0=\pi/2$), Eqs.\ (\ref{eq:sum[I1]}) and (\ref{eq:sum[I2]}) reduce to 
\begin{eqnarray}\label{JumpIrec}
\left[{\cal I}_{1}^{{\rm rec}}\right](\theta;r_0)&=&-\frac{2 Er_0^4}{3M\Delta_0^2},
\nonumber \\
\left[{\cal I}^{{\rm rec}}_2\right](\theta;r_0)&=&\frac{3L\cos\theta}{r_0^4 },
\end{eqnarray}
where distributional contributions with support only on the particle have been omitted. That such an omission is justified, for our purpose, is shown in Appendix \ref{a:ProofInterchange}.

We see that the contribution from the reconstructed metric to the invariant quantities ${\cal I}_{1,2}$ has a finite discontinuity at $r=r_0$, even away from the particle's location. We further notice that the discontinuity in ${\cal I}_{1}^{{\rm rec}}$ is purely monopolar ($\theta$-independent), while the discontinuity in ${\cal I}_{2}^{{\rm rec}}$ is purely dipolar. Below we will establish that both discontinuities can be removed with a suitable choice of the perturbation's completion piece.

\subsection{Completion piece}\label{ss:Sch_comp}

We write the completion piece of the metric perturbation as a sum of mass and angular-momentum perturbations, as in Eq.\ (\ref{h_comp}), copied here for easy reference:
\begin{equation}\label{h_comp_again}
h^{\rm comp\pm}_{\alpha\beta}= {\cal E}^{\pm}h^{(\delta M)}_{\alpha\beta}+{\cal J}^{\pm}h^{(\delta J)}_{\alpha\beta}.
\end{equation}
As usual, $\pm$ indicates values in the corresponding domains ${\cal S}^{\pm}$. $h^{(\delta M)}_{\alpha\beta}$ and $h^{(\delta J)}_{\alpha\beta}$ are homogeneous perturbations that represent trivial variations of the background geometry with respect to its mass and angular momentum parameters, as prescribed below; each is a solution of the linearized vacuum Einstein's equations. The constant amplitude coefficients ${\cal E}^{\pm}$ and ${\cal J}^{\pm}$ are to be determined.


Following \citep{PriceThesis}, we choose to construct $h^{(\delta M)}_{\alpha\beta}$ and $h^{(\delta J)}_{\alpha\beta}$ in a ``Boyer--Lindquist'' gauge, using
\begin{eqnarray}\label{eq:hdeltaM}
h_{\alpha\beta}^{(\delta M)}(r)&=&\left.\frac{\partial g^{(0)}_{\alpha\beta}(x^\mu;M,J)}{\partial M}\right|_{J\to 0},
\\ \label{eq:hdeltaJ}
h_{\alpha\beta}^{(\delta J)}(r,\theta)&=&
\left.\frac{\partial g^{(0)}_{\alpha\beta}(x^\mu;M,J)}{\partial J}\right|_{J\to 0},
\end{eqnarray}
where $g^{(0)}_{\alpha\beta}$ is the Kerr metric, parametrized by mass $M$ and angular-momentum $J=aM$, and the partial derivatives are taken with fixed Boyer-Lindquist coordinates. Explicitly, 
\begin{equation}\label{completionmetric}
h^{(\delta M)}_{tt}=\frac{2}{r},
\quad
h^{(\delta M)}_{rr}=\frac{2r^3}{\Delta^2},
\quad
h_{t\varphi}^{(\delta J)}=-\frac{2\sin^2\theta}{r},
\end{equation}
and all other independent components vanish. Our goal now is to calculate the contribution from $h^{\rm comp\pm}_{\alpha\beta}$ to the two invariants ${\cal I}_{1,2}^{\pm}$, which we shall call ${\cal I}_{1,2}^{{\rm comp}\pm}$.

We start with the perturbation in $\psi_2$, which can be derived either from the perturbation in the Weyl curvature associated with (\ref{completionmetric}) (making sure to take into account the perturbation in the null tetrad); or, much more simply, by varying $\psi_2^{(0)}$ in Eq.\ (\ref{psi20}) with respect to $M$ (at fixed $J=aM$ and $r$) and with respect to $J$ (at fixed $M$ and $r$). Either way, the result is
\begin{equation}
\psi_2^{(1){\rm comp}\pm}=-\frac{{\cal E}^\pm}{r^3}-\frac{3i{\cal J}^\pm\cos\theta}{r^4},
\end{equation}
from which we obtain, using (\ref{xir}), (\ref{eq:hrr}), and (\ref{eq:I_Schw}) successively, with $\psi_2^{(1)}$ replaced with $\psi_2^{(1){\rm comp}\pm}$,
\begin{equation}
{\cal I}_{1}^{{\rm comp}\pm}=\frac{2{\cal E}^{\pm}r^4}{3M\Delta^2},
\quad\quad
{\cal I}_{2}^{{\rm comp}\pm}=-\frac{3{\cal J}^{\pm}\cos\theta}{r^4}.
\end{equation}
The jumps across $\cal S$, defined as in (\ref{difference}), are thus
\begin{equation}\label{JumpIcomp}
[{\cal I}_{1}^{{\rm comp}}]=\frac{2[{\cal E}]r_0^4}{3M\Delta_0^2},
\quad\quad
[{\cal I}_{2}^{{\rm comp}}]=-\frac{3[{\cal J}]\cos\theta}{r_0^4},
\end{equation}
with $[{\cal E}]$ and $[{\cal J}]$ as defined in Eq.\ (\ref{diffEJ}).




\subsection{Determination of the completion amplitudes}\label{ss:Sch_jumps}

The jumps $[{\cal E}]$ and $[{\cal J}]$ are determined from the two continuity conditions 
\begin{eqnarray}\label{eq:regularityCirc}
0=[{\cal I}_1]=[{\cal I}_{1}^{{\rm rec}}]+[{\cal I}_{1}^{{\rm comp}}]&=&\frac{2r_0^4}{3M\Delta_0^2}([{\cal E}]-E),
\\ {}
0=[{\cal I}_2]=[{\cal I}_{2}^{{\rm rec}}]+[{\cal I}_{2}^{{\rm comp}}]&=&-\frac{3\cos\theta}{r_0^4}([{\cal J}]-L)
\end{eqnarray}
for $\theta\ne \pi/2$, where we have substituted from Eqs.\ (\ref{JumpIrec}) and (\ref{JumpIcomp}).
We immediately find
\begin{equation}\label{jumps_Sch}
[{\cal E}]=E, \quad\quad [{\cal J}]=L . 
\end{equation}
Namely, the jumps $[{\cal E}]$ and $[{\cal J}]$ are simply the conserved energy and angular momentum of the particle's geodesic orbit. 

Let us make a few simple observations. First, it is evident from Eqs.\ (\ref{JumpIrec}) and (\ref{JumpIcomp}) that, for each of $n=1,2$, the jumps $[{\cal I}_{n}^{{\rm rec}}]$ and $[{\cal I}_{n}^{{\rm comp}}]$ share the same dependence on the angle $\theta$. This means that imposing the continuity condition $[{\cal I}_n]=0$ at any particular value of $\theta(\ne\pi/2)$ automatically guarantees continuity across the entire of $\check{\cal S}$. That this is the case is an important consistency test for our method and calculation. (This test appears somewhat trivial in the Schwarzschild case; it will take a less trivial form in Kerr, as we shall see.) One should be able to check that, with our chosen completion, the full invariant fields ${\cal I}_n={\cal I}_{n}^{{\rm rec}}+{\cal I}_{n}^{{\rm comp}}$ are not only continuous but also {\em smooth} across $\check{\cal S}$. They are, in fact, smooth everywhere outside the black hole, except (possibly) on the ring $(r,\theta)=(r_0,\pi/2)$ containing the particle. 

Second, as it turned out, our specific choice of auxiliary invariants was such that $[{\cal I}_1]$ involved $[\cal E]$ alone (and not $[\cal J]$), while $[{\cal I}_2]$ involved $[\cal J]$ alone (and not $[\cal E]$). In consequence, the equations for $[\cal E]$ and $[\cal J]$ automatically decoupled. This is merely an artefact of our choice of invariants (combined with the special symmetry of the Schwarzschild background), and in general need not be the case for our method to work. Indeed, in the Kerr case, as we shall see, the continuity condition for either ${\cal I}_1$ or ${\cal I}_2$ will yield an algebraic equation involving both $[\cal E]$ and $[\cal J]$. 

Third, and most important, we see that the jumps $[\cal E]$ and $[\cal J]$ are completely and uniquely determined by imposing the field equations with usual regularity conditions (i.e., that geometrical invariants should be regular anywhere outside physical singularities). This conclusion carries over to the Kerr case, to be considered in subsequent sections. However, the individual amplitudes ${\cal E}^{\pm}$ and ${\cal J}^{\pm}$ remain undetermined: One can always add arbitrary homogeneous mass or angular-momentum perturbations without violating either the field equations or regularity. 

To fix ${\cal E}^{\pm}$ and ${\cal J}^{\pm}$ requires additional information, alluding to suitable notions of ``mass'' and ``angular momentum'' defined in the full perturbed spacetime.
Given such notions, one can fix the amplitudes ${\cal E}^{\pm}$ and ${\cal J}^{\pm}$ in a number of ways. For instance, prescribing the total mass and angular momentum of the perturbed spacetime (as measured at spatial infinity) should fix ${\cal E}^{+}$ and ${\cal J}^{+}$, with the amplitudes ${\cal E}^{-}$ and ${\cal J}^{-}$ then determined from the known jumps $[\cal E]$ and $[\cal J]$. Or, alternatively, prescribing the mass and angular momentum of the black hole (as measured on the horizon) should fix ${\cal E}^{-}$ and ${\cal J}^{-}$, with ${\cal E}^{+}$ and ${\cal J}^{+}$ now determined from the known jumps. The first route seems advantageous in that it requires only global notions of mass and angular momentum. However, even following that route, one would ideally wish to have a supplementary semi-local notion of mass and angular momentum in order to verify that the completed geometry in the inner region ${\cal S}^-$ corresponds to that of a black hole with the desired properties (in our case, a Schwarzschild black hole of mass $M$). 
In Sec.\ \ref{s:asymp} we will employ the Abbott-Deser notion of quasi-local mass and angular momentum, in combination with our results for $[\cal E]$ and $[\cal J]$, in order to determine the individual amplitudes ${\cal E}^{\pm}$ and ${\cal J}^{\pm}$ (in the more general Kerr case).\footnote{Refs.\ \citep{Keidlhom,friedman3} instead employ the Komar notion of mass and angular momentum in their discussion of the completion problem. These, however, are not defined in the full perturbed spacetime, which lacks any Killing symmetry.}

It should be said that, in the Schwarzschild case considered above, the completion amplitudes may also be determined from a simple argument, as follows.  Thanks to the spherical symmetry of the Schwarzschild background, multipole modes of the metric perturbation are globally well defined (in terms of tensorial spherical harmonics) and satisfy decoupled evolution equations. Mass and angular momentum perturbations of the Schwarzschild geometry have a pure monopolar and dipolar profile and are entirely contained in the $\ell=0,1$ modes of the metric perturbation. Crucially, the (Teukolsky) $\ell$-mode $\psi_{4\ell}$ can be be shown to contribute, via the reconstruction procedure, only to the corresponding (tensor-harmonic) $\ell$-mode of the metric perturbation. It follows that the reconstructed piece $h_{\alpha\beta}^{\rm rec}$, which is made up of $\ell\geq 2$ Teukolsky modes only, adds no contribution to the mass and angular momentum of the full (retarded) perturbation $h_{\alpha\beta}$.\footnote{This was first pointed out by Stewart in \cite{Stewart:1978tm}, referring to general, asymptotically flat vacuum perturbations in Schwarzschild.} This is true in both ${\cal S}^+$ and ${\cal S}^-$. The entire contribution to the mass and angular momentum of $h_{\alpha\beta}$ is contained in the completion piece $h_{\alpha\beta}^{\rm comp}$. If we then impose that the black hole has a mass $M$ and no spin, we immediately find $h_{\alpha\beta}^{\rm comp-}=0$, i.e.,
\begin{equation}\label{ampIn}
{\cal E}^- =0, \quad\quad {\cal J}^- =0.
\end{equation}
From (\ref{diffEJ}) and (\ref{jumps_Sch}) it then follows that
\begin{equation}\label{ampOut}
{\cal E}^+ = E, \quad\quad {\cal J}^+ =L,
\end{equation}
consistent with a total mass $M+E$ and an angular momentum $L$, as expected. [In fact, the values of the jumps $[{\cal E}]$ and $[{\cal J}]$ themselves follow immediately, in the Schwarzschild case, from the requirement that the $\ell=0,1$ modes satisfy the field equations on $r=r_0$, so one need not actually rely on Eq.\ (\ref{jumps_Sch}) to obtain (\ref{ampOut}).]

The above argument does not work in the Kerr case, where multipole modes of the perturbation couple, the contribution from each individual Teukolsky $\ell$-mode spreads over infinitely many tensorial $\ell$-modes of the reconstructed metric perturbation, and mass and angular momentum perturbations do not have simple monopole-dipole structures (except in the limit $r\to\infty$). Under these circumstances, it may appear unlikely that the above results---in particular, ${\cal E}^-=0={\cal J}^-$---should carry over to Kerr. In the proceeding sections we will establish that this, remarkably, is precisely the case.

\section{Circular equatorial orbits in Kerr spacetime}\label{s:Kerrcirc}

As a first generalization of the above analysis, we now replace the background geometry with that of a Kerr black hole of mass $M\gg\mu$ and spin parameter $a$, and consider the completion problem for a particle of mass $\mu\ll M$ moving on a circular geodesic of radius $r=r_0$ in the equatorial plane ($\theta=\pi/2$) of the black hole. The particle's energy-momentum again takes the form (\ref{Tmunu}), with conserved energy $E=-\mu u_t$ and angular momentum $L=\mu u_{\varphi}$ that are now given explicitly by
\begin{eqnarray}
E&=& \mu\,  \frac{1-2v^2+\tilde a v^3}{\sqrt{1-3v^2+2\tilde a v^3}},
\nonumber\\
L&=&\mu\, \frac{r_0v (1-2\tilde a v^3+\tilde a^2 v^4)}{\sqrt{1-3v^2+2\tilde a v^3}} ,
\end{eqnarray}
with $v:=\sqrt{M/r_0}$ and $\tilde a:=a/M$.  Our convention is that $a>0$ ($a<0$) refers to prograde (retrograde) orbits, i.e.\ the orbital angular momentum being aligned (anti-aligned) with the black hole's spin direction.

Our completion procedure will follow closely and generalize that of the Schwarzschild case, and many of our intermediate results can be checked against their Schwarzschild counterparts by setting $a=0$. To enable this, and for notational simplicity, we use the same notation for the various Kerr quantities (like $E$ and $L$ above) as for the Schwarzschild quantities they generalize, overriding the notation of Section \ref{s:Schcir}.


\subsection{Stationary and axisymmetric piece of $\psi_4$}\label{ss:Kerr_psi4}

For a generic perturbation in Kerr, the Teukolsky equation governing the Weyl scalar $\psi_4$ is only separable in terms of (spin-weighted) {\em spheroidal}-harmonic functions, which are frequency-dependent. However, for the purely SAS perturbations of relevance to us here, the spheroidal harmonics reduce to (spin-weighted) {\em spherical} harmonics, and Teukolsy's equation becomes separable in terms of ${}_sY_{\ell0}(\theta)$, just as in the Schwarzschild case. More precisely, the master equation for $\psi_4^{\rm SAS}$ is separable using 
\begin{equation}\label{psi4_separation}
\psi_4^{\rm SAS}=\varrho^4\sum_{\ell=2}^\infty R_{\ell}(r)\, {}_{-2}\!Y_{\ell 0}(\theta),
\end{equation}
where [recall Eq.\ (\ref{varrho})] $\varrho=-(r-ia\cos\theta)^{-1}$.
The modal radial functions $R_{\ell}(r)$ then satisfy the radial Teukolsky equation (\ref{eq:TeukolskyCirc}), where now
\begin{equation}
\Delta=r^2-2Mr+a^2.
\end{equation}
The source $T_\ell(r;r_0)$ again has the form (\ref{eq:sourceCirc}), but with the coefficients $s_n^\ell(r_0)$ now given by
\begin{eqnarray}\label{snKerr}
s_0^{\ell}&=& \frac{4\pi\mu r_0^2 u_n^2}{\Delta_0^2 u^t}\left[{}_{-2}\!Y''_{\ell0}(\theta_0)-2{}_{-2}\!Y_{\ell0}(\theta_0)\right]
\nonumber\\ && \quad\quad
-\frac{8\pi\mu u_n(\sqrt{2}\,\Delta_0 u_{\bar m}+i ar_0 u_n)}{\Delta_0^2 u^t}\, {}_{-2}\!Y'_{\ell0}(\theta_0) , \nonumber \\
s_1^{\ell}&=& -\frac{4\sqrt{2}\,\mu\pi r_0 u_n u_{\bar m}}{\Delta_0 u^t}\, {}_{-2}\!Y'_{\ell0}(\theta_0)
		+\frac{4\pi\mu u_{\bar m}^2}{r_0 u^t}\, {}_{-2}\!Y_{\ell0}(\theta_0),\nonumber\\
s_2^{\ell}&=& \frac{2\pi\mu u_{\bar m}^2}{u^t}\, {}_{-2}\!Y_{\ell0}(\theta_0).
\end{eqnarray}
Here we have introduced 
\begin{eqnarray}\label{numu}
u_n&:=&u_{\alpha}n^{\alpha} = \frac{1}{2\mu r_0^2}\left[aL-(r_0^2+a^2)E\right],
\nonumber\\ 
u_{\bar m}&:=u_{\alpha}&\bar m^{\alpha}=\frac{-i}{\sqrt{2}\,\mu r_0}(L-aE),
\end{eqnarray}
where $n^{\alpha}$ and $\bar m^{\alpha}$ are two of the legs of the Kinnersley null tetrad  (\ref{eq:kerrtetrad}) (here evaluated on the orbit), and $u^t$ is the $t$ component of the particle's four-velocity, given explicitly by
\begin{equation}
u^t=\frac{1+\tilde a v^3}{\sqrt{1-3v^2+2\tilde a v^3}}.
\end{equation}
It can be checked that (\ref{snKerr}) reduces to the Schwarzschild expressions (\ref{sn}) for $a=0$.

A suitable basis of radial homogeneous solutions, generalizing that of (\ref{eq:PlQl4}) to Kerr, is 
\begin{eqnarray}\label{eq:PlQl4Kerr}
R_{\ell}^{-}&=& (\kappa\lambda_2)^{-1/2} \Delta(r) {\sf P}_{\ell}^{m=2}\left((r-M)/\kappa\right),
\nonumber\\
R_{\ell}^{+}&=& (\kappa\lambda_2)^{-1/2} \Delta(r) {\sf Q}_{\ell}^{m=2}\left((r-M)/\kappa\right),
\end{eqnarray}
where $\kappa:=\sqrt{M^2-a^2}$, and the normalization is such that the Wronskian, defined as in Eq.\ (\ref{W}), is $W=-\Delta$, just as in the Schwarzschild case. Consequently, the identities (\ref{Widentities}) apply as they are in the Kerr case too. The inhomogeneous solution of the radial Teukolsky equation, with physical boundary conditions, has the same form as in Eqs.\ (\ref{Rlvariation})--(\ref{Cpm}), and one then constructs the one-sided fields $\psi_{4\ell}^{\pm}$ and $\psi_4^{\rm SAS\pm}$ using (\ref{psi4ell}) and (\ref{psi4SAS}) respectively, just as in the Schwarzschild case, only replacing the prefactor $r^{-4}$ in (\ref{psi4SAS}) with $\varrho^4$.


\subsection{Metric reconstruction and perturbation in $\psi_2$}\label{ss:Kerr_cck}

We start by introducing the one-sided Hertz potentials $\Psi^{\pm}$, whose axially-symmetric parts are each required to satisfy the $s=-2$ vacuum Teukolsky equation as well as an ``inversion'' formula, which now reads
\begin{equation}\label{inversionKerr}
\bar\eth^4 \bar\Psi^{\pm} = 8\varrho^{-4}\psi_4^{\rm SAS\pm}.
\end{equation}
We again expand $\Psi^{\pm}$ in ${}_{-2}\!Y_{\ell 0}(\theta)$ as in Eq.\ (\ref{Psiexpansion}), and expand its complex conjugate $\bar\Psi^{\pm}$ in ${}_{+2}\!Y_{\ell 0}(\theta)$ as in Eq.\ (\ref{Psibarexpansion}). Proceeding as in the Schwarzschild case to solve for the modal functions $\Psi_\ell^{\pm}$, one arrives at the unique solution 
\begin{equation}\label{Psiell_Kerr}
\Psi_\ell^{\pm}=8\lambda_2^{-1}\bar\psi_{4\ell}^{\pm},
\end{equation}
whose simple form is identical to that of its Schwarzschild counterpart (\ref{Psiell}). The total (complex-conjugated) Hertz potentials on either sides of $r=r_0$ are thus 
\begin{equation}\label{barPsi}
\bar\Psi^{\pm}=\sum_{\ell=2}^{\infty}(8/\lambda_2)C_{\ell}^{\pm}(r_0)R_{\ell}^{\pm}(r){}_{+2}\!Y_{\ell 0}(\theta),
\end{equation}
where $R_{\ell}^{\pm}$ are the homogeneous solutions given in Eq.\ (\ref{eq:PlQl4Kerr}), and the coefficients $C_{\ell}^{\pm}(r_0)$ are just as in Eq.\ (\ref{Cpm}) but with the source coefficients $s^{\ell}_n$ now as given in Eq.\ (\ref{snKerr}).



For our calculation of the invariants ${\cal I}_{1,2,3}$ we require the $rr$, $r\theta$ and $\theta\theta$ components of the metric perturbation reconstructed from $\Psi^\pm$, as well as the associated perturbation $\psi_2^{(1){\rm rec}\pm}$.   Specializing the reconstruction formula (\ref{eq:kerrh}) to a SAS perturbation in Kerr gives, after some manipulation, 
\begin{eqnarray}\label{hrr_Kerr}
h_{rr}^{{\rm rec}\pm}&=&-{\rm Re}\, \frac{1}{\Delta^{2}\bar\varrho^{4}}\, \bar\eth_1\left(\bar\varrho^2 \bar\eth_2 \bar\Psi^{\pm}\right),
\\ \label{hrtheta_Kerr}
h_{r\theta}^{{\rm rec}\pm}&=&-{\rm Re}\,\frac{1}{\Delta\bar\varrho^{3}\varrho}\left[\left(\bar\varrho\varrho \bar\eth_2\bar\Psi^{\pm}\right)_{,r}-\left(\bar\varrho\varrho\right)_{,\theta}\bar\Psi^{\pm}_{,r}\right],
\\ \label{hthetatheta_Kerr}
h_{\theta\theta}^{{\rm rec}\pm}&=&-{\rm Re}\,\frac{1}{\bar\varrho^{4}}\left(\bar\varrho^2 \bar\Psi^{\pm}_{,r}\right)_{,r}.
\end{eqnarray}
Here the operator $\bar\eth_s:=-\left(\partial_\theta+s\cot\theta\right)$ is the usual spin-lowering operator $\bar\eth$ whenever it acts on ${}_sY_{\ell 0}(\theta)$. Note, however, how in Eqs.\ (\ref{hrr_Kerr})--(\ref{hthetatheta_Kerr}) the reconstructed metric components fail in general to be of a pure spin, due to the dependence of $\varrho$ and $\bar\varrho$ on $\theta$ [this dependence disappears only in the Schwarzschild case, where all three components become manifestly pure-spin ($s=0,1,2$, respectively), with (\ref{hrr_Kerr}) reducing to (\ref{hrr})]. As for $\psi_2^{(1){\rm rec}\pm}$, the reduction of Eq.\ (\ref{eq:deltapsi2}) to a SAS perturbation yields
\begin{align}\label{Psi2_Kerr}
\psi_2^{(1){\rm rec}\pm}=\frac{1}{4}\left(\varrho^2 \bar\eth_1 \bar\eth_2\bar\Psi^{\pm}\right)_{,rr}
-\frac{\varrho_{,\theta}}{\varrho}\left[\varrho\left(\varrho\,\bar\eth_2\bar\Psi^{\pm}\right)_{,r}\right]_{,r}
\nonumber\\
+\frac{3}{2}\varrho_{,\theta}\left(\varrho_{,\theta}\bar\Psi^{\pm}_{,r}\right)_{,r}.
\end{align}
In the Schwarschild limit the last two terms drop (note $\varrho_{,\theta}=ia\varrho^2\sin\theta$) and Eq.\ (\ref{deltapsi2}) is recovered. 

\subsection{Auxiliary invariants}\label{subsec}

The fields ${\cal I}^{{\rm rec}\pm}_{n}$ ($n=1,2,3$) are now obtained as sums over $\ell$-modes by substituting (\ref{barPsi}) in Eqs.\ (\ref{hrr_Kerr})--(\ref{Psi2_Kerr}) and then using Eqs.\ (\ref{eq:gauge9})--(\ref{eq:I_Kerr}). 
The outcome has the form 
\begin{multline}\label{Irec}
{\cal I}_n^{{\rm rec}\pm}=\sum_{\ell=2}^{\infty}\sum_{j=0}^{3}\sum_{k=0}^{3}
(1/\lambda_2){}_2\! Y^{(j)}_{\ell 0}\!(\theta) R_{\ell}^{\pm(k)}(r)
\\ \times
\left\{{\rm Re}[C_{\ell}^{\pm}(r_0)]f_{njk}(r,\theta) +
{\rm Im}[C_{\ell}^{\pm}(r_0)]g_{njk}(r,\theta)
\right\},
\end{multline}
where parenthetical superscripts denote differentiation with respect to the argument.
The coefficients $f_{njk}(r,\theta)$ and $g_{njk}(r,\theta)$ are certain real-valued, $\ell$-independent functions that are simple but many, so we will not list them here but rather proceed directly to evaluating the jumps $\left[{\cal I}^{{\rm rec}}_{n}\right](\theta;r_0)$ across $r=r_0$.  [We only point out one property of these coefficients, namely that, for each $\ell jk$, the entire summand in Eq.\ (\ref{Irec}) is a {\em smooth} function of $r$ and of $\cos\theta$---multiplied by $(\sin\theta)^{-2}$ for $n=2$ and by $(\sin\theta)^{-1}$ for $n=3$. (These singular factors trace back simply to the singular nature of the background Boyer-Lindquist coordinates at the poles; recall ${\cal I}_2=\tilde h_{\theta\theta}$ and ${\cal I}_3=\tilde h_{r\theta}$.) This smoothness property will play a role in the proof of Appendix \ref{a:ProofInterchange}.]

Recall that $C_{\ell}^{\pm}(r_0)$, given in Eq.\ (\ref{Cpm}), are linear combinations (with complex, $r_0$-dependent coefficients) of the real functions $R_{\ell}^{\mp}(r)$ and their first and second derivatives, all evaluated at $r=r_0$. Thus, the jump $\left[{\cal I}^{{\rm rec}}_{n}\right]$ involves the homogeneous solutions $R_{\ell}(r_0)$ only through the combinations listed in Eq.\ (\ref{Widentities})---the same combinations as in the Schwarzschild case. Each of these combinations depends on $\ell$ in a simple way: it is proportional to either $\lambda=(\ell+2)(\ell-1)$, $\lambda_1=\ell(\ell-1)$ or $\lambda_2=\lambda \lambda_1$, or it is $\ell$-independent. Also note, recalling the form of the source coefficients $s_n$ in Eq.\ (\ref{snKerr}), that $C_{\ell}^{\pm}(r_0)$ are linear combinations of ${}_{-2}Y_{\ell0}(\theta_0)$, ${}_{-2}Y_{\ell0}'(\theta_0)$, and ${}_{-2}Y_{\ell0}''(\theta_0)$. Altogether, we therefore have the form 
\begin{multline}\label{JumpForm}
\left[{\cal I}^{{\rm rec}}_{n}\right](\theta;r_0)
= \sum_{j=0}^3\sum_{i=0}^2 \sum_{k=0}^{3}
h_{njik}(\theta;r_0) 
\\ \times
\sum_{\ell=2}^{\infty} \Lambda_{\ell k} \,
{}_2\! Y^{(j)}_{\ell0}\!(\theta){}_{2}\! Y^{(i)}_{\ell0}\!(\theta_0),
\end{multline} 
where $\Lambda_{\ell k}:=\{1,\lambda_1^{-1},\lambda^{-1},\lambda_2^{-1}\}$ respectively for $k=\{0,1,2,3\}$, and we have used the fact that ${}_{-2}Y_{\ell0}\equiv {}_{+2}Y_{\ell0}$. The coefficient $ h_{njik}(\theta;r_0)$ are smooth (except, possibly, at the poles) and independent of $\ell$; they are simple but numerous so we will not list them here. We find it more convenient here to work directly with spin-2 spherical harmonics rather than re-express them in terms of spin-0 harmonics as we did in the Schwarzschild case.

The four sums over $\ell$ in Eq.\ (\ref{JumpForm}) (one for each $k$) can now be evaluated explicitly via term-by-term differentiation of the completeness relation 
\begin{equation} \label{closure}
\sum_{\ell=2}^{\infty} {}_{2}\! Y_{\ell0}\!(\theta){}_{2}\! Y_{\ell0}\!(\theta_0)
=(2\pi)^{-1}\delta(\cos\theta-\cos\theta_0),
\end{equation}
and the summation formulas
\begin{eqnarray}\label{summation1}
\sigma_1:&=&\sum_{\ell=2}^{\infty}\lambda_1^{-1} {}_{2}\! Y_{\ell0}\!(\theta){}_{2}\! Y_{\ell0}\!(\theta_0)
\nonumber\\
&=&\frac{1}{8\pi}\tan^2\!\left(\frac{\theta_{<}}{2}\right)\cot^2\!\left(\frac{\theta_{>}}{2}\right),
\end{eqnarray}
\begin{eqnarray}\label{summation2}
\sigma_2:&=&\sum_{\ell=2}^{\infty}\lambda^{-1} {}_{2}\! Y_{\ell0}\!(\theta){}_{2}\! Y_{\ell0}\!(\theta_0)
\nonumber\\
&=&\frac{1}{3}\sigma_1(2+\cos\theta_{<})(2-\cos\theta_{>}),
\end{eqnarray}
\begin{equation}\label{summation3}
\sigma_3:=\sum_{\ell=2}^{\infty}\lambda_2^{-1} {}_{2}\! Y_{\ell0}\!(\theta){}_{2}\! Y_{\ell0}\!(\theta_0)
=\frac{1}{2}(\sigma_2-\sigma_1).
\end{equation}
Here $\theta_{>}:=\max\{\theta,\theta_0\}$ and $\theta_{<}:=\min\{\theta,\theta_0\}$. A derivation of (\ref{summation1}) and (\ref{summation2}) is presented in Appendix \ref{a:sums}, and (\ref{summation3}) follows directly from $\lambda_2^{-1}=\frac{1}{2}(\lambda^{-1}-\lambda_1^{-1})$.
With the sums over $\ell$ (and $k$) in Eq.\ (\ref{JumpForm}) now explicitly evaluated, we next drop all terms proportional to $\delta(\cos\theta-\cos\theta_0)$ and derivatives thereof (cf.\ Appendix \ref{a:ProofInterchange} once more for a justification), and algebraically simplify the resulting expressions using computer algebra. The final results are remarkably simple:
\begin{equation}\label{JumpI1rec_Kerr}
\left[{\cal I}^{{\rm rec}}_{1}\right](\theta;r_0)=
-\frac{2\Sigma_0\left[(r_0^2+5a^2)E-3aL\right]}{3M\Delta_0^2},
\end{equation}
\begin{equation}\label{JumpI2rec_Kerr}
\left[{\cal I}^{{\rm rec}}_{2}\right](\theta;r_0)=
\frac{2\Sigma_0\left[6L -aE(9-\cos 2\theta)\right]}{6aM\sin^2\theta},
\end{equation}
where $\Sigma_0:=\Sigma(r_0)=r_0^2+a^2\cos^2\theta$, and we find $\left[{\cal I}^{{\rm rec}}_{3}\right]\equiv 0$.
It can be checked that $\left[{\cal I}^{{\rm rec}}_{1}\right]$ reduces to its Schwarzschild value, given in Eq.\ (\ref{JumpIrec}), for $a=0$.

\subsection{Completion piece}

The completion piece of the metric perturbation again has the form (\ref{h_comp}), with amplitudes ${\cal E}^{\pm}$ and ${\cal L}^{\pm}$ to be determined on either sides of $\cal S$. The homogeneous perturbations $h_{\alpha\beta}^{(\delta M)}$ and $h_{\alpha\beta}^{(\delta J)}$ are obtained via Eqs.\ (\ref{eq:hdeltaM}) and (\ref{eq:hdeltaJ}), respectively---this time without taking $J\to 0$. Explicitly, we find 
\begin{eqnarray}\label{eq:dMexplicit}
h_{tt}^{(\delta M)}&=& \frac{2r}{\Sigma ^2} \left(r^2+3a^2\cos^2\theta \right), 
\nonumber\\
h_{t\varphi}^{(\delta M)}&=& -\frac{r a^3 \sin^22\theta}{\Sigma ^2},
\nonumber\\
h_{rr}^{(\delta M)}&=&\frac{2r}{M \Delta ^2} \left[Mr^2+3a^2M+a^2(r-3M)\sin^2\theta)\right],
\nonumber\\
h_{\theta\theta}^{(\delta M)}&=&-(2/M)a^2\cos^2\theta,
\nonumber\\
h_{\varphi\varphi}^{(\delta M)}&=& -\frac{2a^2\sin^2\theta}{M\Sigma^2}
\left[\Sigma^2+Mr(r^2-a^2\cos^2\theta)\sin^2\theta\right],
\nonumber\\
\end{eqnarray}

\begin{eqnarray}\label{eq:dJexplicit}
h_{tt}^{(\delta J)}&=& -\frac{4ar\cos^2\theta}{\Sigma ^2}, 
\nonumber\\
h_{t\varphi}^{(\delta J)}&=& -\frac{2r\sin^2\theta}{\Sigma ^2}\left(r^2-a^2\cos^2\theta\right),
\nonumber\\
h_{rr}^{(\delta J)}&=&-\frac{ar}{M \Delta ^2} \left[r+2M-(r-2M)\cos2\theta\right],
\nonumber\\
h_{\theta\theta}^{(\delta J)}&=&(2/M)a\cos^2\theta,
\nonumber\\
h_{\varphi\varphi}^{(\delta J)}&=& \frac{2a^2\sin^2\theta}{M\Sigma^2}
\left(\Sigma^2+2Mr^3\sin^2\theta\right),
\end{eqnarray}
with all other components vanishing.
The corresponding perturbation in $\psi_2$ on ${\cal S}^{\pm}$ is
\begin{equation}
\psi_2^{(1){\rm comp}\pm}=-\varrho^{4}\left[(r-4ia\cos\theta){\cal E}^{\pm}
+3i{\cal L}^{\pm}\cos\theta \right],
\end{equation}
and the contributions to our auxiliary invariants work out to give 
\begin{equation}
{\cal I}^{{\rm comp}\pm}_{1}(r, \theta)=
\frac{2\Sigma\left[(r^2+5a^2){\cal E}^{\pm}-3a{\cal J}^{\pm}\right]}{3M\Delta^2},
\end{equation}
\begin{equation}
{\cal I}^{{\rm comp}\pm}_{2}(r,\theta)=
-\frac{2\Sigma\left[6{\cal J}^{\pm} -a{\cal E}^{\pm}(9-\cos 2\theta) \right]}{6aM\sin^2\theta},
\end{equation}
with ${\cal I}^{{\rm comp}\pm}_{3}\equiv 0$.
Thus
\begin{equation}\label{JumpI1comp_Kerr}
\left[{\cal I}^{{\rm comp}}_{1}\right](\theta;r_0)=
\frac{2\Sigma_0\left[(r_0^2+5a^2)[{\cal E}]-3a[{\cal J}]\right]}{3M\Delta_0^2},
\end{equation}
\begin{equation}\label{JumpI2comp_Kerr}
\left[{\cal I}^{{\rm comp}}_{2}\right](\theta;r_0)=
-\frac{2\Sigma_0\left[6[{\cal J}] -a[{\cal E}](9-\cos 2\theta)\right]}{6aM\sin^2\theta},
\end{equation}
with $\left[{\cal I}^{{\rm comp}}_{3}\right]\equiv 0$.

\subsection{Determination of $[{\cal E}]$ and $[{\cal J}]$}\label{ss:Kerr_jumps}

The jumps $[{\cal E}]$ and $[{\cal J}]$ can now be determined from the continuity conditions 
$0=[{\cal I}_n]=[{\cal I}_{n}^{{\rm rec}}]+[{\cal I}_{n}^{{\rm comp}}]$. For $n=3$ the condition is 
satisfied trivially and gives us no useful information (besides providing a consistency check). However, the combination of the two conditions $[{\cal I}_1]=0$ and $[{\cal I}_2]=0$ (evaluated at some $\theta\ne 0,\pi$)\footnote{Note ${\cal I}_2=\tilde h_{\theta\theta}$ has a singularity at the poles, which is due to the singular nature of the background Boyer-Lindquist coordinates there. This does not pose a problem to us here.} uniquely determines $[{\cal E}]$ and $[{\cal J}]$: 
\begin{equation}\label{jumps_KerrCirc}
[{\cal E}]=E, \quad\quad [{\cal J}]=L ,
\end{equation}
as immediately seen by comparing Eqs.\ (\ref{JumpI1rec_Kerr}) and (\ref{JumpI2rec_Kerr}) to Eqs.\ (\ref{JumpI1comp_Kerr}) and  (\ref{JumpI2comp_Kerr}). Note that the condition $[{\cal I}_2]=0$ alone uniquely determines both $[{\cal E}]$ and $[{\cal J}]$ if it is to hold for any value of $\theta$. 

We find that the jumps $[{\cal E}]$ and $[{\cal J}]$ are simply the conserved energy and angular momentum of the particle's geodesic orbit, just as in the Schwarzschild case.

\section{Eccentric equatorial orbits in Kerr spacetime}\label{s:ecceqorbits}

As a final generalization, we consider the two-parameter family of bound (eccentric) geodesic orbits in the equatorial plane of a Kerr black hole. The position of the particle is described by $x^\alpha=\{t_{\rm p}(\tau),r_{\rm p}(\tau),\pi/2,\varphi_{\rm p}(\tau)\}$ (Boyer-Lindquist coordinates), where $\tau$ is proper time along the orbit, and the radius is bounded as 
$
r_+<\rmin\leq r_{\rm p}(\tau)\leq\rmax<\infty .
$
The orbits may be parametrized by the pair $\{r_{\rm min},r_{\rm max}\}$, or, alternatively, by the conserved energy $E=-\mu u_t$ and angular momentum $L=\mu u_\varphi$, where we have again written
$u^{\alpha}=dx^{\alpha}/d\tau$ and $u_{\alpha}=g_{\alpha\beta}^{(0)}u^{\beta}$. 
The period of radial libration (i.e., the $t$ interval between two successive periastron crossings at $r_p=r_{\rm min}$) is $P=\int u^t d\tau$, where the integral is taken over a full radial cycle. 
The particle's energy-momentum is given by the distribution (\ref{Tmunu_general}), which in the current case reduces to
\begin{equation}\label{Tmunu_ecc}
T_{\alpha\beta}=\frac{\mu u_{\alpha}u_{\beta}}{r_{\rm p}^2(t) u^t}\delta(r-r_{\rm p}(t))\delta(\cos\theta)\delta(\varphi-\varphi_{\rm p}(t)),
\end{equation}
where by $r_{\rm p}(t)$ we hereafter mean $r_{\rm p}(\tau(t))$, with $\tau(t)$ obtained by inverting $t=t_{\rm p}(\tau)$. 

Our ultimate goal is to determine the completion amplitudes ${\cal E}^{+}$ and ${\cal L}^{+}$ in the vacuum domain ${\cal S}^+:r>r_{\rm p}(t)$, and ${\cal E}^{-}$ and ${\cal L}^{-}$ in the vacuum domain ${\cal S}^-: r_+<r<r_{\rm p}(t)$. For our purpose it will be useful to think of the separating surface ${\cal S}:r=r_{\rm p}(t)$ as a ``pulsating'' 2-sphere, periodically expanding and contracting between $r=r_{\rm min}$ and $r=r_{\max}$.  In this section we will determine the jumps $[{\cal E}]$ and $[{\cal J}]$ across $\cal S$, leaving the determination of the individual amplitudes ${\cal E}^{\pm}$ and ${\cal L}^{\pm}$ to section \ref{s:asymp}.


\subsection{``Partial-ring'' decomposition}

Since the completion piece of the metric perturbation is stationary and axially symmetric, we again concentrate on the SAS part of the reconstructed metric. The SAS part of the energy-momentum source $T_{\alpha\beta}$ (i.e., its $\omega=0=m$ mode) is given by 
\begin{eqnarray}\label{Tmunu_SAS}
\!\!\!\!\!\!
T_{\alpha\beta}^{\rm SAS}&=&\frac{1}{2\pi P}\int_{0}^P \!\! dt \int_{0}^{2\pi}\!\! d\varphi \, T_{\alpha\beta}
\nonumber\\
&=& \frac{\mu U_{\alpha\beta}(r)}{\pi P  r^2 \dot{r}(r)}\Theta(r-r_{\rm min})\Theta(r_{\rm max}-r)\delta(\cos\theta),
\end{eqnarray}
where $\Theta(\cdot)$ is the Heaviside step function,
\begin{equation}\label{U}
U_{\alpha\beta}(r):=\left\{
\begin{array}{ll}
0,	& \alpha\beta\in\{rt,r\varphi,tr,\varphi r\}, \\
u_{\alpha}u_{\beta}	,& \text{otherwise},
\end{array}
\right.
\end{equation}
and we have defined $\dot{r}(r):=|u^r(r)|$. In both this last expression and in Eq.\ (\ref{U}), the four-velocity components $u^\alpha$ are regarded as functions of $r$ along the ``outbound'' part of the orbit going from $r_{\rm min}$ out to $r_{\rm max}$. The second line of (\ref{Tmunu_SAS}) ``folds over'' the contribution from each point on the inbound part [$u^r(r)<0$] onto that of the corresponding outbound point [$u^r(r)>0$ with same $r$].
We see that the SAS source is supported on an equatorial annulus of inner radius $r_{\rm min}$ and outer radius $r_{\rm max}$.

To proceed, it would now be tempting to consider $T_{\alpha\beta}^{\rm SAS}$ as a linear superposition of static, circular-ring sources, each with a radius $r_{\rm min}\leq R\leq r_{\rm max}$ and energy-momentum of (say)
\begin{equation}\label{TR}
T_{\alpha\beta}^{(R)}=\frac{\mu U_{\alpha\beta}(R)}{\pi P R^2 \dot{r}(R)}\delta(r-R)\delta(\cos\theta),
\end{equation}
so that
\begin{equation}
T_{\alpha\beta}^{\rm SAS}=\int_{r_{\rm min}}^{r_{\rm max}} T_{\alpha\beta}^{(R)}dR.
\end{equation}
Then, perhaps, one could proceed precisely as in the circular-orbit case, constructing the jumps $[{\cal E}]$ and $[{\cal J}]$ for each such ``partial ring'' individually, and then integrating over ring contributions to obtain the total jumps. However, here one must exercise caution. A naively constructed $R$-ring is not necessarily an admissible, ``conserved'' physical source: it can be easily checked that $\nabla^\beta T_{\alpha\beta}^{(R)}\ne 0$ for the example in (\ref{TR}). It is then unclear whether invariant fields constructed from (completed) perturbations in the vacuum regions $r>R$ and $r<R$ are to be expected to match continuously across $r=R$ away from the equator (as they do for a physical, circular geodesic). In fact, our explicit calculation below will demonstrate that invariant fields sourced by the $T_{\alpha\beta}^{(R)}$ of Eq.\ (\ref{TR}) can be {\it dis}continuous on the sphere $r=R$. 

We resolve this difficulty by designing a modified decomposition of $T_{\alpha\beta}^{\rm SAS}$ into partial rings, each with energy-momentum $\tilde T_{\alpha\beta}^{(R)}$ satisfying $\nabla^\beta \tilde T_{\alpha\beta}^{(R)}=0$. Different choices of such ``conserved'' $R$-rings are possible. One that we find particularly convenient (because it leads to a particularly simple source for the Teukolsky equation; see below) is
\begin{equation}\label{tildeT}
\tilde T_{\alpha\beta}^{(R)}=T_{\alpha\beta}^{(R)}+
\frac{\mu }{\pi P}\dot{r}(r)A_{\alpha\beta}(r)\delta'(r-R)\delta(\cos\theta),
\end{equation}
with
\begin{equation}
A_{tt}=\frac{1}{r}, \quad
A_{t\varphi}=-\frac{r^2+2a^2}{2ra}=A_{\varphi t}, \quad
A_{\varphi\varphi}=\frac{r^2+a^2}{r}
\end{equation}
($A_{\alpha\beta}=0$ for all other components), where a prime denotes a derivative with respect to the argument, and $T_{\alpha\beta}^{(R)}$ is the original, ``non-conserved'' source given in Eq.\ (\ref{TR}). Physically, the added $\delta'$ term supplies the differential pressure necessary to balance the overall pressure on the static $R$-ring. Yet this added term contributes nothing when integrated over all rings (note $\dot{r}=0$ at the turning radii $r=r_{\rm min},r_{\rm max}$, while $A_{\alpha\beta}$ is bounded there), so that
\begin{equation}
T_{\alpha\beta}^{\rm SAS}=\int_{r_{\rm min}}^{r_{\rm max}} \tilde T_{\alpha\beta}^{(R)}dR
\end{equation}
as required.

(We note that our choice of $\tilde T^{(R)}$ is unsuitable for $a=0$, where $A_{t\varphi}$ becomes indefinite.  This will turn out not to be a problem: our final expressions for the jumps $[{\cal E}]$ and $[{\cal J}]$ will appear to have perfectly regular limits $a\to 0$, meaning the Schwarzschild case is also accessible to our analysis, in effect. There exist choices of $\tilde T^{(R)}$ that avoid the irregularity at $a=0$, but among these we could not find one that was as simple to work with as ours.)

Our plan of action now is as follows. Considering an individual, conserved partial ring with a particular (but arbitrary) value of $R$, we will reconstruct the physical metric perturbation and corresponding invariant fields ${\cal I}^{\rm rec}_n$ in the vacuum domains $r>R$ and $r<R$, in exactly the same manner as for a circular geodesic orbit. We will then impose that the completed invariant fields corresponding to the $R$-ring are continuous on $r=R$ (away from the equator, and excluding the poles), and use this condition to determine the partial $R$-ring contributions to the jumps $[{\cal E}]$ and $[{\cal J}]$---call these $[{\cal E}]^{(R)}$ and $[{\cal J}]^{(R)}$, respectively; hereafter we use superscripts `$(R)$' to label $R$-ring contributions to relevant quantities: $\psi_4^{(R)\pm}$, $\Psi^{(R)\pm}$, $h_{\alpha\beta}^{(R){\rm rec}\pm}$, etc.
From linearity, the completion pieces of the {\it total} metric perturbation at $r>r_{\rm max}$ and $r<r_{\rm min}$ are given by, respectively,  
\begin{align}
h_{\alpha\beta}^{{\rm comp}\pm}(r,\theta)=
h_{\alpha\beta}^{(\delta M)}(r)\int_{r_{\rm min}}^{r_{\rm max}}{\cal E}^{(R)\pm}dR
\nonumber \\
+
h_{\alpha\beta}^{(\delta J)}(r,\theta)\int_{r_{\rm min}}^{r_{\rm max}}{\cal J}^{(R)\pm}dR,
\end{align}
so, recalling Eq.\ (\ref{h_comp}), the total jumps across $\cal S$ are finally obtained using
\begin{eqnarray}\label{totalEL}
[{\cal E}]&=&\int_{r_{\rm min}}^{r_{\rm max}}[{\cal E}]^{(R)}dR,
\nonumber\\
{}[{\cal J}]&=&\int_{r_{\rm min}}^{r_{\rm max}}[{\cal J}]^{(R)}dR.
\end{eqnarray}

\subsection{Metric reconstruction and auxiliary invariants for a partial ring}

We start by writing $\psi_4^{(R)}$ as a sum over $\ell$-modes as in Eq.\ (\ref{psi4_separation}). The radial functions $R^{(R)}_\ell(r)$ satisfy the modal Teukolsky equation 
\begin{equation}\label{eq:TeukolskyEcc}
\Delta^2 \frac{d}{dr}\left(\Delta^{-1}\frac{R^{(R)}_{\ell}}{dr}\right)-\lambda\,R^{(R)}_{\ell}= \tilde T^{(R)}_{\ell}(r;R),
\end{equation}
where the source corresponds to the $R$-ring energy-momentum $\tilde T_{\alpha\beta}^{(R)}$. This source is derived as prescribed in Appendix \ref{a:KerrBack}---see, in particular, Eqs.\ (\ref{eq:kerrsource-2}) and (\ref{Tslm}). The derivation is straightforward albeit tedious and we will not review it here but simply state the result:
\begin{multline}\label{eq:sourceEcc}
\tilde T^{(R)}_{\ell}(r;R)=\frac{\Delta^2(r)}{P}
\left[\tilde s^{\ell}_0(R)\delta(r-R) + \right.
\\ 
\left. \tilde s^{\ell}_1(R)\delta'(r-R)
+\tilde s^{\ell}_2(R)\delta''(r-R)\right],
\end{multline}
analogous in form to the circular-orbit source [compare with Eq.\ (\ref{snKerr})].
The fact that no third derivatives of $\delta(r-R)$ occur [despite the presence of a $\delta'(r-R)$ term in $\tilde T_{\alpha\beta}^{(R)}$] owes itself to our particular choice of coefficients $A_{\alpha\beta}$ in Eq.\ (\ref{tildeT}); indeed, avoiding such a term in $\tilde T^{(R)}_{\ell}$ was our prime motivation in making that choice. The coefficients $\tilde s^{\ell}_n$ in Eq.\ (\ref{eq:sourceEcc}) read
\begin{eqnarray}\label{snKerrEcc}
\tilde s_0^{\ell}&=& \frac{4\mu R^2 U_{nn}^2}{\Delta^2 \dot{r}(R)}\left[{}_{-2}\!Y''_{\ell0}(\theta_0)-2{}_{-2}\!Y_{\ell0}(\theta_0)\right]
\nonumber\\ && \quad\quad
-\frac{8\mu (\sqrt{2}\,U_{n\bar m} \Delta +i a R U_{nn})}{\Delta^2\dot{r}}\, {}_{-2}\!Y'_{\ell0}(\theta_0) \nonumber\\
		&& \quad\quad
		+\frac{2i\mu R\left[2(a^2-MR)\dot{r}+R\dot{r}'\Delta\right]}{a\Delta^2}\, {}_{-2}\!Y'_{\ell0}(\theta_0),
	\nonumber\\
\tilde s_1^{\ell}&=& -\frac{4\sqrt{2}\,\mu R U_{n\bar m}}{\dot{r}\Delta}\, {}_{-2}\!Y'_{\ell0}(\theta_0)
		+\frac{4\mu U_{\bar m\bar m}}{R \dot{r}}\, {}_{-2}\!Y_{\ell0}(\theta_0)\nonumber\\
		&& \quad\quad
		+\frac{i\mu R^2\left[(a^2-R^2)\dot{r}+R\dot{r}'\Delta\right]}{a\Delta^2}\, {}_{-2}\!Y'_{\ell0}(\theta_0),
	\nonumber\\
\tilde s_2^{\ell}&=&  \frac{2\mu U_{\bar m\bar m}}{\dot{r}}\, {}_{-2}\!Y_{\ell0}(\theta_0)
-\frac{i\mu R^3 \dot{r}}{a\Delta}\, {}_{-2}\!Y'_{\ell0}(\theta_0),
\end{eqnarray}
where $U_{nn}:=U_{\alpha\beta}n^{\alpha}n^{\beta}$ (etc.), $\Delta=\Delta(R)$, $\dot{r}=\dot{r}(R)$, and $\dot{r}'=d\dot{r}(R)/dR$. 
In each of the expressions (\ref{snKerrEcc}), the last term ($\propto {}_{-2}\!Y'_{\ell0}$) is due to the $\delta'(r-R)$ term of $\tilde T_{\alpha\beta}^{(R)}$.

From this point onward, the calculation proceeds just as for circular orbits (Sec.\ \ref{s:Kerrcirc}), simply replacing the source coefficients $s_n^\ell$ of Eq.\ (\ref{snKerr}) with the coefficients $\tilde s_n^\ell$ of Eq.\ (\ref{snKerrEcc}) (and, of course, replacing $r_0$ with $R$). For a given value of $R$, we construct the vacuum solutions $\psi_{4\ell}^{(R)\pm}$ and the corresponding Hertz potentials $\Psi_\ell^{(R)\pm}$, and then reconstruct the $R$-ring perturbation in the metric and in $\psi_2$. From these we finally obtain the (reconstructed piece of the) invariants, ${\cal I}_n^{(R){\rm rec}\pm}$. These have precisely the form (\ref{Irec}) they had for a circular geodesic, with the same coefficients $f_{njk}$ and $g_{njk}$ (but replacing $r_0\to R$). The replacement $s_n^\ell\to\tilde s_n^\ell$ affects only the explicit values of the functions $C_{\ell}^{\pm}(R)$ [via Eq.\ (\ref{Cpm})]. Given ${\cal I}_n^{(R){\rm rec}\pm}$, we then proceed as described in subsection \ref{subsec} to obtain an expression analogous to (\ref{JumpForm}) for the jumps $[{\cal I}_n^{\rm rec}]^{(R)}(\theta;R)$ across the sphere ${\cal S}^{(R)}$ corresponding to the $R$-ring. Exactly the same sums over $\ell$ occur in this expression, and they are again evaluated analytically using the summations formulas (\ref{summation1})--(\ref{summation3}). 

We thus obtain 
\begin{equation}\label{JumpI1rec_ecc}
\left[{\cal I}^{{\rm rec}}_{1}\right]^{(R)}=
\Sigma(R,\theta)f_1(R),
\end{equation}
\begin{equation}\label{JumpI2rec_ecc}
\left[{\cal I}^{{\rm rec}}_{2}\right]^{(R)}=
\frac{\Sigma(R,\theta)\left[f_2(R)+f_{2c}(R)\cos 2\theta\right]}{\sin^2\theta},
\end{equation}
with $[{\cal I}_3^{(R)\rm rec}](\theta;R)\equiv 0$. Here $\Sigma(R,\theta)=R^2+a^2\cos^2\theta$. The coefficients $f_1$, $f_2$ and $f_{2c}$ depend on $R$, as well as on $E$, $L$ and $a$, but in terms of these variables they take a rather complicated form---especially in comparison with the simple circular-orbit counterparts (\ref{JumpI1rec_Kerr}) and (\ref{JumpI2rec_Kerr}). Simplification is achieved by replacing (some, but not all occurrences of) $L$ and $L^2$ in favour of $\dot{r}(R)$ and $\dot{r}'(R)$ [using the normalization $u_{\alpha}u^{\alpha}=-1$ and equation of motion $d(u_{\alpha}u^{\alpha})/dR=0$ as a coupled set, and solving for $\{L,L^2\}$ in terms of $\{\dot{r}(R),\dot{r}'(R)\}$, treating the two variables in each pair as mutually independent for that purpose]. Anticipating the next step of our calculation, we further manipulate the expressions for $f_n$ to bring them to a form more readily amenable to integration over $R$. We obtain
\begin{eqnarray}
f_{2}(R)&=& \mu\left(A(R)\dot{r}(R)\right)'+\frac{2t'(R)}{PM}(2L/a-3E),
\nonumber\\
f_{2c}(R) &=& \mu\left(A_c(R)\dot{r}(R)\right)'+\frac{2t'(R)E}{3PM},
\nonumber\\
f_1(R) &=& \left[a^2f_2(R)-(2R^2+a^2)f_{2c}(R)\right]/\Delta^2(R),
\nonumber\\
\end{eqnarray}
where $t'(R):=u^t(R)/\dot{r}(R)$,
\begin{eqnarray}
A&=& -\frac{2R(R^4+RMa^2-a^4)}{a^2M\Delta(R)P},
\nonumber\\
A_{c}&=& -\frac{2R(R^2-RM+a^2)}{3M\Delta(R)P},
\end{eqnarray}
and a prime denotes $d/dR$.

\subsection{Completion amplitudes for a partial ring}

We write the completion piece of the $R$-ring metric perturbation in the form
\begin{equation}\label{h_comp_again2}
h^{(R){\rm comp\pm}}_{\alpha\beta}= {\cal E}^{(R)\pm}h^{(\delta M)}_{\alpha\beta}+{\cal J}^{(R)\pm}h^{(\delta J)}_{\alpha\beta},
\end{equation}
where $h^{(\delta M)}_{\alpha\beta}$ and $h^{(\delta J)}_{\alpha\beta}$ are the homogeneous mass and angular-momentum perturbations given in Eqs.\ (\ref{eq:dMexplicit}) and (\ref{eq:dJexplicit}), and ${\cal E}^{(R)\pm}$ and ${\cal J}^{(R)\pm}$ are amplitudes to be determined. The corresponding contribution to the jumps in the invariant fields can be read off Eqs.\ (\ref{JumpI1comp_Kerr}) and (\ref{JumpI2comp_Kerr}), simply replacing $r_0\to R$:
\begin{equation}\label{JumpI1comp_R}
\left[{\cal I}^{{\rm comp}}_{1}\right]^{(R)}=
\frac{2\Sigma(R,\theta)\left((R^2+5a^2)[{\cal E}]^{(R)}-3a[{\cal J}]^{(R)}\right)}{3M\Delta^2(R)},
\end{equation}
\begin{equation}\label{JumpI2comp_R}
\left[{\cal I}^{{\rm comp}}_{2}\right]^{(R)}=
-\frac{2\Sigma(R,\theta)\left(6[{\cal J}]^{(R)} -a(9-\cos 2\theta)[{\cal E}]^{(R)}\right)}{6aM\sin^2\theta},
\end{equation}
with $\left[{\cal I}^{{\rm comp}}_{3}\right]^{(R)}\equiv 0$.

We now require, for each individual $R$-ring, that
\begin{equation}\label{continuity_R}
\left[{\cal I}^{{\rm rec}}_{n}\right]^{(R)}+\left[{\cal I}^{{\rm comp}}_{n}\right]^{(R)} = 0
\end{equation}
identically for all $\theta\ne \pi/2$ (with the exclusion of the poles).
This continuity condition is satisfied trivially for $n=3$, while for $n=1$ it determines a certain linear combination of the sought-for amplitudes $[{\cal E}]^{(R)}$ and $[{\cal J}]^{(R)}$. Considering Eq.\ (\ref{JumpI2comp_R}) in conjunction with (\ref{JumpI2rec_ecc}), we see that, for $n=2$, the continuity condition determines $[{\cal E}]^{(R)}$ and $[{\cal J}]^{(R)}$ individually. For all $n$, the manifest consistency in the angular profile, between $\left[{\cal I}^{{\rm rec}}_{n}\right]^{(R)}$ and $\left[{\cal I}^{{\rm comp}}_{n}\right]^{(R)}$,  guarantees that (\ref{continuity_R}) can be satisfied on the entire of $\check{\cal S}^{(R)}$, as required. This consistency of angular profile constitutes a strongly non-trivial test of our method and calculation. We point out, for example, that a calculation based on the non-conserved $R$-rings with energy momentum $T_{\alpha\beta}^{(R)}$ as in Eq.\ (\ref{TR}) yields $\left[{\cal I}^{{\rm rec}}_{1,2}\right]^{(R)}$ whose $\theta$ dependence is inconsistent with that of $\left[{\cal I}^{{\rm comp}}_{1,2}\right]^{(R)}$ (specifically, $\left[{\cal I}^{{\rm comp}}_{1}\right]^{(R)}$ picks up an additional term $\propto \Sigma|\cos\theta|$, and $\left[{\cal I}^{{\rm comp}}_{2}\right]^{(R)}$ picks up an additional term $\propto \Sigma|\cos\theta|/\sin^2\theta$). No values of $[{\cal E}]^{(R)}$ and $[{\cal J}]^{(R)}$ then satisfy the continuity conditions (\ref{continuity_R}) on the entire of $\check{\cal S}^{(R)}$. This is a reassuring evidence in validation of our procedure for constructing ``conserved'' $R$-rings.

Solving Eq.\ (\ref{continuity_R}) with $n=2$ for $[{\cal E}]^{(R)}$ and $[{\cal J}]^{(R)}$, we now obtain
\begin{eqnarray}\label{JER}
[{\cal E}]^{(R)}&=& 3M f_{2c}
\nonumber\\
&=& 3\mu M\left(A_{c}(R)\dot{r}(R)\right)'+\frac{2t'(R)}{P}E,
\end{eqnarray}
\begin{eqnarray}\label{JJR}
[{\cal J}]^{(R)}&=& \frac{1}{2}aM(f_2+9f_{2c})
\nonumber\\
&=&   \mu\left(B(R)\dot{r}(R)\right)'+\frac{2t'(R)}{P}L,
\end{eqnarray}  
with 
\begin{equation}
B=-\frac{R(R^4+3R^2 a^2-2R M a^2+2a^4)}{a\Delta(R)P}.
\end{equation}
It can be checked that this solution satisfies Eq.\ (\ref{continuity_R}) for $n=1$ as well.
Interestingly, we find that, for each and every $R$-ring, the jumps $[{\cal E}]^{(R)}$ and $[{\cal J}]^{(R)}$ are simply $E$ and $L$, respectively (multiplied by $2t'/P$), up to terms that are total derivatives along the orbit. 


\subsection{Determination of $[{\cal E}]$ and $[\cal J]$}

The sought-for jumps across $\cal S$ in the total completion amplitudes $\cal E$ and $\cal J$ are now obtained by integrating $[{\cal E}]^{(R)}$ and $[{\cal J}]^{(R)}$ over all $R$-rings, using Eqs.\ (\ref{totalEL}). Since $A_{c}(R)$ and $B(R)$ are bounded at the integration boundaries, $R=r_{\rm min},r_{\rm max}$, while $\dot{r}(R)=0$ there, we find that the total-derivative terms in Eqs.\ (\ref{JER}) and (\ref{JJR}) do not contribute to the integrals. We are left with, simply,
\begin{eqnarray}\label{jumps_Kerr}
[{\cal E}]&=&\frac{2}{P}\int_{r_{\rm min}}^{r_{\rm max}} E\, t'(R)dR = E,
\nonumber \\
{}
[{\cal J}]&=&\frac{2}{P}\int_{r_{\rm min}}^{r_{\rm max}} L\, t'(R)dR = L,
\end{eqnarray}
remarkably generalizing the simple result (\ref{jumps_KerrCirc}) to any eccentric orbit in Kerr spacetime. 


\section{Mass and angular-momentum contents of $h^{{\rm rec}\pm}_{\alpha\beta}$ and $h^{{\rm comp}\pm}_{\alpha\beta}$}\label{s:asymp}

As we have seen, the field equations, with the regularity condition for invariant fields off the particle, uniquely determine the jumps $[\cal E]$ and $[\cal J]$ across $\cal S$. However, they alone do not determine the individual amplitudes ${\cal E}^{\pm}$ and ${\cal L}^{\pm}$ on ${\cal S}^\pm$. These remain arbitrary, since one is free to add to the metric any vacuum mass or angular-momentum perturbations, i.e any perturbation of the form ${\cal E}h_{\alpha\beta}^{(\delta M)}+{\cal J}h_{\alpha\beta}^{(\delta L)}$, with arbitrary $\cal E$ and $\cal J$, without violating either the field equations or the regularity assumption.  To specify the individual amplitudes ${\cal E}^{\pm}$ and ${\cal L}^{\pm}$ requires additional information, as discussed at the end of Sec.\ \ref{s:Schcir}. 

In this section we determine the individual amplitudes ${\cal E}^{\pm}$ and ${\cal L}^{\pm}$ from conditions on the total mass and angular-momentum contents of spacetime, combined with the now-known jumps $[{\cal E}]$ and $[{\cal J}]$. Referring to a specific (perturbative) notion of quasi-local mass and angular momentum, we then also discuss a restatement of our main result (\ref{jumps_Kerr}) in terms of the mass and angular-momentum contents of the {\em reconstructed} piece of the perturbation: We show that {\it the reconstructed perturbation has no mass or angular momentum} in either ${\cal S}^+$ or ${\cal S}^-$. This may be seen as, effectively, a corollary of (\ref{jumps_Kerr}).

For the discussion in this section we adopt the quasilocal notions of mass and angular momentum introduced by Abbott and Deser \citep{AbbDes} in the context of linear perturbation theory. The Abbott-Deser formulation will serve us well here, for several reasons. First, it can be applied to the full metric perturbation---as opposed to the Komar definitions, which require a Killing symmetry and are thus only applicable to the SAS piece of spacetime. Second, it can be applied at spatial infinity to obtain the total mass and angular momentum of the full perturbation even for an ``eternally'' periodic radiating source (as ours is assumed to be)---compared to the ADM quantities, which are ill-defined in that case (at least formally). Finally, Abbott-Deser definitions apply quasilocally, unlike the ADM or Bondi notions, which are defined at infinity only. We emphasize, however, that all above definitions---Komar's, ADM's, Bondi's and Abbott--Deser's---coincide and agree when applied to the SAS part of the perturbed spacetime at infinity. The Abbott-Deser definitions also agree with Komar's quasilocally when applied to the SAS part.

The structure of the rest of this section is as follows. In Sec.\ (\ref{AD}) we review the Abbott-Deser definitions of mass and angular momenum in perturbation theory. In Sec.\ (\ref{EJpm}) we determine the individual amplitudes ${\cal E}^{\pm}$ and ${\cal J}^{\pm}$ from conditions on the mass and angular momentum at infinity. Finally, in Sec.\ (\ref{NoMass}) we discuss the implications of our results with regard to the mass and angular-momentum contents of the reconstructed perturbation.

\subsection{The Abbott-Deser formulation}\label{AD}

The Abbott--Deser construction (\citep{AbbDes}, as reviewed in \citep{dolan3}) applies to a generic metric perturbation $h_{\alpha\beta}$ of a vacuum background metric $g_{\alpha\beta}^{(0)}$ admitting a Killing vector field $k^{\alpha}$. (We emphasize that, unlike in the Komar definitions, the full spacetime $g_{\alpha\beta}^{(0)}+h_{\alpha\beta}$ need not have any symmetry; only symmetry in the background is required.) One introduces the antisymmetric two-form
\begin{equation}
F_{\alpha \beta} := - \frac{1}{8 \pi} \left( k^\lambda \bar{h}_{\lambda [\alpha ; \beta]} + {k^\lambda}_{;[\alpha} \bar{h}_{\beta] \lambda} + k_{[\alpha} \bar{h}_{\beta]\lambda}^{\ \ ;\lambda}  \right),  \label{AD_F}
\end{equation}
where
$\bar{h}_{\alpha \beta} := h_{\alpha \beta} - \frac{1}{2} g^{(0)}_{\alpha \beta} h^{\ \lambda}_\lambda$
is the trace-reversed metric perturbation, a semicolon denotes a covariant derivative compatible with $g_{\alpha\beta}^{(0)}$, and square brackets indicate antisymmetrization of indices, as in  $\bar{h}_{[\alpha ; \beta]} = \frac{1}{2} (\bar h_{\alpha;\beta} - \bar h_{\beta;\alpha})$. The key property of $F_{\alpha \beta}$ is that ${F_{\alpha \beta}}^{;\beta} = T_{\alpha \beta} k^\beta=:j_{\alpha}$, where $T_{\alpha \beta}$ is the energy momentum-tensor appearing on the right-hand side of the linearized Einstein's equations. Assuming ${T_{\alpha\beta}}^{;\beta}=0$, and since $k_{(\alpha;\beta)}=0$, we have that the ``current'' $j^{\alpha}$ is divergence-free. This allows us to formulate a conservation law for a ``charge'' $Q$ defined by integrating $j^{\alpha}$ over a spacelike 3-volume $\Sigma$ (assuming $j^{\alpha}=0$ on the boundary $\partial\Sigma$ of that volume). Furthermore, using Stokes' theorem, it is possible to relate $Q$ to the surface integral 
\begin{equation}
{\cal F}(h_{\alpha\beta};k^\alpha,\partial\Sigma):= \frac{1}{2}  \int_{\partial\Sigma} F^{\alpha \beta} d\Sigma_{\alpha \beta} ,  \label{stokes}
\end{equation}
in which $d\Sigma_{\alpha \beta}$ is an appropriate 2-surface element on $\partial\Sigma$ 
(see \citep{dolan3} for details).

Specializing now to a Kerr background, we have the two Killing vector fields $k^\alpha_{(t)}:=\partial x^{\alpha}/\partial t$ and $k^\alpha_{(\varphi)}:=\partial x^{\alpha}/\partial\varphi$ associated, respectively, with the stationarity and axial symmetry of the Kerr geometry. To each of these there corresponds a quasilocally conserved integral:
\begin{eqnarray}\label{AD_ML}
{\cal M}_{\rm AD}(h_{\alpha\beta};\partial\Sigma)&:=&{\cal F}(h_{\alpha\beta};k^\alpha_{(t)},\partial\Sigma) ,
\nonumber\\
{\cal L}_{\rm AD}(h_{\alpha\beta};\partial\Sigma)&:=&{\cal F}(h_{\alpha\beta};k^\alpha_{(\varphi)},\partial\Sigma) 
\end{eqnarray}
(again, assuming $j^{\alpha}=0$ on $\partial\Sigma$). 
We henceforth refer to ${\cal M}_{\rm AD}$ and ${\cal L}_{\rm AD}$ as the Abbott--Deser (AD) mass and angular momentum, and note that they depend only on the value of the metric perturbation $h_{\alpha\beta}$ on the surface $\partial\Sigma$. These quantities may be interpreted as the total mass and angular-momentum contents of the metric perturbation in the volume enclosed within $\partial\Sigma$. In Ref.\ \citep{dolan3}, Dolan and Barack establish that ${\cal M}_{\rm AD}$ and ${\cal L}_{\rm AD}$ (unlike $F_{\alpha\beta}$ itself) are gauge invariant, as required. 

To further illustrate that the above interpretation makes physical sense (note, for instance, the sensitivity of ${\cal M}_{\rm AD}$ and ${\cal L}_{\rm AD}$ to the choice of normalization for the Killing vector fields), Ref.\ \citep{dolan3} considered the example of a point particle moving on a bound geodesic orbit around the Kerr black hole. It showed that, for any solution $h_{\alpha\beta}$ of the inhomogeneous linearized Einstein's equations,
\begin{eqnarray}\label{AD_ML_particle}
{\cal M}_{\rm AD}(h_{\alpha\beta};\partial\Sigma_2)-{\cal M}_{\rm AD}(h_{\alpha\beta};\partial\Sigma_1)&=&E ,
\nonumber\\
{\cal L}_{\rm AD}(h_{\alpha\beta};\partial\Sigma_2)-{\cal L}_{\rm AD}(h_{\alpha\beta};\partial\Sigma_1)&=&L,
\end{eqnarray}
where $\Sigma_{1,2}$ are any 2-spheres defined by $t=\text{const}$ and $r=r_{1,2}$ for some constant Boyer-Lindquist radii satisfying  $r_+<r_1<r_{\rm p}(t)$ and $r_2>r_{\rm p}(t)$. Thus, ${\cal M}_{\rm AD}$ and ${\cal L}_{\rm AD}$ have constant values on each of the separate vacuum domains ${\cal S}^\pm$, and these values ``jump'' across $\cal S$ by amounts precisely equal to (respectively) the geodesic energy and angular momentum of the particle. This further reinforces the interpretation of ${\cal M}_{\rm AD}$ and ${\cal L}_{\rm AD}$ as energy and angular momentum.

\subsection{Determination of ${\cal E}^{\pm}$ and ${\cal J}^{\pm}$}\label{EJpm}

Let us now return to the question of determining the individual amplitudes ${\cal E}^{\pm}$ and ${\cal J}^{\pm}$ in the completion piece $h_{\alpha\beta}^{{\rm comp}\pm}$ [recall Eq.\ (\ref{h_comp})], given the jumps $[{\cal E}]$ and $[{\cal J}]$. For the following discussion, we write the completed metric perturbation outside of ${\cal S}$ as
\begin{eqnarray}\label{h+}
h_{\alpha\beta}^{+}&=&h_{\alpha\beta}^{{\rm rec}+}+h_{\alpha\beta}^{{\rm comp}+}
\nonumber\\
&=& (h_{\alpha\beta}^{{\rm rec}+})^{\rm SAS}+
(h_{\alpha\beta}^{{\rm rec}+})^{\rm nonSAS}
+{\cal E}^{+}h^{(\delta M)}_{\alpha\beta}+{\cal J}^{+}h^{(\delta J)}_{\alpha\beta},
\nonumber\\
\end{eqnarray}
where we have split the reconstructed piece into its SAS part $(h_{\alpha\beta}^{{\rm rec}+})^{\rm SAS}$ and its non-SAS part $(h_{\alpha\beta}^{{\rm rec}+})^{\rm nonSAS}:=h_{\alpha\beta}^{{\rm rec}+}-(h_{\alpha\beta}^{{\rm rec}+})^{\rm SAS}$. Our strategy will be as follows. 
We will calculate the total AD mass and angular momentum in the completed metric perturbation, by explicitly evaluating ${\cal M}_{\rm AD}^{\infty}(h_{\alpha\beta}^+):={\cal M}_{\rm AD}(h_{\alpha\beta}^+;\partial\Sigma_\infty)$ and ${\cal L}_{\rm AD}^{\infty}(h_{\alpha\beta}^+):=(h_{\alpha\beta}^+;\partial\Sigma_\infty)$, where $\Sigma_\infty$ is the surface  $t,r={\rm const}$ with $r\to\infty$; we will separately evaluate the contributions to ${\cal M}_{\rm AD}^{\infty}$ and ${\cal L}_{\rm AD}^{\infty}$ from each of the four terms in the second line of (\ref{h+}) and then add them up. The result will be an expression for  ${\cal M}_{\rm AD}^{\infty}$ and ${\cal L}_{\rm AD}^{\infty}$ in terms of the (yet unknown) amplitudes ${\cal E}^+$ and ${\cal J}^+$. We will then {\it impose}
\begin{equation}\label{AD_total}
{\cal M}_{\rm AD}^{\infty}(h_{\alpha\beta}^+)=E,\quad\quad
{\cal L}_{\rm AD}^{\infty}(h_{\alpha\beta}^+)=L,
\end{equation} 
 and solve the resulting set of equations for ${\cal E}^+$ and ${\cal J}^+$. The amplitudes ${\cal E}^-$ and ${\cal J}^-$ will follow  immediately from the known jumps, ${\cal E}^+-{\cal E}^-=E$ and ${\cal J}^+-{\cal J}^-=L$.

Our choice (\ref{AD_total}) is equivalent, by virtue of (\ref{AD_ML_particle}), to setting 
\begin{equation}\label{AD_in}
{\cal M}_{\rm AD}^{\cal H}(h_{\alpha\beta}^-)=0,\quad\quad
{\cal L}_{\rm AD}^{\cal H}(h_{\alpha\beta}^-)=0,
\end{equation}
where ${\cal M}_{\rm AD}^{\cal H}$ and ${\cal L}_{\rm AD}^{\cal H}$ are the AD integrals evaluated on the event horizon, $r=r_+$. This amounts to choosing the central black hole to be of Kerr mass $M$ and spin $aM$. We should remain mindful, though, of the fact that the choice (\ref{AD_total}) is, to an extent, arbitrary, and should be considered in relation to the specifics of the problem at hand.\footnote{As an example: The retarded, asymptotically-flat and horizon-regular Lorenz-gauge metric perturbation associated with an orbiting particle in Schwarzschild spacetime is known to have a nonzero value of ${\cal M}_{\rm AD}^{\cal H}$ \cite{dolan3}, which may be absorbed into a redefinition of the background mass. This appears to be a convenient strategy in second-order self-force calculations \cite{2ndorder}.}


Let us now implement the above strategy, starting with the evaluation of ${\cal M}_{\rm AD}^{\infty}(h_{\alpha\beta}^{+})$ and ${\cal L}_{\rm AD}^{\infty}(h_{\alpha\beta}^{+})$. We will consider one by one the contributions coming from each of the four terms in the second line of (\ref{h+}). For each term, we will calculate the tensor $F_{\alpha\beta}$ via Eq.\ (\ref{AD_F}) (first for $k^{\alpha}=k^{\alpha}_{(t)}$ and then for $k^{\alpha}=k^{\alpha}_{(\varphi)}$), and then evaluate the corresponding surface integral (\ref{stokes}) at infinity, where it simplifies to 
\begin{equation}
{\cal F}(h_{\alpha\beta};k^\alpha,\partial\Sigma_{\infty})= 
\lim_{r\to\infty}\int F^{rt}(h_{\alpha\beta};k^\alpha) r^2 d\Omega ,  \label{stokesKerr}
\end{equation}
with $d\Omega=\sin^2\theta d\theta d\varphi$.

Starting with the term  $(h_{\alpha\beta}^{{\rm rec}+})^{\rm SAS}$, we show that it falls off sufficiently fast at infinity to have a vanishing contribution to the surface integrals $\cal F$ at infinity, and hence no contribution to ${\cal M}_{\rm AD}^{\infty}$ or to ${\cal L}_{\rm AD}^{\infty}$. To see this, it suffices to keep track of the asymptotic scaling in $r$ of the various fields involved in the reconstruction procedure described in Secs.\ \ref{s:Kerrcirc} and \ref{s:ecceqorbits}. First, using ${\sf Q}_{\ell}^{m=2}(r)\sim r^{-\ell-1}$ and $\Delta\sim r^2$ in Eq.\ (\ref{eq:PlQl4Kerr}) (where henceforth in the current discussion `$\sim$' indicates the leading-order scaling with $r$ at $r\to\infty$), we observe $R_{\ell}^{+}\sim r^{-\ell+1}$. Thus, using Eq.\ (\ref{psi4_separation}) and recalling $\varrho\sim 1/r$, we find $\psi_4^{+}\sim r^{-5}$ (dominated by the slowest-decaying, $\ell=2$ mode). This, in turn, gives $\bar \Psi^+\sim 1/r$ for the Hertz potential [Eq.\ (\ref{inversionKerr})], leading to $h^{{\rm rec}+}_{rr}\sim r^{-3}$ [Eq.\ (\ref{hrr_Kerr})], and a similar scaling for the other nonvanishing components of the perturbation (in suitably normalized, Cartesian-like coordinates). Turning now to $F^{rt}$ in Eq.\ (\ref{AD_F}), we see it falls off at least as fast as $\sim 1/r^3$, for either $k^\alpha= k^{\alpha}_{(t)}$ or $k^\alpha= k^{\alpha}_{(\varphi)}$. It follows that the corresponding surface integrals in Eq.\ (\ref{stokesKerr}) vanish in the limit $r\to\infty$, and one concludes
\begin{eqnarray}\label{AD_SAS}
{\cal M}_{\rm AD}^{\infty}((h_{\alpha\beta}^{{\rm rec}+})^{\rm SAS})&=& 0 ,
\nonumber\\
{\cal L}_{\rm AD}^{\infty}((h_{\alpha\beta}^{{\rm rec}+})^{\rm SAS})&=&0.
\end{eqnarray}

We next turn to the term $(h_{\alpha\beta}^{{\rm rec}+})^{\rm nonSAS}$. It is easy to see that the surface integral in Eq.\ (\ref{stokesKerr}) vanishes trivially for any perturbation with an azimuthal dependence $\sim e^{im\varphi}$ with $m\ne 0$. It remains to consider axisymmetric ($m=0$) modes that are  nonstationary (these may occur in the case of noncircular orbits). For periodic orbits, such modes will have a time dependence of the form $\sim e^{i\omega t}$ (with some frequency $\omega$), which would naively imply a similar time dependence for the corresponding $F^{rt}$ and hence for ${\cal M}_{\rm AD}^{\infty}$ and ${\cal L}_{\rm AD}^{\infty}$, in violation of the fact that these AD quantities are conserved (time-independent). This immediately tells us that nonstationary axisymmetric modes cannot possibly contribute to $F^{rt}$, and they must give a zero contribution to the AD mass and angular momentum.    
Thus we conclude
\begin{eqnarray}\label{AD_nonSAS}
{\cal M}_{\rm AD}^{\infty}((h_{\alpha\beta}^{{\rm rec}+})^{\rm nonSAS})&=& 0 ,
\nonumber\\
{\cal L}_{\rm AD}^{\infty}((h_{\alpha\beta}^{{\rm rec}+})^{\rm nonSAS})&=&0.
\end{eqnarray}

Next we consider the mass-perturbation term ${\cal E}^+h^{(\delta M)}_{\alpha\beta}$, where, recall, $h^{(\delta M)}_{\alpha\beta}$ is the homogeneous solution given explicitly in Eq.\ (\ref{eq:dMexplicit}). 
It is straightforward to evaluate the 2-sphere integrals ${\cal F}$ for this explicit solution even without taking the limit $r\to\infty$. The result, for any surface of constant $t,r$, is
\begin{eqnarray}\label{AD_deltaM}
{\cal M}_{\rm AD}(h^{(\delta M)}_{\alpha\beta})&=&1 ,
\nonumber\\ 
{\cal L}_{\rm AD}(h^{(\delta M)}_{\alpha\beta})&=&0.
\end{eqnarray}
Therefore, in particular,
\begin{eqnarray}\label{AD_deltaMinf}
{\cal M}_{\rm AD}^{\infty}({\cal E}^+ h^{(\delta M)}_{\alpha\beta})&=&{\cal E}^+ ,
\nonumber\\
{\cal L}_{\rm AD}^{\infty}({\cal E}^+ h^{(\delta M)}_{\alpha\beta})&=&0.
\end{eqnarray}
Similarly, for the homogeneous angular-momentum perturbation $h^{(\delta J)}_{\alpha\beta}$, given explicitly in Eq.\ (\ref{eq:dJexplicit}) (and for any 2-sphere), one obtains 
\begin{eqnarray}\label{AD_deltaJ}
{\cal M}_{\rm AD}( h^{(\delta J)}_{\alpha\beta})&=& 0 ,
\nonumber\\  
{\cal L}_{\rm AD}( h^{(\delta J)}_{\alpha\beta})&=& 1 ,
\end{eqnarray}
leading to
\begin{eqnarray}\label{AD_deltaJinf}
{\cal M}_{\rm AD}^{\infty}({\cal J}^+ h^{(\delta J)}_{\alpha\beta})&=& 0 ,
\nonumber\\
{\cal L}_{\rm AD}^{\infty}({\cal J}^+ h^{(\delta J)}_{\alpha\beta})&=&{\cal J}^+.
\end{eqnarray}

Collecting our results (\ref{AD_SAS}), (\ref{AD_nonSAS}), (\ref{AD_deltaMinf}) and (\ref{AD_deltaJinf}), we finally obtain 
\begin{equation}
{\cal M}_{\rm AD}^{\infty}(h_{\alpha\beta}^+)={\cal E}^+,\quad\quad
{\cal L}_{\rm AD}^{\infty}(h_{\alpha\beta}^+)={\cal J}^+.
\end{equation} 
Hence, with the total AD mass and angular momentum fixed as in Eq.\ (\ref{AD_total}), we arrive at
the simple result
\begin{equation}\label{EJ+}
{\cal E}^+=E, \quad\quad {\cal J}^+=L,
\end{equation} 
which, by virtue of Eq.\ (\ref{jumps_Kerr}), also gives 
\begin{equation}\label{EJ-}
{\cal E}^-=0, \quad\quad {\cal J}^-=0.
\end{equation} 

Equations (\ref{EJ+}) and (\ref{EJ-}) are our main results in this subsection, fixing the completion piece of the metric perturbation both outside $\cal S$ and inside it. We remind that these results apply to any bound (circular or eccentric) equatorial geodesic orbit in Kerr geometry.

\subsection{Mass and angular-momentun contents of the reconstructed perturbation }\label{NoMass}

We now discuss an interesting implication of our results, namely that the {\it reconstructed} piece of the metric perturbation has no AD mass or angular momentum either outside or inside of $\cal S$:
\begin{equation}\label{MLin}
{\cal M}_{\rm AD}(h_{\alpha\beta}^{{\rm rec}\pm})=0,\quad\quad
{\cal L}_{\rm AD}(h_{\alpha\beta}^{{\rm rec}\pm})=0,
\end{equation} 
where the surface integrals can be evaluated on any closed spatial 2-surface.
This holds regardless of one's choice of total AD mass ${\cal M}_{\rm AD}^{\infty}$ and angular momentum ${\cal L}_{\rm AD}^{\infty}$ in Eq.\ (\ref{AD_total}). 

That (\ref{MLin}) applies outside of $\cal S$ follows directly from the combination of (\ref{AD_SAS}) and (\ref{AD_nonSAS}), recalling that ${\cal M}_{\rm AD}$ and ${\cal L}_{\rm AD}$ are constant across the entire vacuum domain ${\cal S}^+$. To see why (\ref{MLin}) holds also inside $\cal S$, let us first introduce the shorthand notation $[{\cal M}_{\rm AD}(h)]$ for the jump across $\cal S$ in the value of the AD mass associated with a field $h$ (and similarly for the angular momentum). Equation (\ref{AD_ML_particle}) implies
\begin{equation}
[{\cal M}_{\rm AD}(h_{\alpha\beta})]=E, \quad\quad 
[{\cal L}_{\rm AD}(h_{\alpha\beta})]=L
\end{equation}
for the jumps in the mass and angular momentum of the full perturbation, $h_{\alpha\beta}=h_{\alpha\beta}^{\rm rec}+h_{\alpha\beta}^{\rm comp}$,
which is a solution of the inhomogeneous field equations. In addition, the combination of Eqs.\ (\ref{h_comp}), (\ref{AD_deltaM}), (\ref{AD_deltaJ}) and (\ref{jumps_Kerr}) gives
\begin{equation}
[{\cal M}_{\rm AD}(h_{\alpha\beta}^{\rm comp})]=E, \quad\quad 
[{\cal L}_{\rm AD}(h_{\alpha\beta}^{\rm comp})]=L
\end{equation}
for the jumps in the mass and angular momentum of the completion piece. Since $h_{\alpha\beta}^{\rm rec}=h_{\alpha\beta}-h_{\alpha\beta}^{\rm comp}$, we conclude
\begin{equation} \label{MLrecjump}
[{\cal M}_{\rm AD}(h_{\alpha\beta}^{\rm rec})]=0, \quad\quad 
[{\cal L}_{\rm AD}(h_{\alpha\beta}^{\rm rec})]=0.
\end{equation}
Since the AD mass and angular momentum of $h_{\alpha\beta}^{\rm rec}$ are both zero outside $\cal S$, it follows from (\ref{MLrecjump}) that they are also zero inside $\cal S$.

That the CCK-reconstructed perturbation carries no mass or angular momentum is almost a trivial statement in the Schwarzschild case, where individual $\ell$-modes of the perturbation have separate dynamics: In this case, mass and angular momentum perturbations have a pure $\ell=0,1$ angular dependence, while the reconstructed piece is made solely of $\ell\geq 2$ modes, meaning it cannot contain mass and angular momentum. However, it is quite remarkable that the same result appears to apply even in the Kerr case, where different $\ell$-modes couple, and mass and angular-momentum perturbations spread over all modes. Even then, we now see, the reconstructed perturbation is devoid of mass and angular momentum (at least for equatorial orbits, but we conjecture that the same applies to any CCK-reconstructed vacuum perturbation). We are not aware of any direct proof of this result.

\section{Summary and Conclusions}\label{s:summary}

We have determined the completion piece of the metric perturbation for any bound geodesic orbit in Schwarzschild spacetime or in the equatorial plane of a Kerr black hole. Recalling (\ref{h_comp}) with  (\ref{EJ+}) and (\ref{EJ-}), our main result is that
\begin{equation}\label{mainresult}
h^{\rm comp\pm}_{\alpha\beta}= 
\left\{
\begin{array}{ll}
E h^{(\delta M)}_{\alpha\beta}+Lh^{(\delta J)}_{\alpha\beta} & \text{in ${\cal S}^+$}, 
\\
0 &  \text{in ${\cal S}^-$},
\end{array}
\right.
\end{equation}
for any such orbit. Here $h^{(\delta M)}_{\alpha\beta}$ and $h^{(\delta J)}_{\alpha\beta}$ are the vacuum perturbations given explicitly in Eqs.\ (\ref{eq:dMexplicit}) and (\ref{eq:dJexplicit}), and $E$ and $L$ are the conserved energy and angular momentum associated with the geodesic orbit. The result (\ref{mainresult}) assumes that the total energy and angular momentum contents of the perturbation are fixed as in Eq.\ (\ref{AD_total}). {\em Independently of this assumption}, we find that the jump across $\cal S$ in the completion piece of the metric perturbation is given by 
\begin{equation}\label{[hcomp]}
[h^{\rm comp}_{\alpha\beta}]=E h^{(\delta M)}_{\alpha\beta}+Lh^{(\delta J)}_{\alpha\beta}.
\end{equation} 
As a consequence (and a corollary) of (\ref{[hcomp]}), we find that the {\it reconstructed} piece of the metric perturbation contains no mass or angular momentum (either in or out of $\cal S$), in a sense expressed in Eq.\ (\ref{MLin}).

Our method consists in demanding that certain gauge-invariant fields constructed from the completed metric perturbation (and its derivatives) are continuous anywhere away from sources. This is a necessary condition that the perturbation must satisfy in order to solve the linear field equations anywhere in the vacuum (the reconstructed piece of the perturbation, by itself, fails to do so). As we have seen, imposing this continuity condition on suitably chosen invariant field(s) determines the completion piece of the perturbation completely and uniquely (up to gauge perturbations). It is expected from uniqueness that our completion renders the invariant fields {\em smooth} (and not just continuous), although we have not confirmed that with an explicit calculation.

Our final results, as expressed in Eqs.\ (\ref{mainresult}), (\ref{[hcomp]}) and (\ref{MLin}), are extremely simple despite the long calculation leading to them. This is striking, and begs an explanation. In particular, one naturally wonders whether the fact that the reconstructed perturbation does not contain mass or angular momentum could be arrived at based on a more general argument (but one that is nonetheless as mathematically rigorous), without resorting to a detailed calculation. We have not been able to devise such an argument so far (except in the trivial, Schwarzschild case). One way to approach the problem would be via a direct evaluation of the Abbott-Deser mass and angular momentum contents of $h_{\alpha\beta}^{{\rm rec}\pm}$ in ${\cal S}^-$, which we have not been able to do analytically for Kerr, so far. If in the future a simple method is found to perform such a calculation in the Kerr case, it could offer a more direct route to the completion problem, and perhaps hint at the reasons for the simplicity of the results. 

The work presented here takes an important step towards a complete formulation of a practical scheme for calculating the gravitational self-force in astrophysically motivated inspiral problems, conveniently starting from solutions of the Teukolsky equation. Two important tasks remain. First, and most obvious, our analysis must be extended to encompass {\it non-equatorial} geodesic orbits in Kerr spacetime. We envisage using a similar methodology to the one applied here. One could start with the special subset of circular inclined (``spherical'') orbits, for which the energy-momentum source is supported on $r=r_0$ and $\pi-\theta_1\leq \theta\leq\theta_1$ with some constant $r_0$ and $0<\theta_1<\pi/2$. In this case, one would require continuity of the invariant fields across $r=r_0$ for $0\leq\theta<\theta_1$ and $\pi-\theta_1\leq\theta<\pi$. For orbits that are both inclined and eccentric, which are generically ergodic, the key step will be the formulation of a suitable decomposition of the energy-momentum source into simple partial elements (spherical sections?) that are each energy conserving, following our strategy in Sec.\ \ref{s:Kerrcirc}. The special cases of polar orbits and of resonant orbits would need to be considered separately. 

The second remaining task is that of {\it gauge regularization}. While our completion procedure guarantees the continuity of invariant fields at vacuum points, it does not guarantee the continuity of the metric perturbation itself. In fact, our completed metric perturbation will generally have a gauge discontinuity across $\cal S$, even off the particle (see, for example, the explicit calculation in Ref.\ \cite{vandeMeent:2015lxa}). This can be a problem in applications that require perturbation information on both sides of $\cal S$, such as a self-force calculation based on the simpler of the two methods formulated in Ref.\ \citep{BMP1}. Typically, for the results of a calculation to have a clear physical interpretation, one must place certain conditions on the gauge. For instance, one usually requires asymptotic flatness, and, for periodic orbits, also a particular periodicity. In the latter case, one must be able to relate the frequency (or frequencies) of the perturbation in and out of $\cal S$, and, for that purpose, one must be able to relate the coordinate times and angles in and out of that surface. A continuity of the perturbation across $\cal S$ is necessary for ``passing on'' such (and other) essential gauge information from the exterior to the interior. The goal of gauge regularization is to locally remove the gauge discontinuity in the neighbourhood of the particle, via a suitable, discontinuous gauge transformation. Optimally, one would aim to construct a perturbation that is entirely continuous across $\check{\cal S}$, at least near the particle. 

However, depending on the application, it might be sufficient to gauge-regularize only certain relevant pieces of the perturbation. For example, a partial gauge regularization of the SAS piece of the completed perturbation was performed recently in Refs.\ \citep{Shah:2015nva} (for circular orbits in Schwarzschild) and \citep{vandeMeent:ISCO} (for circular equatorial orbits in Kerr), sufficient for the purpose of calculating ``invariant'' frequencies (that is, frequencies with respect to asymptotic time $t$). This gauge regularization should now be extended to more general orbits; our partial-ring approach should offer an easy route. Other applications may require further gauge regularization of other pieces of the perturbation. For instance, one may need to work in a ``center-of-mass'' gauge (as defined via a condition on the mass dipole moment of the perturbed spacetime) in order to allow comparison with certain results from the post-Newtonian theory. This would require going beyond the SAS part, and gauge-regularizing also the $m=\pm 1$ azimuthal modes of the completed perturbation. Such a calculation is yet to be done. Other pieces of the perturbation may need to be gauge-regularized for other foreseeable applications.

\section*{Acknowledgements}

CM acknowledges support from CONACyT. LB, AP and MvdM gratefully acknowledge support from the European Research Council under the European Union's Seventh Framework Programme FP7/2007-2013/ERC, Grant No.\ 304978. LB additionally acknowledges support from STFC through grant number PP/E001025/1.

\appendix
\section{Background material: Teukolsky's equation and metric reconstruction}\label{a:KerrBack}

We review here the essential elements of formalism that go into our analysis: the Newman--Penrose (NP) formalism, Teukolsky's equation and metric reconstruction in vacuum. This will also serve as a convenient all-in-one-place summary of our notation and conventions. For historical reasons, much of the NP literature uses the $(+---)$ metric signature, opposite to the one used in our paper, which may bring confusion.  For that reason, we carefully describe our sign conventions and note where they differ from common choices. 

In Boyer-Lindquist coordinates, the line element for the Kerr geometry with mass $M$, spin $J=aM$, and signature $(-+++)$ is given by
\begin{equation}
\begin{split}
\label{eq:ds2kerr}
ds^2=&-\SP{1-\frac{2 M r}{\Sigma}}dt^2+\frac{\Sigma}{\Delta}dr^2+\Sigma d\theta^2\\
&+\SP{r^2+a^2+\frac{2M a^2 r \sin^2\theta}{\Sigma}}\sin^2\theta d\varphi^2 \\
&-\frac{4 M a r \sin^2\theta}{\Sigma}dtd\varphi ,
\end{split}
\end{equation}
with
\begin{align}
\Delta &:= r^2-2Mr+a^2, \\ 
\Sigma &:= r^2+a^2\cos^2\theta.
\end{align}

\subsection{The Newman--Penrose null-tetrad formalism}
Much of black hole perturbation theory can be conveniently formulated using the NP formalism, which expresses geometric quantities in terms of a conveniently chosen tetrad of null vectors. In this paper we use Kinnersley's tetrad for the Kerr metric, whose four legs are given by
\begin{subequations} \label{eq:kerrtetrad}
\begin{align}
\tet{1}{\alpha} = \ell ^\alpha &=\frac{1}{\Delta}\SP{r^2+a^2,\Delta,0,a},\\
\tet{2}{\alpha} = n^\alpha &=\frac{1}{2\Sigma}\SP{r^2+a^2,-\Delta,0,a },\\
\tet{3}{\alpha} = m ^\alpha &=\frac{1}{\sqrt{2}(r+ia\cos\theta)}\bP{ia\sin\theta,0,1,\frac{i}{\sin\theta}}, \\
\tet{4}{\alpha} = \bar m^\alpha &=\frac{-1}{\sqrt{2}(r-ia\cos\theta)}\bP{ia\sin\theta,0,-1,\frac{i}{\sin\theta}}
\end{align}
\end{subequations}
(in Boyer-Lindquist coordinates), with overbars denoting complex conjugation. These legs are all null and mutually orthogonal, except $\ell^\alpha n_\alpha=-1$ and  $m^\alpha\bar{m}_\alpha=1$. In what follows, Greek indices refer to spacetime components while Latin indices denote tetrad components. The directional derivatives along the tetrad legs are denoted $\boldsymbol D=\ell^\mu\partial_\mu$, $\boldsymbol\Delta=n^\mu\partial_\mu$ and $\boldsymbol \delta=m^\mu\partial_\mu$.

The NP formalism expresses the equations of general relativity in terms of Ricci coefficients
\begin{equation}
\gamma_{abc} := g_{\mu\lambda}\tet{a}{\mu}\tet{c}{\nu}\nabla_{\nu}\tet{b}{\lambda},
\end{equation}\\
referred to as \emph{spin coefficients} and customarily given special individual symbols:
\begin{equation}
\begin{aligned}
\kappa &:= -\gamma_ {311}, 
	& \varpi &:= -\gamma_{241},
		& \epsilon &:= -\frac{\gamma_ {211}+\gamma_ {341}}{2}, \\
 \tau &:= -\gamma_{312} ,
	& \nu &:= -\gamma_ {242}, 	
		& \gamma &:= -\frac{\gamma_ {212}+\gamma_ {342}}{2},\\
\sigma &:= -\gamma_ {313},
	&	\mu &:= -\gamma_{243},
		& \beta &:= -\frac{\gamma_{213}+\gamma_ {343}}{2},\\
 \varrho &:= -\gamma_{314} ,
 	& \lambda &:= -\gamma_ {244} ,
 		& \alpha &:= -\frac{\gamma_ {214}+\gamma_ {344}}{2} .
\end{aligned}
\end{equation}
Note these definitions have opposite signs compared to the usual ones applied with a $(+---)$ signature [see, e.g., \cite{chand}]. This ensures that the coordinate expressions for the various spin coefficients (below) remain as conventional.
In Kerr spacetime with the Kinnersley tetrad \eqref{eq:kerrtetrad}, the spin-coefficients $\kappa$, $\lambda$, $\nu$, $\sigma$ and $\epsilon$ all vanish, while the rest take the values
\begin{subequations}
\begin{align}
\varrho &= \frac{-1}{r- i a \cos\theta},\\
\varpi &= \frac{i a \varrho^2\sin\theta}{\sqrt{2}},\\
\tau &= -\frac{i a\sin\theta}{\sqrt{2}\Sigma},\\
\mu &= \frac{\varrho\Delta}{2\Sigma},\\
\gamma &= \frac{\varrho\Delta+r-M}{2\Sigma},\\
\beta &= -\frac{\bar\varrho \cot\theta}{2\sqrt{2}},\\
\alpha &= \varpi-\bar\beta.
\end{align}
\end{subequations}

The Weyl curvature scalars are defined in terms of the components of the Weyl tensor $C_{\alpha\beta\gamma\delta}$  as
\begin{subequations}\label{eq:psi}
\begin{align}
\psi_0=&C_{\alpha\beta\gamma\delta}\,\ell^\alpha m^\beta \ell^\gamma m^\delta , \label{eq:psi0}\\
\psi_1=&C_{\alpha\beta\gamma\delta}\,\ell^\alpha m^\beta \ell^\gamma n^\delta,\\
\psi_2=&C_{\alpha\beta\gamma\delta}\,\ell^\alpha m^\beta \bar m^\gamma n^\delta , \label{eq:psi2}\\
\psi_3=&C_{\alpha\beta\gamma\delta}\,\ell^\alpha n^\beta \bar m^\gamma n^\delta,\\
\psi_4=&C_{\alpha\beta\gamma\delta}\, n^\alpha \bar m^\beta n^\gamma \bar m^\delta ,\label{eq:psi4}
\end{align}
\end{subequations}
where we have chosen the overall signs such that the NP form of the field equations for the various curvature scalars remains unchanged (with respect to the standard form, as given, e.g., in \cite{chand}). 
In Kerr spacetime with the tetrad \eqref{eq:kerrtetrad}, the Weyl scalars $\psi_0$, $\psi_1$, $\psi_3$ and $\psi_4$ all vanish, implying that Kerr spacetime is of Petrov type D, with $l^\alpha$ and $n^\alpha$ as the (double) principal null vectors. The remaining, non-zero Weyl scalar $\psi_2$ takes the value shown in Eq.\ (\ref{psi20}).

The requirement that the tetrad legs are null, orthogonal and appropriately normalized fixes the tetrad (given a metric) only up to local $SO(1,3)$ rotations. Consequently, linear perturbations to the NP quantities, denoted $\psi^{(1)}_n$, have an additional, non-physical, $so(1,3)$ gauge freedom corresponding to infinitesimal tetrad rotations (on top of the usual gauge freedom associated with infinitesimal coordinate transformations). As shown in (e.g.) \cite{chand}, 
on a Kerr background with the Kinnersley tetrad, the perturbations $\psi^{(1)}_0$, $\psi^{(1)}_2$, and $\psi^{(1)}_4$ are invariant under such tetrad rotations. Since $\psi_0$ and $\psi_4$ are scalar fields that vanish on the background, their perturbations are also invariant under coordinate gauge transformations. This means $\psi^{(1)}_0$ and $\psi^{(1)}_4$ are true gauge-invariant fields. Similarly, as we point out in Sec.\ \ref{s:InvSch}, in the case $a=0$ the imaginary part of $\psi^{(1)}_2$ is also a true gauge-invariant field, since $\mathrm{Im}(\psi_2)=0$ on the Schwarzschild background.

\subsection{Teukolsky equation}
Teukolsky showed \citep{Teuk,Teukolsky:1973ha} that the NP field equations for the perturbations $\phi_{+2}=\psi_0^{(1)}$ and $\phi_{-2}=\varrho^{-4}\psi_4^{(1)}$ decouple from the rest of the NP equations. The master Teukolsky equation for a general ``spin-weight'' $s$, reads
\begin{equation}
\begin{aligned}
\label{eq:kerrteuk} 
&-\BP{\frac{(r^2+a^2)^2}{\Delta} -a^2\sin^2\theta}\frac{\partial^2\phi_s}{\partial t^2}
-\frac{4Mar}{\Delta}\frac{\partial^2\phi_s}{\partial t\partial\varphi}
\\
&-\BP{\frac{a^2}{\Delta}-\frac{1}{\sin^2\theta}}\frac{\partial^2\phi_s}{\partial\varphi^2}
+\Delta^{-s}\frac{\partial}{\partial r}\BP{\Delta^{s+1}\frac{\partial\phi_s}{\partial r}}
\\
&+\frac{1}{\sin\theta}\frac{\partial}{\partial\theta}\BP{\sin\theta \frac{\partial\phi_s}{\partial\theta}}
\\ 
&+2s\BP{\frac{M(r^2-a^2)}{\Delta}-r-ia\cos\theta}\frac{\partial\phi_s}{\partial t} 
\\
&+2s\BP{\frac{a(r-M)}{\Delta}+\frac{i\cos\theta}{\sin^2\theta}}\frac{\partial\phi_s}{\partial\varphi}
\\
&\hspace{10.0em}-(s^2\cot^2\theta -s)\phi_s = T_s.
\end{aligned}
\end{equation}
For $s=\pm 2$, the source term $T_s$ is obtained from the energy-momentum tensor using
\begin{align}
T_{+2}&=8\pi\Sigma\begin{aligned}[t]\BB{
(\boldsymbol{\delta }  -2\beta -4\tau )(\boldsymbol{\delta }  -\bar\varpi )&T_{\boldsymbol{11}}
\\
 - (\boldsymbol{D} -4\varrho-\bar\varrho )(\boldsymbol{\delta } +2\bar\alpha ) &T_{\boldsymbol{13}}
 \\
-(\boldsymbol{\delta }  -2\beta -4\tau )(\boldsymbol{D} -2\bar\varrho )&T_{\boldsymbol{13}}
\\
 +(\boldsymbol{D} -4\varrho-\bar\varrho )(\boldsymbol{D} -\bar\varrho )&T_{\boldsymbol{33}} 
 },
 \end{aligned}
\label{eq:kerrsource}
\\
T_{-2}&=
\hspace{-3.0em}\begin{aligned}[t]
\frac{8\pi\Sigma}{\varrho^{4}}\BB{
(\boldsymbol{\bar\delta}+2\alpha +5\varpi  -\bar\tau )(\boldsymbol{\bar\delta} +2\varpi -\bar\tau )&T_{\boldsymbol{22}} 
\\
-(\boldsymbol{\Delta } +3\gamma -\bar\gamma +4\mu +\bar\mu )(\boldsymbol{\bar\delta} +2\alpha -2\bar\tau )&T_{\boldsymbol{24}}
\\
 -(\boldsymbol{\bar\delta} +2\alpha  +5\varpi  -\bar\tau)(\boldsymbol{\Delta } +2\gamma +2\bar\mu ) &T_{\boldsymbol{24}}
\\
+(\boldsymbol{\Delta } +3\gamma -\bar\gamma +4\mu +\bar\mu )(\boldsymbol{\Delta } +2\gamma -2\bar\gamma +\bar\mu )&T_{\boldsymbol{44}} 
},
\end{aligned} \label{eq:kerrsource-2}
\end{align}
where $T_{\boldsymbol{11}}=T_{\alpha\beta}e_{1}^{\alpha}e_1^{\beta}$, etc.

Moreover, Teukolsky showed that (\ref{eq:kerrteuk}) admits a full separation of variables. Solutions of the Teukolsky equation can be written as
\begin{align}\label{eq:teuksplit}
\phi_s  =  \int d\omega \sum_{\ell m} {_sR_{\ell m\omega}}(r)\,{_sS_{\ell m\omega}}(\theta)e^{i(m\varphi-\omega t)},
\end{align}
where the function ${_sR_{\ell m\omega}}(r)$ satisfies the radial Teukolsky equation
\begin{align}
&
\SP{
	\Delta^{-s}\frac{d}{d r}\BP{\Delta^{s+1}\frac{d}{d r}}
	-V_{s\ell m\omega}
}{_sR_{\ell m\omega}} 
= T_{s\ell m\omega},
\label{eq:kerrR}
\end{align}
with potential
\begin{align}
&V_{s\ell m\omega} =\lambda_{s\ell m\omega}-4i s\omega r -\frac{K^2-2i s (r-M)K}{\Delta},
\end{align}
where
\begin{align}
 K_{m\omega} &:= (r^2+a^2)\omega- a m,
\end{align}
and $\lambda_{s\ell m\omega}$ is the eigenvalue of the angular equation (see below).
The source in Eq.\ (\ref{eq:kerrR}) is given by
\begin{equation}\label{Tslm}
 T_{s\ell m\omega} = \frac{1}{2\pi}
 \int_{-\infty}^{\infty}\hspace{-1.0em} dt
 \int_{-1}^{1}\hspace{-0.9em} d\cos\theta \hspace{-0.3em}
 \int_{-\pi}^\pi\hspace{-1.0em} d\varphi\, T_s e^{i(\omega t-m\varphi)}{_sS_{\ell m\omega}}. 
\end{equation}

The functions ${_sS_{\ell m\omega}}(\theta)$ are \emph{spin-weighted spheroidal harmonics}, which satisfy the angular equation,
\begin{align}
\label{eq:kerrang}
\SP{\frac{1}{\sin\theta}\frac{d}{d\theta}\SP{\sin\theta\frac{dS}{d\theta}}-U_{s\ell m\omega}}{_sS_{\ell m\omega}}=0.
\end{align}
Here the potential is 
\begin{equation}
\begin{split}
U_{s\ell m\omega}=
 &
 \frac{(m+s\cos\theta)^2}{\sin^2 \theta}
 +a^2\omega^2\sin^2\theta 
\\
 &
 +2sa\omega\cos\theta
 -2m a\omega  
 -s
 -\lambda_{s\ell m\omega}.
\end{split}
\end{equation}
We follow the convention that the spheroidal functions are normalized according to
\begin{equation}
\int_{-1}^{1} {_sS_{\ell m\omega}^2}(\theta)\, d\cos\theta = \frac{1}{2\pi},
\end{equation}
with
\begin{subequations}
\begin{align}
{_{-s}S_{\ell m\omega}}(\theta) &= (-1)^{s+m}{_sS_{\ell(-m)(-\omega)}}(\theta),\\
{_{s}S_{\ell m\omega}}(\pi-\theta) &= (-1)^{s+\ell}{_sS_{\ell(-m)(-\omega)}}(\theta).
\end{align}
\end{subequations}
When $a\omega=0$ the eigenvalue $\lambda_{s\ell m\omega}$ becomes $\ell(\ell+1)-s(s+1)$, and the spin-weighted spheroidal harmonics reduce to spin-weighted spherical harmonics: $_sS_{\ell m}(\theta)e^{im\varphi}=\,_sY_{\ell m}(\theta ,\varphi)$.

Using the spin-weight raising and lowering operators 
\begin{subequations}
\begin{align}
\eth_s&=-\partial_\theta-{i\csc\theta}\partial_\varphi+s \cot\theta,\\
\label{edth}
\bar\eth_s&=-\partial_\theta+{i\csc\theta}\partial_\varphi-s \cot\theta,
\end{align}
\end{subequations}
spin-weighted spherical harmonics can be rewritten as the derivatives of harmonics with a different spin-weight using
\begin{subequations}
\begin{align}\label{raising}
\eth_s\,{_sY_{\ell m}} &= +\sqrt{(\ell-s)(\ell+s+1)} \, {_{s+1}Y_{\ell m}},\\
\label{lowering}
\bar\eth_s\,{_sY_{\ell m}} &= -\sqrt{(\ell+s)(\ell-s+1)} \, {_{s-1}Y_{\ell m}}.
\end{align}
\end{subequations}
In particular, by repeated application of the above identities, any spin-weighted spherical harmonic can be written in terms of derivatives of ordinary spherical harmonics with the same $\ell$ and $m$. 

\subsection{Metric reconstruction}

For vacuum perturbations, a procedure to obtain the metric perturbations starting from the curvature scalars $\psi_0$ or $\psi_4$ was first proposed by Chrzanowski \citep{chrza} and also by Cohen and Kegeles \citep{cohen79}. The CCK reconstruction formula gives the metric perturbation in either the ingoing or outgoing (traceless) radiation gauge. In this paper we choose to work in the ingoing radiation gauge (IRG), satisfying 
\begin{equation}\label{gaugecondition}
\ell^\alpha h_{\alpha\beta}^{\mathrm{rec}}= 0
\quad \text{and} \quad 
g^{\alpha\beta}_{(0)}h_{\alpha\beta}^{\mathrm{rec}}=0,
\end{equation} 
where $g^{\alpha\beta}_{(0)}$ is the (inverse of the) background Kerr metric. The CCK reconstruction formula for the IRG perturbation is given by 
\begin{align}
\label{eq:kerrh}
\begin{split}
h_{\alpha\beta}^{\mathrm{rec}}=& -\BP{
\ell_\alpha \ell_\beta
\SP{\boldsymbol{\delta}+\bar\alpha +3\beta-\tau }\SP{\boldsymbol{\delta}+4\beta +3\tau} 
\\
&+m_\alpha m_\beta \SP{\boldsymbol{D}-\varrho}\SP{\boldsymbol{D} +3\varrho}
\\
 & -\ell_{(\alpha} m_{\beta)} \BB{
 \SP{\boldsymbol{\delta}-2\bar\alpha +2\beta-\tau}\SP{\boldsymbol D+3\varrho}
\\
&+\SP{\boldsymbol{D} +\bar\varrho-\varrho}\SP{\boldsymbol{\delta} +4\beta+3\tau}
}
}\Psi +c.c.,
\end{split}
\end{align}
where $\Psi$ is the IRG ``Hertz potential''. The latter satisfies the homogeneous Teukolsky equation \eqref{eq:kerrteuk} with spin-weight $s=-2$. In addition it satisfies a fourth-order differential equation linking it to $\psi_4$ \citep{Lousto:2002em}:
\begin{align}
8\varrho^{-4}\psi_4 =&\bar{\mathfrak L}_{-1}\bar{\mathfrak L}_{0}\bar{\mathfrak L}_{1}\bar{\mathfrak L}_{2}\bar\Psi -12 M \partial_t \Psi, \label{eq:Heqkerra}
\end{align}
with $\bar{\mathfrak L}_s:=\bar\eth_s+ia\sin\theta\partial_t$. 

The perturbation to $\psi_2$ can be calculated from the metric perturbation (also taking account of the perturbations to the tetrad legs). We quote here the particularly compact expression obtained by Sano and Tagoshi \citep{SaTa2}: 
\begin{equation}\label{eq:deltapsi2}
\begin{split}
\psi_2^{(1)}=&\frac{1}{2}\BP{
\boldsymbol{D}\boldsymbol {D}\varrho(\boldsymbol{\bar\delta}+2\bar\beta)\frac{1}{\varrho}(\boldsymbol{\bar\delta}+4\bar\beta)
\\
&-4\varpi(\boldsymbol{D}+\varrho)\boldsymbol{D}(\boldsymbol{\bar\delta}+4\bar\beta)
 +6\varpi\boldsymbol{D}\varpi\boldsymbol{D} }\bar\Psi.
\end{split}\end{equation}

\section{Validity of procedure for evaluating jumps via interchange of mode-sum and limit}\label{a:ProofInterchange}

At the basis of our method is the requirement that, at any vacuum point, the invariant fields ${\cal I}_n$ must be continuous. In practice, we construct (the SAS part of) these four-dimensional fields as ${\cal I}^\pm_n(r,\theta)=\sum_\ell {\cal I}^{\rm rec\pm}_{n\ell}(r,\theta)+{\cal I}^{\rm comp\pm}_n(r,\theta)$, meaning that the continuity condition at $r=r_0$ (off the particle) reads 
\begin{multline}\label{continuity condition}
\lim_{r\to r_0^+}\sum_\ell {\cal I}^{\rm rec+}_{n\ell}(r,\theta)+{\cal I}^{\rm comp+}_n(r_0,\theta) 
\\
= \lim_{r\to r_0^-}\sum_\ell {\cal I}^{\rm rec-}_{n\ell}(r,\theta)+{\cal I}^{\rm comp-}_n(r_0,\theta).
\end{multline}
Imposing this condition requires first evaluating the sums and then taking the limit to $r_0$. If the summands are first evaluated at $r_0$, the sums diverge for all $\theta$; this is also true of their difference, $\sum_\ell [ {\cal I}^{\rm rec+}_{n\ell}(r_0,\theta) - {\cal I}^{\rm rec-}_{n\ell}(r_0,\theta)]$. This is easily seen by counting powers of $\ell$ in Eqs.\ (\ref{Irec}) or (\ref{JumpForm}), for example.

However, in our calculations in the body of the paper, we allow ourselves to bring the limit inside the sums, take their difference, and then discard any terms that, while pointwise divergent, can be interpreted as distributions supported at $\theta=\theta_0=\pi/2$. In this Appendix, we prove the validity of that method: If the jump $[{\cal I}_n]$ that we define by this procedure vanishes for all $\theta\neq\theta_0$, then the continuity condition~\eqref{continuity condition} is satisfied.\footnote{We do not prove the converse. However, this one-way implication suffices for us because of uniqueness. Since $[{\cal I}_n]=0$ uniquely determines $[\cal E]$ and $[{\cal J}]$ and guarantees that ${\cal I}_n$ is continuous, and since there can be only one pair of values $([\cal E],[{\cal J}])$ that satisfies Eq.~\eqref{continuity condition}, the determined values of $[{\cal E}]$ and $[{\cal J}]$ are the unique, correct ones.}

To establish this result, we re-express the invariants in a more convenient form. We first introduce the rescaled invariants 
\begin{equation}\label{pn}
I^\pm_n:=(\sin\theta)^{p_n} {\cal I}^\pm_n, 
\end{equation}
where $p_n$ is chosen to make $I^\pm_n$ go smoothly to zero at the poles; the reason for this, and a concrete choice for $p_n$, will become apparent in the course of the proof. We also eliminate $\theta$ in favor of $z:=\cos\theta$ and absorb $I^{\rm comp\pm}_n$ into the sum over $\ell$, giving us $I^\pm_n(r,z)=\sum_\ell I^{\pm}_{n\ell}(r,z)$; the particular way in which this is done is immaterial, and we will largely ignore the completion terms in the arguments below. Finally, we work with the ``jump function'' $\Delta I_{n\ell}(\delta r,z):=I^+_{n\ell}(r_0+\delta r,z)-I^-_{n\ell}(r_0-\delta r,z)$ and
\beq
\Delta I_n(\delta r,z):=I^+_n(r_0+\delta r,z)-I^-_n(r_0-\delta r,z).
\eeq 
In terms of this quantity, our goal will be to show that our procedure ensures 
\begin{equation}\label{DeltaI=0}
\lim_{\delta r\to0}\Delta I_n(\delta r,z)=0
\end{equation}
(except, possibly, at the particle's position $z=z_0=0$).

Our proof is based on the following more general result: 
\begin{lemma}
Let $g(\delta r,z)=\sum_{\ell=0}^\infty g_{\ell}(\delta r,z)$ on $(0,c]\times[-1,1]$, with some constant $c>0$, and let $\check S=[-1,1]-\{z_0\}$. If 
\begin{enumerate}
\item[(i)] $\sum_{\ell=0}^\infty g_{\ell}(\delta r,z)$ converges uniformly on $[b,c]\times [-1,1]$ for all $b\in(0,c)$, 
\item[(ii)] each $g_\ell(\delta r,z)$ is continuous on $[0,c]\times[-1,1]$, 
\item[(iii)] $\lim_{(\delta r,z)\to(0,z^*)}g(\delta r,z)$ yields a continuous function of $z^*$ for all paths in $(0,c]\times \check S$ and all $z^*\in \check S$, 
and
\item[(iv)] the sequence of functions $\tilde g_N(\delta r):=\int_{-1}^1 dz\, \phi(z)\sum_{\ell=0}^N g_\ell(\delta r,z)$ converges uniformly on $[0,c]$ to a function $\lim_{N\to\infty} \tilde g_N = \tilde g_{\phi}$ given by
\begin{equation}
\tilde g_\phi(\delta r) = \begin{cases}0 & \text{if } \delta r = 0\\
\int_{-1}^1 dz\, \phi(z) g(\delta r,z) & \text{if } \delta r\in(0,c]
\end{cases}
\end{equation}
for all smooth test functions $\phi$ whose support is wholly contained in $\check S$,
\end{enumerate}
then $\lim_{\delta r\to0}g(\delta r,z)=0$ for all $z\in \check S$.
\end{lemma}
To get more quickly to our main result, we delay the proof of this lemma until the end of the section.

We use the lemma by letting $\Delta I_n$ and $\Delta I_{n\ell}$ play the roles of $g$ and $g_{\ell}$, and we choose $c$ to be any constant small enough to avoid evaluating any functions at (or behind) the horizon. Our desired conclusion  then follows if we can show that $\Delta I_n$ and $\Delta I_{n\ell}$ satisfy the four conditions of the lemma. We do this for the case of circular orbits in Kerr; the same arguments apply for each partial ring in the case of eccentric orbits.

Refer to Eq.\ (\ref{Irec}). Note that at large $\ell$, $R^{\pm(k)}_\ell(r)C^\mp_\ell (r_0)$ behaves as $\sim\ell^{k-2} \P_\ell^{m=2}(r_</M-1)\Q_\ell^{m=2}(r_>/M-1)$, where $r_\lessgtr:={\rm min/max}\{r,r_0\}$. For all $r\neq r_0$, $\P_\ell^{m=2}(r_</M-1)\Q_\ell^{m=2}(r_>/M-1)$ decays faster than any power of $1/\ell$, and all other factors in Eq.~(80) grow no faster than a finite power of $\ell$, guaranteeing exponential convergence of the sum~(80). However, this holds {\em only} for $r\neq r_0$; for $r=r_0$, the sums diverge for all $z$. Hence, $\sum_\ell I^{\pm}_{n\ell} (r,z)$ does not converge uniformly on the open region $(r_0,r_0\pm c]\times[-1,1]$ (where the $\pm$ signs correspond to the superscripts in $I^{\pm}_{n}$), but they do converge uniformly in each closed region $[r_0\pm b,r_0\pm c]\times [-1,1]$ with $0<b\leq c$. This carries over immediately to $\sum_\ell \Delta I_{n\ell} (\delta r,z)$ in all regions $[b,c]\times[-1,1]$ with $0<b\leq c$, establishing condition (i) of the lemma. 

Next, inspection of Eq.\ (\ref{Irec}) reveals that each $\Delta I_{n\ell}(\delta r,z)$ is continuous on $[0,c]\times[-1,1]$, establishing condition (ii); this is manifestly true away from the boundaries of that region, and it could only be violated on the boundaries if $I^\pm_{n\ell}(r,z)$ became singular at $r\to r_0^\pm$ or at $z=\pm 1$---which, manifestly, it does not [for sufficiently large values of $p_n$ in Eq.\ (\ref{pn})].

Third, note that if condition (iii) were violated, it would never be possible to make $I_n$ continuous on the sphere $r=r_0$. But we know that the linearized Einstein equation with a conserved point-particle source {\em does} have a solution, and in particular, it has a solution with our choice of boundary conditions. That solution will necessarily have smooth invariants at points away from the particle, meaning condition (iii) is satisfied.
 
This leaves condition (iv). To show that it is satisfied, we begin by establishing pointwise convergence of $\widetilde{\Delta I}_{nN}(\delta r):=\int_{-1}^1 dz\, \phi(z)\sum_{\ell=0}^N \Delta I_{n\ell}(\delta r,z)$. For $\delta r=0$, the result is exactly the condition we impose in the body of the paper: we find the unique pair $([{\cal E}],[{\cal J}])$ for which $\lim_{N\to\infty}\widetilde{\Delta I}_{nN}(0)=0$.\footnote{Note that here we use the precise  mathematical definition of what it means for a sum $\sum_\ell f_{\ell} (z,z_0)$ to converge to a delta function $\delta(z-z_0)$:  
 $\lim_{N\to\infty}\sum_{\ell=0}^N\int_{-1}^1 dz\, \phi(z)f_{\ell}(z,z_0) = \phi(z_0)$, and similarly for derivatives of a delta function.} For each $\delta r>0$, we can straightforwardly move the limit inside the integral. This is made legal by the fact that the integrand $\phi(z)\sum_{\ell=0}^N \Delta I_{n\ell}(\delta r,z)$ appearing in $\widetilde{\Delta I}_{n\ell}(\delta r)$ is bounded by an integrable function for all $N$; for example, for each given $\delta r>0$ take the dominating function to be $\sup_{N\in\mathbb{N},z\in \check S}|\phi(z)\sum_{\ell=0}^N \Delta I_{n\ell}(\delta r,z)|$. The dominated convergence theorem then guarantees that we can pass the limit into the integral, yielding
\begin{equation}
\lim_{N\to\infty}\widetilde{\Delta I}_{nN}(\delta r) = \int_{-1}^1 dz\, \phi(z)\Delta I_{n}(\delta r,z)
\end{equation}
for all $\delta r\in(0,c]$ and for all smooth test functions. Therefore, $\widetilde{\Delta I}_{nN}(\delta r)$ converges pointwise to the function
\begin{equation}
\widetilde{\Delta I}_{n}(\delta r) = \begin{cases}0 & \text{if } \delta r=0\\
\int_{-1}^1 dz\, \phi(z) \Delta I_n(\delta r,z) & \text{if } \delta r\in(0,c]
\end{cases}
\end{equation}
for all smooth test functions $\phi$ whose support is contained in $\check S$.

We must now show that this convergence is uniform. To this end, using the property described below Eq.\ (\ref{Irec}), we write $\Delta I_{n\ell}$ in the form
\beq
\Delta I_{n\ell}(\delta r,z) = \sum_{j=0}^3 F_{nj\ell}(\delta r,z) \frac{d^j}{dz^j}{}_2Y_{\ell0}(\theta(z)),
\eeq
where each $F_{nj\ell}$ is a smooth function of $z$ that vanishes at least as $(1-z^2)^{p_n/2-q_n/2}$ at the poles, with $q_1=0$, $q_2=2$, and $q_3=1$ [recall Eq.\ (\ref{pn})]. Now examine the separate integrals $\int^1_{-1}dz\, \phi(z) F_{nj\ell}(\delta r,z)\frac{d^j}{dz^j}{}_2Y_{\ell0}$. If we write the harmonic explicitly as ${}_2Y_{\ell0} = \sqrt{\frac{2\ell+1}{4\pi\lambda_2}}(1-z^2)\frac{d^2}{dz^2}\P_\ell(z)$ and repeatedly integrate by parts, we see that all the integrals can be expressed in the form $\int^1_{-1}dz\, G_{nj\ell}(\delta r,z)\P_\ell (z)$ plus boundary terms, where $G_{nj\ell}$ is a smooth function of $z$. We eliminate the boundary terms by choosing sufficiently large values for $p_n$: since the boundary terms have the form $\P_\ell(z)\frac{d^{j+1}}{dz^{j+1}}[(1-z^2)F_{nj\ell}(z)]\bigr|^1_{-1}$ plus lower derivatives, we may choose, for example, $p_n=q_n + 2+2{\rm max}(j) =q_n+8$. 

After eliminating the boundary terms, we are left with the following sum:
\beq\label{sum}
\lim_{N\to\infty}\widetilde{\Delta I}_{nN}(\delta r)=\sum_{\ell=0}^\infty \sum_{j=0}^3 \int^1_{-1}dz\, G_{nj\ell}(\delta r,z)\P_\ell (z).
\eeq 
We note, again referring to Eq.\ (\ref{Irec}) that $G_{nj\ell}(\delta r,z)$ can be written as a sum of a few terms, each of the form $K_{nj\ell}(\delta r)\tilde G_{nj}(z)$, where $\tilde G_{nj}(z)$ is smooth and independent of $\ell$, and $K_{nj\ell}(\delta r)$ can be uniformly bounded at large $\ell$ by $|K_{nj\ell}|<\ell^\alpha$ with some power $\alpha$. Consequently, each of the Legendre integrals in (\ref{sum}) is guaranteed to decay faster than any power of $1/\ell$, even at $\delta r=0 $. Uniform convergence then follows from the Weierstrass $M$-test: take $M_\ell$ to be 
\beq
M_\ell=\sum_{j=0}^3\sup_{\delta r\in [0,c]}\left|\int^1_{-1}dz\, G_{nj\ell}(\delta r,z)\P_\ell (z)\right|,
\eeq
which, because of the exponential decay of the Legendre integrals for all $\delta r\in [0,c]$, has a convergent sum $\sum_\ell M_\ell$. The $M$-test then implies that the sum~\eqref{sum} converges uniformly on the interval $\delta r\in[0,c]$. 

This establishes the last of the conditions of the lemma, thereby proving (\ref{DeltaI=0}).

{\em Proof of lemma}. We now provide the proof of the lemma. We begin by showing $\lim_{\delta r\to0}g(\delta r,z)=0$ on a sequence of closed subsets of $\check S$, and then we take the union of these sets to show the result on the whole of $\check S$. 

Let $\check S_k=[-1,z_0-1/k]\cup[z_0+1/k,1]$, where $k\in\mathbb{N}^+$, and let $D_k=\{\phi_k\}$ be the set of smooth test functions with support $\text{supp}(\phi_k)\subset \check S_k$. Since each $g_\ell(\delta r,z)$ is continuous on $[0,c]\times \check S_k$, each $\tilde g_N(\delta r)$ is as well. From this fact and the uniform convergence of $\tilde g_N\to \tilde g$, it follows that $\tilde g(\delta r)$ is continuous on that same interval of $\delta r$, and in particular, at $\delta r=0$. Ergo, if we specialize to any of the test functions $\phi_k\in D_k$,
\begin{align}
\tilde g(0) &= \lim_{\delta r\to0}\tilde g(\delta r)= \lim_{\delta r\to0}\int_{\check S_k} dz\, \phi_k(z)g(\delta r,z).\label{continuity}
\end{align}
We now bring the limit inside the integral by appealing to the dominated convergence theorem, which in the present case states that 
\begin{equation}\label{interchange}
\lim_{\delta r\to0}\int_{\check S_k} dz\, \phi_k(z)g(\delta r,z)=\int_{\check S_k} dz\, \phi_k(z)\lim_{\delta r\to0}g(\delta r,z)
\end{equation}
if two criteria are met: $\lim_{\delta r\to0}g(\delta r,z)$ exists and is finite almost everywhere in $\check S_k$; and there exists an integrable function $f(z)$ satisfying $|g(\delta r,z)|\leq f(z)$ for almost all $z\in \check S_k$ and for all $\delta r\in(0,c]$. Condition (iii) of the lemma guarantees that the first criterion is met. To see that the second criterion is also met, consider the function 
\begin{equation}
g^*(\delta r, z) = \begin{cases} g(\delta r,z) & \text{if } \delta r\in (0,c] \text{ and } z\in \check S_k\\
																	\lim_{\delta r\to 0}g(\delta r,z) & \text{if } \delta r=0 \text{ and } z\in \check S_k
\end{cases} 
\end{equation}
and take the dominating function to be the constant function $f(z)=\sup|g^*(\delta r,z)|$, which by construction satisfies $f(z)\geq |g(\delta r,z)|$ for all $\delta r\in(0,c]$. The finiteness of the supremum can be proved as follows: Since $\sum_{\ell=0}^\infty g_{\ell}(\delta r,z)$ converges uniformly on $[b,c]\times \check S_k$ for all $b\in(0,c)$, and each $g_{\ell}(\delta r,z)$ is continuous on that domain, $g(\delta r,z)$ is continuous on all such sets as well. Therefore $\lim_{(\delta r,z)\to(\delta r^*,z^*)}|g^*(\delta r,z)|$ is finite for all $\delta r^*\in(0,c]$ and $z^*\in \check S_k$. And $\lim_{(\delta r,z)\to(0,z^*)}|g^*(\delta r,z)|$ is finite by hypothesis. Hence, $\lim_{(\delta r,z)\to(\delta r^*,z^*)}|g^*(\delta r,z)|$ is finite for all $(\delta r,z)\in[0,c]\times \check S_k$. But if $\sup|g^*(\delta r,z)|$ were not finite, then there would exist $(\delta r^*,z^*)$ such that $\lim_{(\delta r,z)\to(\delta r^*,z^*)}|g^*(\delta r,z)|=\infty$. Therefore, $\sup|g^*(\delta r,z)|<\infty$. Because the integration domain $\check S_k$ is finite, $f(z)$ is also integrable, and the criteria for the dominated convergence theorem have been met.

Now, since $\tilde g(0)=0$ for all $\phi_k\in D_k$, the equalities \eqref{continuity} and \eqref{interchange} together show
\begin{equation}
\int_{\check S_k} dz\, \phi_k(z)\lim_{\delta r\to0} g(\delta r,z) = 0
\end{equation}
for all test functions $\phi_k\in D_k$. It follows that $\lim_{\delta r\to0} g(\delta r,z)=0$ for almost all $z\in \check S_k$. This leaves the possibility that $\lim_{\delta r\to0} g(\delta r,z)$ is nonzero on some set of measure zero in $\check S_k$. But by hypothesis, $\lim_{\delta r\to0} g(\delta r,z)$ is continuous in $\check S_k$. Therefore, $\lim_{\delta r\to0} g(\delta r,z)=0$ for all $z\in \check S_k$.

Since this result holds in each $\check S_k$, it also holds in their union $\bigcup_{k\in\mathbb{N}^+}\check S_k=[-1,z_0)\cup(z_0,1]=\check S$, which completes the proof.

\section{Summation formulas}\label{a:sums}

We derive here the summation formulas (\ref{summation1}) and (\ref{summation2}). The sums in question are 
\begin{eqnarray}\label{sums}
\sigma_1(\theta,\theta_0):&=&\sum_{\ell=2}^{\infty}\frac{{}_{2}\! Y_{\ell}\!(\theta){}_{2}\! Y_{\ell}\!(\theta_0)}{\ell(\ell+1)},
\nonumber \\ 
\sigma_2(\theta,\theta_0):&=&\sum_{\ell=2}^{\infty}\frac{{}_{2}\! Y_{\ell}\!(\theta){}_{2}\! Y_{\ell}\!(\theta_0)}{(\ell+2)(\ell-1)},
\end{eqnarray}
where ${}_2\!Y_{\ell}\!(\theta)\equiv{}_2\!Y_{\ell0}\!(\theta)$ are spin-weighted spherical harmonics with spin $s=0$ and azimuthal number $m=0$, $\theta\in[0,\pi]$, and $\theta_0\in[\theta_1,\pi-\theta_1]$ for some $0<\theta_1<\pi/2$. (For our completion calculation we require $\theta_0$ in the immediate neighbourhood of $\pi/2$ only, but our derivation will apply equally for any $\theta_1$ in the above domain; the requirement $\theta_1>0$ is non-essential but will simplify our analysis somewhat.) For easy reference, we call the above two-dimensional domain $S_1$. We note that each of the two sums converges {\em uniformly} on $S_1$ (a proof will be provided at the end of this appendix), so the sums $\sigma_1(\theta,\theta_0)$ and $\sigma_2(\theta,\theta_0)$ are continuous functions in this domain.

Starting with $\sigma_1$, we first recall that ${}_2\!Y_\ell(\theta)$ satisfies the differential equation 
\begin{equation}\label{Y2Equation}
\frac{1}{\sin\theta}\frac{d}{d\theta}\left(\sin\theta\, \frac{{}_2\!Y_\ell(\theta)}{d\theta}\right)+
\left(\ell(\ell+1)-\frac{4}{\sin^2\theta}\right){}_2\!Y_\ell(\theta)=0,
\end{equation}
which is the reduction of (\ref{eq:kerrang}) to $m=\omega=0$ with $s=2$. Applying the operator $(\sin\theta)^{-1}\partial_\theta\left(\sin\theta\partial_\theta\right)$ to $\sigma_1$ thus gives 
\begin{multline}\label{sigma1Eq0}
\frac{1}{\sin\theta}\frac{d}{d\theta}\left(\sin\theta\, \frac{\sigma_1}{d\theta}\right)
=\\
\sum_{\ell=2}^{\infty}\frac{{}_2\!Y_\ell(\theta_0){}_2\!Y_\ell(\theta)}{\ell(\ell+1)}\left(\frac{4}{\sin^2\theta}-\ell(\ell+1)\right)
=\\
 \frac{4}{\sin^2\theta}\, \sigma_1-\sum_{\ell=2}^{\infty}{}_2\!Y_\ell(\theta_0){}_2\!Y_\ell(\theta),
\end{multline}
or, using the completeness relation (\ref{closure}),
\begin{equation}\label{sigma1Eq}
\frac{1}{\sin\theta}\frac{d}{d\theta}\left(\sin\theta\, \frac{\sigma_1}{d\theta}\right)
-\frac{4}{\sin^2\theta}\, \sigma_1 = -(2\pi)^{-1}\delta(\cos\theta-\cos\theta_0).
\end{equation}
Here we are considering $\sigma$ as a {\it distribution}, necessary for making sense of the term-by-term differentiation applied in the second line of (\ref{sigma1Eq0}) [even though the sums in Eqs.\ (\ref{sums}) converge uniformly, the sums of the derivatives of the summands with respect to either $\theta$ or $\theta_0$ do not converge at all as functions]. Equation (\ref{sigma1Eq}) is a simple ordinary differential equation for $\sigma_1(\theta)$ (with $\theta_0$ regarded as a fixed parameter), and we seek a solution that is continuous on $S_1$. 

Two independent homogeneous solutions of (\ref{sigma1Eq}) are $\tan^2(\theta/2)$ and $\cot^2(\theta/2)$, the first of which blows up at $\theta=\pi$ and the other at $\theta=0$. It follows that a {\em unique} globally continuous solution is given by the distribution
\begin{multline}
\sigma_1=A_1(\theta_0)\tan^2(\theta/2)\Theta(\cos\theta-\cos\theta_0)
\\
+B_1(\theta_0)\cot^2(\theta/2)\Theta(\cos\theta_0-\cos\theta),
\end{multline}
where $\Theta(\cdot)$ is the Heaviside step function, and the coefficients $A_1$ and $B_1$ are determined from the continuity condition $\sigma(\theta\to\theta_0^+)=\sigma(\theta\to\theta_0^-)$ together with the jump condition $\sigma(\theta\to\theta_0^+)-\sigma(\theta\to\theta_0^-)=-(2\pi\sin\theta_0)^{-1}$. We find $A_1=(8\pi)^{-1}\cot^2(\theta_0/2)$ and $B_1=(8\pi)^{-1}\tan^2(\theta_0/2)$, and thus obtain 
\begin{multline}\label{sigma1}
\sigma_1=\frac{1}{8\pi}\cot^2(\theta_0/2)\tan^2(\theta/2)\Theta(\cos\theta-\cos\theta_0)
\\
+\frac{1}{8\pi}\tan^2(\theta_0/2)\cot^2(\theta/2)\Theta(\cos\theta_0-\cos\theta),
\end{multline}
which (as a check) is symmetric under $\theta\leftrightarrow\theta_0$ as it should be. The summation formula (\ref{summation1}) reexpresses (\ref{sigma1}) in a more compact form.

The evaluation of the sum $\sigma_2$ follows analogously. Writing $(\ell+2)(\ell-1)=\ell(\ell+1)-2$ and applying $(\sin\theta)^{-1}\partial_\theta\left(\sin\theta\partial_\theta\right)$ to $\sigma_2$, we obtain the differential equation  
\begin{multline}\label{sigma2Eq}
\frac{1}{\sin\theta}\frac{d}{d\theta}\left(\sin\theta\, \frac{\sigma_2}{d\theta}\right)
-\left(\frac{4}{\sin^2\theta}-2\right)\sigma_2 
=\\
 -(2\pi)^{-1}\delta(\cos\theta-\cos\theta_0).
\end{multline}
Two independent homogeneous solutions are $(2+\cos\theta)\tan^2(\theta/2)$ and $(2-\cos\theta)\cot^2(\theta/2)$, and the unique globally continuous solution has the form 
\begin{multline}\label{sigma2AB}
\sigma_2=A_2(\theta_0)(2+\cos\theta)\tan^2(\theta/2)\Theta(\cos\theta-\cos\theta_0)
\\
+B_2(\theta_0)(2-\cos\theta)\cot^2(\theta/2)\Theta(\cos\theta_0-\cos\theta),
\end{multline}
where $A_2$ and $B_2$ are determined from the same two continuity and jump conditions at $\theta=\theta_0$ as above. After substituting back in (\ref{sigma2AB}), the result is 
\begin{multline}\label{sigma2}
\sigma_2=\frac{1}{24\pi}(2-\cos\theta_0)\cot^2(\theta_0/2)(2+\cos\theta)\tan^2(\theta/2)
\\
\times \Theta(\cos\theta-\cos\theta_0)
\\
+\frac{1}{8\pi}(2+\cos\theta_0)\tan^2(\theta_0/2)(2-\cos\theta)\cot^2(\theta/2)
\\
\times \Theta(\cos\theta_0-\cos\theta),
\end{multline}
expressed more compactly in Eq.\ (\ref{summation2}).

\subsection{Proof of uniform convergence}

The above derivation relied on the assumption that $\sigma_1$ and $\sigma_2$ are each continuous on $S_1$, which, in turn, relied on a statement of uniform convergence of the sums in Eq.\ (\ref{sums}). We now prove that statement. 

First, let us note the relation 
\begin{equation}
 {}_{2}\! Y_{\ell}\!(\theta)=\sqrt{\frac{2\ell+1}{4\pi\lambda_2}}\, {\sf P}_\ell^{m=2}(\cos\theta),
\end{equation}
where, recall, ${\sf P}_\ell^{m}$ is the associated Legendre function of the first kind, and $\lambda_2=(\ell+2)!/(\ell-2)!$. The functions ${\sf P}_\ell^{m=2}$ admit the global, $\theta$-independent upper bound $|{\sf P}_\ell^{m=2}|\leq (\lambda_2/2)^{1/2}$, valid for all $\theta\in[0,\pi]$ \cite{Lohoefer}.
Thus
\begin{equation}\label{bound1}
\left|{}_{2}\! Y_{\ell}\!(\theta)\right|\leq\sqrt{(2\ell+1)/(8\pi)}<\sqrt{\ell}
\end{equation}
for all $\theta\in S_1$ and $\ell\geq 2$. For ${}_{2}\! Y_{\ell}\!(\theta_0)$ we instead invoke the more standard bound $|{\sf P}_\ell^{m=2}|<\sqrt{8/(\pi \ell)}(\ell+2)(\ell+1)(\sin\theta)^{-5/2}$
\cite{Lohoefer}, from which we obtain 
\begin{equation}\label{bound2}
\left|{}_{2}\! Y_{\ell}\!(\theta_0)\right|<2(\sin\theta_1)^{-5/2}
\end{equation}
for all $\theta_0\in S_1$ and $\ell\geq 2$. It follows from the combination of (\ref{bound1}) and (\ref{bound2}) that each of the two summands in Eq.\ (\ref{sums}) is bounded from above by the numerical sequence $a_\ell=2\ell^{-3/2}(\sin\theta_1)^{-5/2}$, which admits a convergent sum. Both sums in Eq.\ (\ref{sums}) are therefore uniformly convergent on $S_1$ by Weierstrass's M-test theorem.

\raggedright
\bibliography{biblio}
\end{document}